%\Overfullrule=0pt
 
\newcount\mgnf  %ingrandimento
\mgnf=0
 
\ifnum\mgnf=0
\def\openone{\leavevmode\hbox{\ninerm 1\kern-3.3pt\tenrm1}}%
\def\*{\vglue0.3truecm}\fi
\ifnum\mgnf=1
\def\openone{\leavevmode\hbox{\ninerm 1\kern-3.63pt\tenrm1}}%
\def\*{\vglue0.5truecm}\fi
 
\ifnum\mgnf=0
   \magnification=\magstep0
   \hsize=14truecm\vsize=24.truecm
   \parindent=0.3cm\baselineskip=18pt
\font\titolo=cmbx12
\font\titolone=cmbx10 scaled\magstep 2
\font\cs=cmcsc10
\font\ottorm=cmr8

\font\msytw=msbm10
\font\msytww=msbm8

\font\indbf=cmbx10 scaled\magstep1
\font\grbold=cmmib10
\fi
\ifnum\mgnf=1
   \magnification=\magstep1\hoffset=0.truecm
   \hsize=14truecm\vsize=24.truecm
   \baselineskip=18truept plus0.1pt minus0.1pt \parindent=0.9truecm
   \lineskip=0.5truecm\lineskiplimit=0.1pt      \parskip=0.1pt plus1pt
\font\titolo=cmbx12 scaled\magstep 1
\font\titolone=cmbx10 scaled\magstep 3
\font\cs=cmcsc10 scaled\magstep 1
\font\ottorm=cmr8 scaled\magstep 1

\font\msytw=msbm10 scaled\magstep1
\font\msytww=msbm8 scaled\magstep1

\font\grbold=cmmib10 scaled\magstep1
\font\indbf=cmbx10 scaled\magstep2
\fi
 
\global\newcount\numsec\global\newcount\numapp
\global\newcount\numfor\global\newcount\numfig\global\newcount\numsub
\numsec=0\numapp=0\numfig=1
\def\veroparagrafo{\number\numsec}\def\veraformula{\number\numfor}
\def\veraappendice{\number\numapp}\def\verasub{\number\numsub}
\def\verafigura{\number\numfig}
 
\def\section(#1,#2){\advance\numsec by 1\numfor=1\numsub=1%
\SIA p,#1,{\veroparagrafo} %
\write15{\string\Fp (#1){\secc(#1)}}%
\write16{ sec. #1 ==> \secc(#1)  }%
\hbox to \hsize{\titolo\hfill \number\numsec. #2\hfill%
\expandafter{\alato(sec. #1)}}\*}
 
\def\appendix(#1,#2){\advance\numapp by 1\numfor=1\numsub=1%
\SIA p,#1,{A\veraappendice} %
\write15{\string\Fp (#1){\secc(#1)}}%
\write16{ app. #1 ==> \secc(#1)  }%
\hbox to \hsize{\titolo\hfill Appendix A\number\numapp. #2\hfill%
\expandafter{\alato(app. #1)}}\*}
 
\def\senondefinito#1{\expandafter\ifx\csname#1\endcsname\relax}
 
\def\SIA #1,#2,#3 {\senondefinito{#1#2}%
\expandafter\xdef\csname #1#2\endcsname{#3}\else
\write16{???? ma #1#2 e' gia' stato definito !!!!} \fi}
 
\def \Fe(#1)#2{\SIA fe,#1,#2 }
\def \Fp(#1)#2{\SIA fp,#1,#2 }
\def \Fg(#1)#2{\SIA fg,#1,#2 }
 
\def\etichetta(#1){(\veroparagrafo.\veraformula)%
\SIA e,#1,(\veroparagrafo.\veraformula) %
\global\advance\numfor by 1%
\write15{\string\Fe (#1){\equ(#1)}}%
\write16{ EQ #1 ==> \equ(#1)  }}
 
\def\etichettaa(#1){(A\veraappendice.\veraformula)%
\SIA e,#1,(A\veraappendice.\veraformula) %
\global\advance\numfor by 1%
\write15{\string\Fe (#1){\equ(#1)}}%
\write16{ EQ #1 ==> \equ(#1) }}
 
\def\getichetta(#1){Fig. \verafigura%
\SIA g,#1,{\verafigura} %
\global\advance\numfig by 1%
\write15{\string\Fg (#1){\graf(#1)}}%
\write16{ Fig. #1 ==> \graf(#1) }}
 
\def\etichettap(#1){\veroparagrafo.\verasub%
\SIA p,#1,{\veroparagrafo.\verasub} %
\global\advance\numsub by 1%
\write15{\string\Fp (#1){\secc(#1)}}%
\write16{ par #1 ==> \secc(#1)  }}
 
\def\etichettapa(#1){A\veraappendice.\verasub%
\SIA p,#1,{A\veraappendice.\verasub} %
\global\advance\numsub by 1%
\write15{\string\Fp (#1){\secc(#1)}}%
\write16{ par #1 ==> \secc(#1)  }}
 
\def\Eq(#1){\eqno{\etichetta(#1)\alato(#1)}}
\def\eq(#1){\etichetta(#1)\alato(#1)}
\def\Eqa(#1){\eqno{\etichettaa(#1)\alato(#1)}}
\def\eqa(#1){\etichettaa(#1)\alato(#1)}
\def\eqg(#1){\getichetta(#1)\alato(fig. #1)}
\def\sub(#1){\0\palato(p. #1){\bf \etichettap(#1)\hskip.3truecm}}
\def\asub(#1){\0\palato(p. #1){\bf \etichettapa(#1)\hskip.3truecm}}
 
\def\equv(#1){\senondefinito{fe#1}$\clubsuit$#1%
\write16{eq. #1 non e' (ancora) definita}%
\else\csname fe#1\endcsname\fi}
\def\grafv(#1){\senondefinito{fg#1}$\clubsuit$#1%
\write16{fig. #1 non e' (ancora) definito}%
\else\csname fg#1\endcsname\fi}
\def\secv(#1){\senondefinito{fp#1}$\clubsuit$#1%
\write16{par. #1 non e' (ancora) definito}%
\else\csname fp#1\endcsname\fi}
 
\def\equ(#1){\senondefinito{e#1}\equv(#1)\else\csname e#1\endcsname\fi}
\def\graf(#1){\senondefinito{g#1}\grafv(#1)\else\csname g#1\endcsname\fi}
\def\secc(#1){\senondefinito{p#1}\secv(#1)\else\csname p#1\endcsname\fi}
\def\sec(#1){{\S\secc(#1)}}
 
\def\BOZZA{
\def\alato(##1){\rlap{\kern-\hsize\kern-1.2truecm{$\scriptstyle##1$}}}
\def\palato(##1){\rlap{\kern-1.2truecm{$\scriptstyle##1$}}}
}
 
\def\alato(#1){}
\def\galato(#1){}
\def\palato(#1){}

{\count255=\time\divide\count255 by 60 \xdef\hourmin{\number\count255}
        \multiply\count255 by-60\advance\count255 by\time
   \xdef\hourmin{\hourmin:\ifnum\count255<10 0\fi\the\count255}}
 
\def\oramin{\hourmin }
 
\def\data{\number\day/\ifcase\month\or gennaio \or febbraio \or marzo \or
aprile \or maggio \or giugno \or luglio \or agosto \or settembre
\or ottobre \or novembre \or dicembre \fi/\number\year;\ \oramin}
\setbox200\hbox{$\scriptscriptstyle \data $}
\footline={\rlap{\hbox{\copy200}}\tenrm\hss \number\pageno\hss}

\let\a=\alpha \let\b=\beta  \let\g=\gamma     \let\d=\delta  \let\e=\varepsilon
\let\z=\zeta  \let\h=\eta   \let\th=\vartheta \let\k=\kappa   \let\l=\lambda
\let\m=\mu    \let\n=\nu    \let\x=\xi        \let\p=\pi      \let\r=\rho
\let\s=\sigma \let\t=\tau        \let\c=\chi
   \let\o=\omega 
 \let\D=\Delta     \let\L=\Lambda  
           
\let\O=\Omega 
 
\def\\{\hfill\break} \let\==\equiv

\let\io=\infty \def\Dpr{\V\dpr\,}

\let\0=\noindent \def\pagina{{\vfill\eject}}

\def\ie{\hbox{\it i.e.\ }}
\let\dpr=\partial 
\let\bs=\backslash
 
\def\tende#1{\,\vtop{\ialign{##\crcr\rightarrowfill\crcr
 \noalign{\kern-1pt\nointerlineskip}
 \hskip3.pt${\scriptstyle #1}$\hskip3.pt\crcr}}\,}
\def\otto{\,{\kern-1.truept\leftarrow\kern-5.truept\to\kern-1.truept}\,}
\def\fra#1#2{{#1\over#2}}
 
\def\PP{{\cal P}}\def\EE{{\cal E}}\def\VV{{\cal V}}
\def\HH{{\cal H}}
\def\TT{{\cal T}}\def\NN{{\cal N}}
\def\RR{{\cal R}}\def\LL{{\cal L}}
\def\DD{{\cal D}}\def\AA{{\cal A}}
 
\def\T#1{{#1_{\kern-3pt\lower7pt\hbox{$\widetilde{}$}}\kern3pt}}
\def\VVV#1{{\underline #1}_{\kern-3pt
\lower7pt\hbox{$\widetilde{}$}}\kern3pt\,}
\def\W#1{#1_{\kern-3pt\lower7.5pt\hbox{$\widetilde{}$}}\kern2pt\,}

\def\mod{{\rm mod}\,}  
\def\indica{\leaders \hbox to 0.5cm{\hss.\hss}\hfill}
\def\guida{\leaders\hbox to 1em{\hss.\hss}\hfill}
\mathchardef\oo= "0521
 
\def\V#1{{\bf #1}}
\def\pp{{\bf p}}\def\xx{{\bf x}}
\def\yy{{\bf y}}\def\kk{{\bf k}}
\def\dd{{\bf d}}\def\zz{{\bf z}}\def\uu{{\bf u}}
\def\xxi{\hbox{\grbold \char24}} \def\bP{{\bf P}}\def\rr{{\bf r}}
\def\tt{{\bf t}}
\def\ss{{\underline \sigma}}\def\oo{{\underline \omega}}
\def\un{{\underline n}}
\def\xxx{{\underline\xx}}

\def\qed{\raise1pt\hbox{\vrule height5pt width5pt depth0pt}}

 \def\bh{{\bar h}}  

\def\indic{\hbox{\raise-2pt \hbox{\indbf 1}}}
\def\bk#1#2{\bar\kk_{#1#2}}

\def\RRR{\hbox{\msytw R}}

\def\NNN{\hbox{\msytw N}} 
 \def\ZZZ{\hbox{\msytw Z}}
\def\zzzz{\hbox{\msytww Z}} 
\def\TTT{\hbox{\msytw T}}

%%% INSERIMENTO FIGURE ( se si usa DVIPS )
%
% Se si vuole utilizzare delle macro postscript personali, contenute
% nel file ini.ps, togliere il commento alla riga seguente
%\special{header=ini.pst}
%
% Il comando seguente inserisce una scatola contenente #3 in modo che
% l'angolo superiore sinistro occupi la posizione (#1,#2)
%
\def\ins#1#2#3{\vbox to0pt{\kern-#2 \hbox{\kern#1 #3}\vss}\nointerlineskip}
%
% Il comando seguente crea una scatola di dimensioni #1x#2 contenente
% il disegno descritto in #4.ps;
% in questo disegno si possono introdurre delle stringhe usando \ins
% e mettendo le istruzioni relative nell'argomento #3.
% Il file #4.ps contiene le istruzioni postscript, che devono essere scritte
% presupponendo che l'origine sia nell'angolo inferiore sinistro della
% scatola, mentre per il resto l'ambiente grafico e' quello standard.
% #5 deve essere della forma \eqg("nome simbolico").
%
% Le istruzioni postscript possono essere inserite nel file che contiene
% l'istruzione \insertplot, racchiudendole fra le istruzioni \initfig{#4}
% e \endfig; inoltre ogni riga deve cominciare con "write13<" e deve finire
% con ">". In questo modo si crea il file #4.ps relativo alla figura.
%
\newdimen\xshift \newdimen\xwidth \newdimen\yshift
 
\def\insertplot#1#2#3#4#5{\par%
\xwidth=#1 \xshift=\hsize \advance\xshift by-\xwidth \divide\xshift by 2%
\yshift=#2 \divide\yshift by 2%
\line{\hskip\xshift \vbox to #2{\vfil%
#3 \includegraphics{#4.ps}}\hfill \raise\yshift\hbox{#5}}}
 
\def\initfig#1{%
\catcode`\%=12\catcode`\{=12\catcode`\}=12
\catcode`\<=1\catcode`\>=2
\openout13=#1.ps}
 
\def\endfig{%
\closeout13
\catcode`\%=14\catcode`\{=1
\catcode`\}=2\catcode`\<=12\catcode`\>=12}
 
%%%%%   FIGURE fig51.ps
 
\initfig{fig51}
\write13<%!>
\write13<%%BoundingBox 0 0 300 150>
\write13<gsave .5 setlinewidth 40 20 260 {dup 0 moveto 140 lineto} for stroke
grestore>
\write13</punto { gsave  % uso: x1 y1 punto>
\write13<2 0 360 newpath arc fill stroke grestore} def>
\write13<40 75 punto>
\write13<60 75 punto>
\write13<80 75 punto>
\write13<100 75 punto 120 68 punto 140 61 punto 160 54 punto 180 47 punto 200
40 punto>
\write13<220 33 punto 240 26 punto 260 19 punto>
\write13<120 82.5 punto>
\write13<140 90 punto>
\write13<160 80 punto>
\write13<160 100 punto>
\write13<180 110 punto>
\write13<180 70 punto>
\write13<200 60 punto>
\write13<200 120 punto>
\write13<220 110 punto>
\write13<220 50 punto>
\write13<240 100 punto>
\write13<240 60 punto>
\write13<120 50 punto>
\write13<260 20 punto>
\write13<240 40 punto>
\write13<240 50 punto>
\write13<260 70 punto>
\write13<200 80 punto>
\write13<260 90 punto>
\write13<260 110 punto>
\write13<220 130 punto>
\write13<40 75 moveto 100 75 lineto 140 90 lineto 200 120 lineto 220 130
lineto>
\write13<200 120 moveto 240 100 lineto 260 110 lineto>
\write13<240 100 moveto 260 90 lineto>
\write13<140 90 moveto 180 70 lineto 200 80 lineto>
\write13<180 70 moveto 220 50 lineto 260 70 lineto>
\write13<220 50 moveto 240 40 lineto>
\write13<220 50 moveto 240 50 lineto>
\write13<100 75 moveto 260 20 lineto>
\write13<100 75 moveto 120 50 lineto stroke>
\write13<grestore>
\endfig

\openin14=\jobname.aux \ifeof14 \relax \else
\input \jobname.aux \closein14 \fi
\openout15=\jobname.aux

%\input fiat
%\BOZZA
%\vskip2.truecm
{\baselineskip=12pt
\centerline{\titolone Renormalization Group, hidden symmetries and}
\vskip.2truecm
\centerline{\titolone approximate Ward identities in the XYZ model, I.}
\vskip1.truecm
\centerline{{\titolo G. Benfatto{${}^\ast$}, V. Mastropietro}%
\footnote{${}^\ast$}{\ottorm Supported by MURST, Italy, and EC HCM contract
number CHRX-CT94-0460.\hfill\break
e-mail: benfatto@mat.uniroma2.it, mastropi@mat.uniroma2.it.}}
\centerline{Dipartimento di Matematica, Universit\`a di Roma ``Tor Vergata''}
\centerline{Via della Ricerca Scientifica, I-00133, Roma}
\vskip1.truecm
\line{\vtop{
\line{\hskip1.5truecm\vbox{\advance \hsize by -3.1 truecm
\0{\cs Abstract.}
{\it Using renormalization group methods, we study the Heis\-enberg-Ising
$XYZ$ chain in an external magnetic field directed as the $z$ axis, in the
case of small coupling $J_3$ in the $z$ direction. We study the
asymptotic behaviour of the spin spa\-ce-time correlation
function in the direction of the magnetic field and the singularities
of its Fourier transform.

The work is organized in two parts. In the present paper an expansion for
the ground state energy and the effective potential is derived, which is
convergent if the running coupling constants are small enough. In the
subsequent paper, by using hidden symmetries of the model, we show that
this condition is indeed verified, if $J_3$ is small enough, and we derive
an expansion for the spin correlation function. We also prove, by means of
an approximate Ward identity, that a critical index, related with the
asymptotic behaviour of the correlation function, is exactly vanishing. }}
\hfill} }}
}
\vskip1.2truecm
\section(1,Introduction)
 
\sub(1.1) If $(S^1_x,S^2_x,S^3_x)={1\over 2}(\s^1_x,\s^2_x,\s^3_x)$,
for $i=1,2,...,L$, $\s^\a_i$, $\a=1,2,3$, being the Pauli matrices, the
Hamiltonian of the {\it Heisenberg-Ising $XYZ$ chain} is given by
$$H=-\sum_{x=1}^{L-1} [J_1 S^1_x S^1_{x+1}+J_2 S^2_x S^2_{x+1}+
J_3 S^3_x S^3_{x+1}+h S^3_x] - h S^3_L + U^1_L \;,\Eq(1.1)$$
where the last term, to be fixed later, depends on the boundary conditions.
The space-time {\it spin correlation function} at temperature $\b^{-1}$
is given by
$$\O^\a_{L,\b}(\xx)=<S^\a_{\xx} S^\a_{\bf 0}>_{L,\b}-
<S^\a_{\xx}>_{L,\b}<S^\a_{\bf 0}>_{L,\b}\;,\Eq(1.2)$$
where $\xx=(x,x_0)$, $S^\a_{\xx}=e^{Hx_0}S^\a_x e^{-Hx_0}$ and $<.>_{L,\b}=
Tr[e^{-\b H} . ]/Tr[e^{-\b H}]$ denotes the expectation in the grand canonical
ensemble. We shall use also the notation
$\O^{\a}(\xx) \= \lim_{L,\b\to\io}\O^{\a}_{L,\b}(\xx)$.
 
The Hamiltonian \equ(1.1) can be written [LSM] as a {\it fermionic
interacting spinless Hamiltonian}. In fact, it is easy to check that the
operators
$$ a_x^\pm \= \left[\prod_{y=1}^{x-1} (-\s_y^3)\right] \s_x^\pm\Eq(1.3)$$
are a set of anticommuting operators and that,
if $\s_x^{\pm}=(\s_x^1\pm i \s_x^2)/2$, we can write
$$\s^-_x=e^{-i\pi \sum_{y=1}^{x-1} a^+_y a^-_y }  a^-_x\;,
\quad \s^+_x= a^+_x e^{i\pi\sum_{y=1}^{x-1} a^+_y a^-_y }\;,\quad
\s^3_x=2 a^+_x a^-_x-1\;.\Eq(1.4)$$
Hence, if we fix the units so that $J_1+J_2=2$ and we introduce
the {\it anisotropy} $u=(J_1-J_2)/(J_1+J_2)$, we get
$$\eqalign{
H &= \sum_{x=1}^{L-1}\left\{-\fra12 [ a^+_{x} a^-_{x+1}+ a_{x+1}^+
 a^-_{x}]-{u\over 2}[ a^+_{x} a^+_{x+1} +
 a^-_{x+1} a^-_{x}]-\right.\cr
&\left.-J_3( a^+_x a^-_x-{1\over 2})( a^+_{x+1} a^-_{x+1}-{1\over 2})
\right\}-h\sum_{x=1}^L ( a^+_x a^-_x-{1\over 2})+U_L^2\;,\cr}\Eq(1.5)$$
where $U_L^2$ is the boundary term in the new variables. We choose it
so that the fermionic Hamiltonian \equ(1.5) coincides with the Hamiltonian
of a fermion system on the lattice with periodic boundary conditions, that
is we put $U^2_L$ equal to the term in the first sum in the r.h.s. of
\equ(1.5) with $x=L$ and $ a_{L+1}^\pm= a_1^\pm$
(in [LMS] this choice for the $XY$ chain is called ``c-cyclic''). It is easy
to see that this choice corresponds to fix the boundary conditions for the
spin variables so that
$$U_L^1= - \fra12 [\s^+_L e^{i\pi\NN}\s^-_1 + \s^-_L e^{i\pi\NN}\s^+_1]
-\fra{u}2 [\s^+_L e^{i\pi\NN}\s_1^+ + \s^-_L e^{i\pi\NN}\s_1^-]
-{J_3\over 4} \s_L^3\s_1^3\;,\Eq(1.6)$$
where $\NN=\sum_{x=1}^L a^+_x a_x$.
Strictly speaking, with this choice $U^1_L$ does not look really like a
boundary  term, because $\NN$ depends on all the spins of the chain. However
$[(-1)^\NN,H]=0$; hence the Hilbert space splits up in
two subspaces on which $(-1)^\NN$ is equal to $1$ or to $-1$
and on each of these subspaces $U_L^1$ really depends only
on the boundary spins. One expects that, in the $L\to\io$ limit,
the correlation functions are independent on the boundary term, but we shall
not face here this problem.
 
\*
\sub(1.2) The Heisenberg $XYZ$ chain has been the subject of a very active
research over many years with a variety of methods.
 
A first class of results is based on the {\it exact solutions}. If one of the
three parameters is vanishing ({\bf e.g.} $J_3=0$), the model is called {\it
$XY$ chain}. Its solution is based on the fact that the hamiltonian, in the
fermionic form \equ(1.5), is quadratic in the fermionic fields, so that it can
be diagonalized (see [LSM], [LSM1]) by a Bogoliubov transformation.
If $u=0$, we get the free Fermi gas with Fermi momentum $p_F=\arccos(-h)$; if
$|u|>0$, it turns out that the energy spectrum has a gap at $p_F$.

The equal time correlation functions $\O^\a(x,0)$ were explicitly
calculated in [Mc] (even at finite $L$ and $\b$), in the case $h=0$, that is
$p_F=\p/2$. Note that, while $\O^3(\xx)$ coincides with the correlation
function of the density in the fermionic representation of the model,
$\O^1(\xx)$ and $\O^2(\xx)$ are given by quite complicated expressions.
It turns out, for example, that, if $|u|<1$, $\O^3(x,0)$ is of the following
form:
$$\O^3(x,0) = -{\a^{|x|}\over \p^2 x^2}\sin^2\left({\p x\over 2}\right)
F(-|x|\log\a,|x|)\;,\quad \a=(1-|u|)/(1+|u|)\;,\Eq(1.7)$$
where $F(\g,n)$ is a bounded function, such that, if $\g\le 1$,
$F(\g,n)=1+O(\g\log\g)+O(1/n)$, while, if $\g\ge 1$ and $n\ge 2\g$,
$F(\g,n)=\p/2+O(1/\g)$.
 
For $|h|>0$, it is not possible to get a so explicit expression for
$\O^3(x,0)$. However, it is not difficult to prove that, if $|u|<\sin p_F$,
$|\O^3(x,0)|\le \a^{|x|}$ and, if $x\not=0$ and $|ux|\le 1$
$$\O^3(x,0) = -{1\over \p^2 x^2}\sin^2(p_F x)
[1+O(|ux|\log |ux|)+O(1/|x|)]\;.\Eq(1.8)$$
Note that, if $u=0$, a very easy calculation shows that $\O^3(x,0)
= -(\p^2 x^2)^{-2} \sin^2 (p_F x)$.
 
We want to stress that the only case in which the correlation functions and
their asymptotic behaviour can be computed explicitly in a rigorous way is
just the $J_3=0$ case.
 
If two parameters are equal (e.g. $J_1=J_2$), but $J_3\not=0$, the model is
called $XXZ$ model. In the case $h=0$, it was solved in [YY] via the {\it
Bethe-ansatz}, in the sense that the Hamiltonian was diagonalized. However, it
was not possible till now to obtain the correlation functions from the exact
solution. Such solution is a particular case of the general solution of the
XYZ model by Baxter [B], but again {\it only in the case of zero magnetic
field}. The ground state energy has been computed and it has been proved that
there is a gap in the spectrum, which, if $J_1-J_2$ and $J_3$ are not too
large, is given approximately by (see [LP])
$$\D=8\pi {\sin\mu\over\mu}|J_1|
\left({|J_1^2-J_2^2|\over 16(J_1^2-J_3^2)}\right)^{\pi\over 2\mu}\Eq(1.9)$$
with $\cos\mu=-J_3/J_1$.
 
The solution is based on the fact that the $XYZ$ chain with periodic
boundary conditions is equivalent to the {\it eight vertex} model, in the
sense that $H$ is proportional to the logarithmic derivative with respect
to a parameter of the eight vertex transfer matrix, if a suitable
identification of the parameters is done, see [S], [B]. The eight vertex
model is obtained by putting arrows in a suitable way on a two-dimensional
lattice with $M$ rows, $L$ columns and periodic boundary conditions.
There are eight allowed vertices, and with each of them an energy is
associated in a suitable way (there are four different values of the energy).
With the above choice of the parameters and $T-T_c<0$ and small,
$u=O(|T-T_c|)$, so that the critical temperature of the eight vertex model
corresponds to no anisotropy in the $XYZ$ chain.
Moreover, see [JKM], the correlation function $C_x$ between two vertical
arrows in a row, separated by $x$ vertices, is given, in the limit $M\to\io$,
by $C_x=<S^2_0 S^2_x>$. However, an explicit expression for the correlation
functions cannot be derived for the $XYZ$ or the eight vertex model. In [JKM]
the correlation length of $C_x$ was computed heuristically under some physical
assumptions (an exact computation is difficult because it does not depend only
on the largest and the next to the largest eigenvalues). The result is
$\x^{-1}=(T-T_c)^{\pi\over 2\mu}$, if $\x$ is the correlation length.
One sees that the critical index of the correlation length is {\it non
universal}.
 
Another interesting observation is that the $XYZ$ model is equivalent to two
interpenetrating two-dimensional Ising lattices with nearest-neighbor coupling,
interacting via a four spins coupling (which is proportional to
$J_3$). The {\it four spin correlation function} is identical to $C_x$.
In the decoupling limit $J_3=0$ the two Ising lattices are independent and one
can see that the Ising model solution can be reduced to the diagonalization,
via a Bogoliubov transformation, of a quadratic Fermi Hamiltonian, see [LSM1].
 
Recent new results using the properties of the transfer matrix can be found in
[EFIK], in which an integro-difference equation for the correlation function
of the XXZ chain is obtained. It is however not clear how to deduce the
physical properties of the correlation function from this equation.
 
\*
\sub(1.3) Since it is very difficult to extract detailed information on the
behaviour of the correlation functions from the above exact solutions, the XYZ
model has been studied by quantum field theory methods, see [LP]. The idea is
to approximate the fermionic hamiltonian \equ(1.5) by the hamiltonian of the
{\it massive Thirring model}, describing a massive relativistic spinning
particle on the continuum $d=1$ space interacting with a local current-current
potential (for a heuristic justification of this approximation, see [A]).
 
As a relativistic field theory, the massive Thirring model is plagued by
ultraviolet divergences, which were absent in the original model,
defined on a lattice; one can heuristically remove this problem by introducing
"by hand" an ultraviolet cut-off. A way to introduce it could be to consider a
short-ranged instead of a local potential; if $J_1=J_2$, this means that we
have approximated the $XXZ$-chain with the {\it Luttinger model}, whose
correlation functions can be explicitly computed, see [ML], [BGM].
 
The Luttinger model is defined in terms of two fields $\psi_{\xx,\o}$, $\o=\pm
1$, and one expects that, if $|h|<1$ and $J_3$ is small enough, the large
distance asymptotic behaviour of $\O^3(\xx)$ is qualitatively similar to that
of the truncated correlation of the operator $\r_\xx=\psi^+_\xx \psi^-_\xx$,
where $\psi^\s_\xx=\sum_\o \exp(i\s\o p_F x) \psi_{\xx,\o}$, if
some ``reasonable'' relationship between the parameters of the two models is
assumed. One can make for instance the substitutions $\l\to-J_3$ and
$p_0^{-1}\to a=1$, if $\l$ is the coupling in the Luttinger model, $a$ is the
chain step and $p_0^{-1}$ is the potential range. Moreover, one expects that
it is possible to choose a constant $\n$ of order $J_3$, so that $h=h_0+\n$
and $p_F={\rm arccos}(J_3-h_0)$, see \sec(1.4) below.

Of course such identification is completely arbitrary, but one can hope that
for large distances the function $\O^3(\xx)$ has something to do with the
truncated correlation of $\r_\xx$, which can be obtained by the general
formula (2.5) of [BGM], based on the exact solution of [ML]. There is
apparently a problem, since the expectation of $\r_\xx$ is infinite; however,
it is possible to see that there exists the limit, as $\e_1,\e_2\to 0^+$, of
$[<\r_{\xx,\e_1} \r_{\yy,\e_2}>-<\r_{\xx,\e_1}> <\r_{\yy,\e_2}>]$, where
$\r_{\xx,\e}=\psi^+_{(x,x_0+\e)}\psi^-_{(x,x_0)}$, and it is natural to take
this quantity, let us call it $G(\xx-\yy)$, as the truncated correlation of
$\r_\xx$.
 
Let us define $v_0=\sin p_F$; from (2.5) of [BGM] (by inserting a missing
$(-\e_i\e_j)$ in the last sum), it follows that, for $|\xx|\to\io$
$$G(\xx)\simeq [1+\l a_1(\l)]{\cos (2 p_F x)\over 2\p^2 [(v_0^*x_0)^2+x^2]^
{1+\l a_3(\l)}}+ {(v_0 x_0)^2-x^2\over 2\p^2[(v_0 x_0)^2+x^2]^2}\;,\Eq(1.10)$$
where $v_0^*=v_0[1+\l a_2(\l)]$ and $a_i(\l)$, $i=1,2,3$, are bounded
functions. Note that, in the second term in the r.h.s. of \equ(1.10), the bare
Fermi velocity $v_0$ appears, instead of the renormalized one, $v^*_0$, as one
could maybe expect.
 
In the physical literature, it is more usual the introduction of other
ultraviolet cutoffs, such that the resulting model is not exactly soluble,
even if $J_1=J_2$; however, it can be studied heuristically, see [LP], and the
resulting density-density correlation function is more or less of the form
\equ(1.10).
 
If $J_1\not=J_2$, there is no soluble model suitable for a similar analysis
of the large distance behaviour of $\O^3(\xx)$. However, one can guess
that the asymptotic behaviour is still of the form \equ(1.10), if
$1<<|\xx|<<1/|u|^\a$, for some $\a$. We shall prove that this is indeed true,
with $\a=1+O(J_3)$.
 
\*
 
\sub(1.4) 
In this paper we develop a rigorous renormalization group analysis for the
$XYZ$ Hamiltonian in its fermionic form (some ``not optimal'' bounds for the
correlation function $\O^3(\xx)$ were already found in [M2]). As we said
before, $\O^3(\xx)$ can be obtained from the exact solution only in the case
$J_3=0$, when the fermionic theory is a non interacting one. In particular, if
$\xx=(x,0)$ and $|ux|<<1$, \equ(1.8) and a more detailed analysis of the
``small'' terms in the r.h.s. (in order to prove that their derivatives of
order $n$ decay as $|x|^{-n}$), show that $\O^3(x,0)$ is a sum of
``oscillating'' functions with frequencies $(n p_F)/\p\,\mod 1$, $n=0,\pm 1$,
where $p_F=\arccos(-h)$; this means that its Fourier transform has to be a
smooth function, even for $u=0$, in the neighborhood of any momentum
$k\not=0,\pm 2p_F$. These frequencies are proportional to $p_F$, so they
depend only on the external magnetic field $h$.

If $J_3\not=0$, a similar property is satisfied for the leading terms
in the asymptotic behaviour, as we shall prove, but the value of $p_F$
depends in general also on $u$ and $J_3$. For example, if $u=0$, the
Hamiltonian \equ(1.5) is equal, up to a constant, to the Hamiltonian of a free
fermion gas with Fermi momentum $p_F=\arccos(J_3-h)$ plus an interaction term
proportional to $J_3$. As it is well known, the interaction modifies the Fermi
momentum of the system by terms of order $J_3$ and it is convenient (see [BG],
for example), in order to study the interacting model, to fix the Fermi
momentum to an interaction independent value, by adding a counterterm to the
hamiltonian. We proceed here in a similar way, that is we fix
$p_F$ and $h_0$ so that
$$h=h_0-\n\;,\qquad \cos p_F=J_3-h_0\;,\Eq(1.10a)$$
and we look for a value of $\n$, depending on $u,J_3,h_0$, such that,
as in the $J_3=0$ case, the leading terms in the asymptotic behaviour
of $\O^3_{L,\b}(\xx)$ can be represented as a sum of oscillating functions
with frequencies $(n p_F)/\pi\,\mod 1$, $n=0,\pm 1$.

As we shall see, we can realize this program only if $J_3$ is small enough and
it turns out that $\n$ is of order $J_3$. It follows that we can only consider
magnetic fields such that $|h|<1$. Moreover, it is clear that
the equation $h=h_0-\n(u,J_3,h_0)$ can be inverted, once the function
$\n(u,J_3,h_0)$ has been determined, so that $p_F$ is indeed a function of the
parameters appearing in the original model.

If $J_1=J_2$, it is conjectured, on the base of heuristic calculations, that
to fix $p_F$ is equivalent to the impose the condition that, in the limit
$L,\b\to\io$, the density is fixed (``Luttinger Theorem'') to the free model
value $\r=p_F/\p$. Remembering that $\r-{1\over 2}$ is the magnetization in
the $3$-direction for the original spin variables, this would mean that to fix
$p_F$ is equivalent to fix the magnetization in the $3$ direction, by suitably
choosing the magnetic field.

If $J_1\not= J_2$, there is in any case no simple relation between $p_F$ and
the mean magnetization, as one can see directly in the case $J_3=0$, where one
can do explicit calculations. The only exception is the case $p_F=\p/2$, where
one can see that, in the limit $L\to\io$, $\nu=J_3$ (so that $h=0$ by
\equ(1.10a)) and that $<S^3_x>=0$. This last property easily follows from the
observation that, if one choose $h=0$ in the original Hamiltonian \equ(1.1),
then the expectation of $S^3_x$ has to be equal to zero, by symmetry reasons,
up to terms which go to $0$ for $L\to\io$.

\*\*

Our main achievement is an expansion of $\O_{L,\b}^3(\xx)$, to be derived in
paper II, which provides a very detailed and explicit description of it. We
state in the following theorem some of its properties, but we stress that many
other interesting properties of $\O_{L,\b}^3(\xx)$ can be extracted from the
expansion.
 
\*
 
\sub(1.5) {\cs Theorem.} {\it Suppose that the equations \equ(1.10a) are
satisfied and that $v_0=\sin p_F\ge \bar v_0>0$, for
some value of $\bar v_0$ fixed once for all, and let us define
$a_0=\min\{p_F/2,(\pi-p_F)/2\}$; then the following is true.
 
\0 a) There exists a constant $\e$, such that, if $(u,J_3)\in \AA$, with
$$\AA=\{(u,J_3):|u|\le {a_0\over 8(1+\sqrt{2})}\;,|J_3|\le \e\}\;,\Eq(1.11)$$
it is possible to choose $\nu$, so that $|\n|\le c|J_3|$, for some constant
$c$ independent of $L$, $\b$, $u$, $J_3$, $p_F$, and the spin correlation
function $\O_{L,\b}^3(\xx)$ is a bounded (uniformly in $L$, $\b$, $p_F$ and
$(u,J_3)\in\AA$) function of $\xx=(x,x_0)$, $x=1,\ldots,L$, $x_0\in [0,\b]$,
periodic in $x$ and $x_0$ of period $L$ and $\b$ respectively, continuous as a
function of $x_0$.
 
\0 b) We can write
$$\O_{L,\b}^3(\xx)=\cos(2 p_F x) \O_{L,\b}^{3,a}(\xx)
+\O_{L,\b}^{3,b}(\xx)+\O_{L,\b}^{3,c}(\xx)\;,\Eq(1.12)$$
with $\O_{L,\b}^{3,i}(\xx)$, $i=a,b,c$, continuous bounded functions,
which are infinitely times differentiable as functions of $x_0$, if
$i=a,b$. Moreover, there exist two constants $\h_1$ and $\h_2$ of the
form
$$\h_1=a_1 J_3+O(J_3^2),\qquad \h_2=-a_2 J_3+O(J_3^2)\;,\qquad\Eq(1.13)$$
$a_1$ and $a_2$ being positive constants, uniformly bounded in $L$, $\b$,
$p_F$ and $(u,J_3)\in\AA$, such that the following is true.
 
Let us define
$$\dd(\xx)=({L\over\pi}\sin({\pi x\over L}),{\b\over\pi}
\sin({\pi x_0\over\b}))\Eq(1.14)$$
and suppose that $|\dd(\xx)|\ge 1$. Then,
given any positive integers $n$ and $N$, there exist positive
constants $\th<1$ and $C_{n,N}$, independent of $L$, $\b$, $p_F$ and
$(u,J_3)\in\AA$, so that, for any integers $n_0, n_1\ge 0$ and putting
$n=n_0+n_1$,
$$|\dpr_{x_0}^{n_0} \bar\dpr_x^{n_1} \O_{L,\b}^{3,a}(\xx)|\le
{1\over |\dd(\xx)|^{2+2\h_1+n}} {C_{n,N}\over 1+[\D |\dd(\xx)|]^{N}}\;,
\Eq(1.15)$$
$$|\dpr_{x_0}^{n_0} \bar\dpr_x^{n_1} \O_{L,\b}^{3,b}(\xx)|\le
{1\over |\dd(\xx)|^{2+n}} {C_{n,N}\over 1+[\D |\dd(\xx)|]^{N}}\;,
\Eq(1.16)$$
$$|\O_{L,\b}^{3,c}(\xx)|\le {1\over |\dd(\xx)|^2} \left[
{1\over |\dd(\xx)|^{\th}} + {(\D|\dd(\xx)|)^{\th} \over
|\dd(\xx)|^{\min \{0, 2\h_1\} } }\right]
{C_{0,N}\over 1+[\D |\dd(\xx)|]^{N}}\;,\Eq(1.17)$$
where $\bar\dpr_x$ denotes the discrete derivative and
$$\D=\max \{|u|^{1+\h_2}, \sqrt{(v_0\b)^{-2} +L^{-2}} \}\;.\Eq(1.18)$$
 
\0 c) There exist the limits $\O^{3,i}(\xx)=\lim_{L,\b\to\io}\O_{L,\b}^{3,i}
(\xx)$, $\xx\in\ZZZ\times\RRR$; they satisfy the bounds \equ(1.15),
with $|\xx|$ in place of $|\dd(\xx)|$.
Moreover, $\O^{3,a}(\xx)$ and $\O^{3,b}(\xx)$ are even functions of
$\xx$ and there exists a constant $\d^*$, of order $J_3$, such that, if
$1\le |\xx|\le \D^{-1}$ and $v^*_0=v_0(1+\d^*)$, given any $N>0$
$$\eqalign{
\O^{3,a}(\xx) &= {1+A_1(\xx)\over 2\p^2[x^2+(v^*_0 x_0)^2]^{1+\h_1}}\;,\cr
\O^{3,b}(\xx) &= {1\over 2\p^2[x^2+(v^*_0 x_0)^2]}\Big\{ {x_0^2-(x/v^*_0)^2
\over x^2+(v^*_0 x_0)^2} + A_2(\xx)\Big\}\;,\cr}\Eq(1.19)$$
$$|A_i(\xx)| \le C_N \left\{ {1\over 1+|\xx|^N} +
|J_3|+(\D|\xx|)^{1/2} \right\}\;,\Eq(1.20)$$
for some constant $C_N$.
 
The function $\O^{3,a}(\xx)$ is the restriction to $\ZZZ\times\RRR$ of
a function on $\RRR^2$, satisfying the symmetry relation
$$\O^{3,a}(x,x_0) = \O^{3,a}\Big(x_0 v_0^*,{x\over v_0^*}\Big)\;.\Eq(1.21)$$
 
\0 d) Let $\hat\O^3(\kk)$, $\kk=(k,k_0)\in [-\p,\p]\times \RRR^1$, the Fourier
transform of $\O^3(\xx)$. For any fixed $\kk$ with $\kk\not=(0,0), (\pm 2
p_F,0)$, $\hat\O^3(\kk)$ is uniformly bounded as $u\to 0$; moreover, for some
constant $c_2$,
$$\eqalign{
|\hat\O^3(0,0)|&\le c_2\left[1+ |J_3| \log{1\over\D}\right]\;,\cr
|\hat\O^3(\pm 2 p_F,0)|&\le c_2 {1-\D^{2\h_1}\over 2\h_1}\;.\cr}\Eq(1.22)$$
Finally, if $u=0$, $|\hat\O^3(\kk)|\le c_2[1+ |J_3| \log |\kk|^{-1}]$ near
$\kk=(0,0)$, and, at $\kk=(\pm 2p_F,0)$, it is singular only if $J_3<0$; in
this case it diverges as $|\kk-(\pm 2p_F,0)|^{2\h_1}/|\h_1|$.
 
\0 e) Let $G(x)=\O^3(x,0)$ and $\hat G(k)$ its Fourier transform. For any
fixed $k\not=0, \pm 2p_F$, $\hat G(k)$ is uniformly bounded as $u\to 0$,
together with its first derivative; moreover
$$\eqalign{
|\dpr_k\hat G(0)|&\le c_2\;,\cr
|\dpr_k\hat G(\pm 2 p_F)|&\le c_2 (1+\D^{2\h_1})\;.\cr}\Eq(1.23)$$
Finally, if $u=0$, $\dpr_k\hat G(k)$ has a first order discontinuity at $k=0$,
with a jump equal to $1+O(J_3)$, and, at $k=\pm 2p_F$, it is singular only if
$J_3<0$; in  this case it diverges as $|k-(\pm 2p_F)|^{2\h_1}$.
}
 
\* \sub(1.6) {\cs Remarks.}
 
a)The above theorem holds for any magnetic field $h$ such that $\sin
p_F>0$; remember that the exact solution given in
[B] is valid only for $h=0$. Moreover $u$ has not to be very small, but we
only need a bound of order $1$ on its value, see \equ(1.11); the only
perturbative parameter is $J_3$. However the interesting (and more
difficult) case is when also $u$ is small.
 
b)A naive estimate of $\e$ is $\e=c(\sin p_F)^\a$, with $c,\a$ positive
numbers; in other words we must take smaller and smaller $J_3$ for $p_F$
closer and closer to $0$ or $\pi$, \ie for magnetic fields of size close to
$1$. It is unclear at the moment if this is only a technical problem or a
property of the model.
 
c)If $J_1\not =J_2$ and $J_3\not =0$, one can distinguish, like in the
$J_3=0$ case \equ(1.7), two different regimes in the asymptotic behaviour
of the correlation function $\O^3(\xx)$, discriminated by an intrinsic
length $\xi$, which is approximately given by the inverse of spectral gap,
whose size, is of order $|u|^{1+\h_2}$, see \equ(1.18), in agreement with
\equ(1.9), found by the exact solution.

If $1<<|\xx|<<\xi$, the bounds for the correlation function are the same
as in the gapless $J_1=J_2$ case; if $\xi<<|\xx|$, there is a faster than
any power decay with rate of order $\xi^{-1}$. In the first region we can
obtain the exact large distance asymptotic behaviour of $\O^3(\xx)$, see
\equ(1.19),\equ(1.20); in the second region only an upper bound is
obtained. Note that, even in the $J_3=0$ case, it is not so easy to obtain
a more precise result, if $h\not=0$, see \sec(1.2).

The spin interaction in the $z$ direction has the effect that the gap
becomes anomalous, in the sense that it acquires a {\it critical index}
$\h_2$; the ratio between the ``renormalized'' and the ``bare'' gap is very
small or very large, if $u$ is small, depending on the sign of $J_3$. 
 
d)It is useful to compare the expression for the large distance behaviour
of $\O^3(\xx)$ in the case $u=0$ with its analogous for the Luttinger
model, see \sec(1.3). A first difference is that, while in the Luttinger
model the Fermi momentum is independent of the interaction, in the $XYZ$
model in general {\it it is changed non trivially} by the interaction,
unless the magnetic external field is zero, \ie $p_F={\pi\over
2}$. The reason is that the Luttinger model has special parity properties
which are not satisfied by the $XYZ$ chain (except if the
magnetic field is vanishing).
 
e)Another peculiar property of the Luttinger model correlation function is
that it depends on $p_F$ only through the factor $\cos(2 p_F x)$; this is true
not only for the asymptotic behaviour \equ(1.10), but also for the complete
expression given in [BGM], and is due to a special symmetry of the Luttinger
model (the Fermi momentum disappears from the Hamiltonian if a suitable
redefinition of the fermionic fields is done, see [BGM]). This property is of
course not true in the $XYZ$ model and in fact the dependence on $p_F$ of
$\O^3(\xx)$ is very complicated. However we prove that $\O^3(\xx)$ can be
written as sum of three terms, see \equ(1.12), and the first two terms are
very similar to the two terms in the r.h.s. of \equ(1.10). In particular, the
functions $\O^{3,a}(\xx)$ and $\O^{3,b}(\xx)$ have the same power decay as the
analogous functions in the Luttinger model and are ``free of oscillations'',
in the sense that each derivative increases the decay power of one unit, see
\equ(1.15),\equ(1.16).

This is not true for the third term $\O^{3,c}(\xx)$, which does not
satisfy a similar bound, because of the presence of oscillating
contributions. However we can prove that such term, if $u=0$, is
negligible for large distances, see \equ(1.17) (note that $\th$ is $J_3$
and $u$ independent, unlike $\h_1$). Of course this is true only for small
$J_3$ and it could be that $\O^{3,c}(\xx)$ plays an important role for
larger $J_3$.

If we compare, in the case $u=0$, the functions $\O^{3,a}(\xx)$ and
$\O^{3,b}(\xx)$, see \equ(1.19), with the corresponding ones in the Luttinger
model, see \equ(1.10), we see that they differ essentially for the non
oscillating functions $A_i(\xx)$, containing terms of higher order in our
expansion. However, this difference is not important in the case of
$\O^{3,a}(\xx)$, which also satisfies the same symmetry property \equ(1.21) as
the analogue in the Luttinger model, of course with different values of
$v_0^*$; note that the validity of \equ(1.21) allows to interpret $v_0^*$ as
the {\it renormalized Fermi velocity}. Guided by the analogy with the
Luttinger model, one would like to prove a similar property for $\O^{3a}(\xx)$
with $v_0$ replacing $v_0^*$; such property holds in fact for the Luttinger
model, see \equ(1.10). However we were not able to prove a similar properties
for $A_2(\xx)$, and this has some influence on our results, see below.
 
f)Another important property of the Luttinger model correlation function is
the fact that the ``not oscillating term'', that is the term corresponding
to $\O^{3,b}(\xx)$, {\it does not acquire a critical index}, contrary to
what happens for the term corresponding to $\cos(2p_F x)\O^{3,b}(\xx)$.
Hence one is naturally led to the conjecture that the critical
index of $\O^{3,b}_{L,\b}(\xx)$ is still vanishing, see for instance [Sp].
In our expansion, the critical index of $\O^{3,b}(\xx)$ is represented as a
convergent series, but, even if an explicit computation of the first order
term gives a vanishing result, it is not obvious that this is true at any
order. However, due to some hidden symmetries of the model (\ie symmetries
approximately enjoyed by the relevant part of the effective interaction),
we can prove a suitable {\it approximate Ward identity}, implying that all
the coefficients of the series are indeed vanishing.

g) The above properties can be used to study the Fourier transform $\hat
G(k)$ of the equal time correlation function $G(x)=\O^3(x,0)$. If $J_3=0$,
$\hat G(k)$ is bounded together with its first order derivative up to
$u=0$; in fact, the possible logarithmic divergence at $k=\pm 2 p_F$ and
$k=0$ (if $u=0$) of $\dpr\hat G(k)$ is changed by the parity properties
of $G(x)$ in a first order discontinuity.

If $J_3\not= 0$, $\dpr\hat G(k)$ behaves near $k=\pm 2 p_F$ in a completely
different way. In fact it is bounded and continuous if $J_3>0$, while it
has a power like singularity, if $u=0$ and $J_3<0$, see item e) of Theorem
\equ(1.5). This is due to the fact that the critical index $\h_1$,
characterizing the asymptotic behaviour of $\O^{3,a}(\xx)$, has the same
sign of $J_3$ (note that $\h_1$ has nothing to do with the critical index
$\h$ related with the two point fermionic Schwinger function, which is
$O((J_3)^2)$).

On the other hand, the behaviour of $\dpr\hat G(k)$ near $k=0$ is the same
for the Luttinger model, the $XYZ$ model and the free fermionic gas
($J_1=J_2$, $J_3=0$) (see also [Sp] for a heuristic explanation). This is
due to the vanishing of the critical index related with $\O^{3,b}(\xx)$ and
to the parity properties of the leading terms, which change, as in the
$J_3=0$ case, the apparent dimensional logarithmic divergence in a first
order discontinuity, 
 
h) If $u=0$, the (two dimensional) Fourier transform can be singular only at
$\kk=(0,0)$ and $\kk=(\pm 2p_F,0)$. If $J_3=0$, the singularity is
logarithmic at $\kk=(\pm 2p_F,0)$; if $J_3\not= 0$, the singularity is
removed if $J_3>0$, while it is enhanced to a power like singularity if
$J_3<0$, see item d) in the Theorem \equ(1.5). Hence, the singularity at
$\kk=(\pm 2p_F,0)$ is of the same type as in the Luttinger model, see
\equ(1.10). 

However, we can not conclude that the same is true for the Fourier
transform at $\kk=0$, which is bounded in the Luttinger model, while we can
not exclude a logarithmic divergence. In order to get such a stronger
result, it would be sufficient to prove that the function $\O^{3,b}(\xx)$
is odd in the exchange of $(x,x_0)$ with $(x_0 v,x/v)$, for some $v$; this
property is true for the leading term corresponding to $\O^{3,b}(\xx)$ in
\equ(1.10), with $v=v_0$, but seems impossible to prove on the base of our
expansion. We can only see this symmetry for the leading term, with
$v=v_0^*$ (or any other value $v$ differing for terms of order $J_3$, since
the substitution of $v_0^*$ with $v$ would not affect the bound
\equ(1.20)), but this is only sufficient to prove that the singularity has
to be of order $J_3$, at least.
 
i) Our theorem cannot be proved by building a multiscale renormalized
expansion, neither by taking as the ``free model'' the $XY$ one and $J_3$
as the perturbative parameter, nor by taking as the free model the $XXY$
one and $u$ as the perturbative parameter. In fact, in order to solve the
model, one cannot perform a single Bogoliubov transformation as in the
$J_3=0$ case; the gap has a non trivial flow and one has to perform a
different Bogoliubov transformation for each renormalization group
integration. This can be seen clearly in \equ(2.66), which is the fermionic
integration of a fermionic theory with gap $\s_h$ and wave function
renormalization $Z_h$. If $J_3=0$, then $\s_h=u$ and $Z_h=1$, but, if
$J_3\not=0$, they are rapidly varying functions of $h$.

l) If $u=0$, the critical indices and $\nu$ can be computed with any
prefixed precision; we write explicitly in the theorem only the first order
for simplicity. However, if $u\not=0$, they are not fixed uniquely; for
what concerns $\n$, this means that, in the gapped case, the system is
insensitive to variations of the magnetic field much smaller than the gap
size.
 
m) There is no reason to restrict the analysis to a nearest-neighbor
Hamiltonian like \equ(1.1); it will be clear in the following that our
results still holds for non nearest-neighbor spin hamiltonians, as such
hamiltonians differ from \equ(1.1) for {\it irrelevant} (in the RG sense)
terms; see also [Spe], where the case $J_3=0$ is studied.
 
n) The same techniques could perhaps be used to study $\O^{1}_{L,\b}(\xx)$
and $\O^{2}_{L,\b}(\xx)$, however this problem is more difficult, as one
has to study the average of the exponential of the sum of fermionic density
operators, see\equ(1.4). In the $J_3=0$ case the evaluation of
$\O^{1}_{L,\b}(\xx)$ and $\O^{2}_{L,\b}(\xx)$ was done in [Mc].

\* \sub(1.7) 
The proof of the theorem is organized into two parts.

In the present paper we define a Renormalization Group expansion for the
effective potential and the ground state energy of the $XYZ$ model, see
\S(2). One has to perform a multiscale analysis with a different Bogoliubov
transformation for each renormalization group integration. A definition of
localization operator is introduced, which is different with respect to the
one suggested by a simple power counting argument. In \S(3) we prove that
such expansion is convergent if the running coupling constants are small
enough.

Despite we are interested in $\O^3_\xx$, we study in detail the convergence
of the effective potential and the ground state energy for pedagogical
reasons as the expansion for $\O^3_\xx$ is clearer once the expansion for
the effective potential is understood. The proof of the convergence
requires some care as the power counting has to be improved. Moreover we
pay attention to perform all the estimates taking finite $L,\b$; this
requires some care, as the preceding analysis of similar problems were not
so careful about this point.

While in this paper we deal essentially with convergence problems of the
renormalized expansions, in the subsequent one we have to analyze carefully
the expansions in order to exploit the cancellations, based on symmetry
properties, which allow to complete the proof of the theorem; the
convergence of the expansion for $\O^3_{L,\b}(\xx)$ is a corollary of the
analogous proof given in this paper for the effective potential or the
ground state energy.

\pagina
\vskip2.truecm
\section(2,Multiscale decomposition and anomalous integration)
 
\sub(2.1) The Hamiltonian \equ(1.5) can be written, if $U_L^2$ is chosen
as explained in \sec(1.1) and the definitions \equ(1.10a) are used,
in the following way (by neglecting a constant term):
$$\eqalign{
H &= \sum_{x\in\L}\left\{(\cos p_F+\n) a^+_x a^-_x\
-\fra12 [ a^+_{x} a^-_{x+1}+ a_{x+1}^+  a^-_{x}]-\right.\cr
&\left. -{u\over 2}[ a^+_{x} a^+_{x+1} +  a^-_{x+1} a^-_{x}]+
\l( a^+_x a^-_x)( a^+_{x+1} a^-_{x+1})
\right\}\;,\cr}\Eq(2.1)$$
where $\L$ is an interval of $L$ points on the one-dimensional lattice of
step one, which will chosen equal to $(-[L/2],[(L-1)/2])$, the fermionic
field $ a^\pm_x$ satisfies periodic boundary conditions and
$$\l=-J_3\;.\Eq(2.2)$$
 
The Hamiltonian \equ(2.1) will be considered as a perturbation of the
Hamiltonian $H_0$ of a system of free fermions in $\L$ with unit mass and
chemical potential $\m=1-\cos p_F$ ($u=J_3=\n=0$); $p_F$ is the {\sl Fermi
momentum}. This system will have, at zero temperature, density $\r=p_F/\p$,
corresponding to magnetization $\r-1/2$ in the $3$-direction for the
original spin system. Since $p_F$ is not uniquely defined at finite volume,
we choose it so that
$$p_F={2\p\over L}(n_F+\fra12)\;,\quad n_F\in \NNN\;,\quad
\lim_{L\to\io} p_F = \p\r\Eq(2.3)$$
This means, in particular, that $p_F$ is not an allowed momentum of the
fermions.
 
We consider also the operators $ a_{\xx}^{\pm}=e^{x_0 H} a_x^{\pm}e^{-Hx_0}$,
with
$$\xx=(x,x_0)\;,\quad -\b/2\le x_0 \le \b/2\;,\Eq(2.4)$$
for some $\b>0$; on $x_0$, which we shall call the
time variable, antiperiodic boundary conditions are imposed.
 
Many interesting physical properties of the fermionic system at inverse
temperature $\b$ can be expressed in terms of the {\sl Schwinger
functions}, that is the truncated expectations in the Grand Canonical
Ensemble of the time order product of the field $ a_{\xx}^{\pm}$ at
different space-time points.
There is of course a relation between these functions
and the expectations of some suitable observables in the spin system.
However, by looking at \equ(1.4), one sees that this relation is simple
enough only in the case of the truncated expectations of the time order
product of the fermionic {\sl density operator}
$\r_\xx= a^+_\xx  a^-_\xx$ at
different space-time points, which we shall call the {\sl density Schwinger
functions}; they coincide with the truncated expectations of the time order
product of the operator $S^3_\xx=e^{x_0 H} S^3_x e^{-Hx_0}$ at different
space-time points.
 
As it is well known, the Schwinger functions can be written as
power series in $\l$ and $u$, convergent for $|\l|,|u| \le \e_\b$, for some
constant $\e_\b$ (the only trivial bound of $\e_\b$ goes to zero, as
$\b\to\io$). This power expansion is constructed in the usual way in terms of
Feynman graphs, by using as {\sl free propagator} the function
$$\eqalign{
g^{L,\b}(\xx-\yy)&=
{{\rm Tr} \left[e^{-\b H_0} {\bf T} ( a^-_\xx  a^+_\yy)\right]
\over {\rm Tr} [e^{-\b H_0}]} = \cr
&={1\over L} \sum_{k\in {\cal D}_L}
e^{-ik(x-y)} \left\{ {e^{-\t e(k)} \over 1+e^{-\b e(k)}}
\indic(\t>0) - {e^{-(\b+\t) e(k)} \over 1+e^{-\b e(k)}} \indic(\t\le 0)
\right\}\; , \cr}\Eq(2.5)$$
where ${\bf T}$ is the time order product, $N=\sum_{x\in\L} a^+_x a^+_x$,
$\t=x_0-y_0$, $\indic(E)$ denotes the indicator function ($\indic(E)=1$, if
$E$ is true, $\indic(E)=0$ otherwise),
$$e(k)=\cos p_F-\cos k\;,\Eq(2.6)$$
and ${\cal D}_L\=\{k={2\pi n/L}, n\in \zzzz, -[L/2]\le n \le [(L-1)/2]\}$.
 
It is also well known that, if $x_0\not= y_0$,
$g^{L,\b}(\xx-\yy)= \lim_{M\to\io} g^{L,\b,M}(\xx-\yy)$, where
$$g^{L,\b,M}(\xx-\yy)= {1\over L\b} \sum_{\kk\in {\cal D}_{L,\b}}
{e^{-i\kk\cdot(\xx-\yy)}\over -ik_0+\cos p_F-\cos k}\; , \Eq(2.7)$$
$\kk=(k,k_0)$, $\kk\cdot\xx=k_0x_0+kx$, ${\cal D}_{L,\b}\={\cal D}_L
\times {\cal D}_\b$, ${\cal D}_\b\=\{k_0=2(n+1/2)\pi/\b, n\in
Z, -M\le n \le M-1\}$. Note that $g^{L,\b,M}(\xx-\yy)$ is real, $\forall M$.
 
Hence, if we introduce a finite set of Grassmanian
variables $\{\hat a^\pm_\kk\}$, one for each $\kk\in\DD_{L,\b}$, and a
linear functional $P(d a)$ on the generated Grassmanian algebra, such that
$$\int P(d a) \hat a^-_{\kk_1}\hat a^+_{\kk_2} = L\b \d_{\kk_1,\kk_2}
\hat g(\kk_1)\;,\quad \hat g(\kk)= {1\over -ik_0+\cos p_F-\cos k}
\; ,\Eq(2.8)$$
we have
$$\lim_{M\to\io}
{1\over L\b} \sum_{\kk\in {\cal D}_{L,\b}} \, e^{-i\kk\cdot(\xx-\yy)} \,
\hat g(\kk) = \lim_{M\to\io} \int P(d a) a^-_\xx  a^+_\yy
\= g^{L,\b}(\xx;\yy) \; ,\Eq(2.9)$$
where the {\sl Grassmanian field} $ a_\xx$ is defined by
$$ a_\xx^{\pm}= {1\over L\b} \sum_{\kk\in {\cal D}_{L,\b}}\hat a_\kk^{\pm}
e^{\pm i\kk\cdot\xx}\; .\Eq(2.10)$$
 
The ``Gaussian measure'' $P(d a)$ has a simple representation in terms of
the ``Lebesgue Grassmanian measure'' $\prod_{\kk\in\DD_{L,\b}}da_\kk^+
da_\kk^-$, defined as the linear functional on the Grassmanian algebra, such
that, given a monomial $Q( a^-, a^+)$ in the variables $a_\kk^-, a_\kk^+$,
$\kk\in\DD_{L,\b}$, its value is $0$, except in the case $Q( a^-, a^+)=
\prod_\kk \hat a^-_\kk \hat a^+_\kk$, up to a permutation of the variables.
In this case the value of the functional is determined, by using the
anticommuting properties of the variables, by the condition
$$\int \;\left[\prod_{\kk\in\DD_{L,\b}}da_\kk^+ da_\kk^-\right]\;
\prod_{\kk\in\DD_{L,\b}} \hat a^-_\kk \hat a^+_\kk = 1\Eq(2.11)$$
We have
$$P(d a) = \Big\{ \prod_\kk (L\b\hat g_\kk) \hat a^+_\kk \hat a^-_\kk\Big\}
\exp \Big\{-\sum_\kk (L\b \hat g_\kk)^{-1} \hat a^+_\kk
\hat a_\kk^- \Big\}\;.\Eq(2.12)$$
Note that, since $(\hat a_\kk^-)^2=(\hat a^+_\kk)^2=0$,
$e^{-z \hat a^+_\kk \hat a_\kk}
=1-z \hat a^+_\kk \hat a_\kk$, for any complex $z$.
\*
 
\0{\bf Remark.} The {\sl ultraviolet cutoff} $M$ on the $k_0$ variable was
introduced so that the Grassman algebra is finite; this implies that the
Grassmanian integration is indeed a simple algebraic operation and all
quantities that appear in the calculations are finite sums. However, $M$
does not play any essential role in this paper, since all bounds
will be uniform with respect to $M$ and they easily imply the
existence of the limit. Hence, we shall not stress the dependence on
$M$ of the various quantities we shall study.
 
\*
By using standard arguments (see, for example, [NO], where a different
regularization of the propagator is used), one can show that
the partition function and the Schwinger functions
can be calculated as expectations of suitable functions of the
Grassmanian field with respect to the ``Gaussian measure'' $P(d a)$.
In particular the partition function ${\rm Tr} [e^{-\b H}]$ is equal to
${\cal Z}_{L,\b} {\rm Tr} [e^{-\b H_0}]$, with
$${\cal Z}_{L,\b} = \int P(d a) e^{-\VV( a)}\; , \Eq(2.13)$$
where
$$\eqalignno{
&\VV( a)=u V_u(a)+\l V_\l( a)+\nu N( a)\;,\cr
V_\l( a)&=\sum_{x,y\in\L}\int_{-\b/2}^{\b/2} dx_0\int_{-\b/2}^{\b/2} dy_0
v_\l(\xx-\yy) a_\xx^+ a_\yy^+ a_\yy^- a_\xx^-\;,\qquad
N( a)=\sum_{x\in\L} \int_{-\b/2}^{\b/2} dx_0 a_\xx^+  a^-_\xx \;,\cr
V_u(a)&=\sum_{x,y\in\L}\int_{-\b/2}^{\b/2} dx_0\int_{-\b/2}^{\b/2} dy_0
v_u(\xx-\yy)\Big[ a_\xx^+  a^+_\yy - a^-_\xx a^-_\yy\Big]&\eq(2.14)\cr}$$
where
$$v_\l(\xx-\yy)=\fra12 \d_{1,|x-y|}\d(x_0-y_0)\;,\quad v_u(\xx-\yy)=
\fra12 \d_{x,y+1}\d(x_0-y_0)\;.\Eq(2.15)$$
 
Note that the parameter $\nu$ has been introduced in order to fix the
singularities of the interacting propagator to the values of the free
model, that is $\kk=(0,\pm p_F)$. Hence $\nu$ is a function of $\l,u,p_F$,
which has to be fixed so that the perturbation expansion is convergent
(uniformly in $L,\b$). This choice of $\n$ has also the effect of fixing
the singularities of the spin correlation function Fourier transform, as we
explained in the introduction, see \sec(1.4).

Note that, if $p_F=\p/2$, one can prove that $\nu=-\l$, by using simple symmetry
properties of our expansion; this implies, by using \equ(1.10a), that $h=0$.

If $u=0$, it is conjectured, on the base of heuristic calculations, that this
condition is equivalent to the condition that, in the limit $L,\b\to\io$, the
density is fixed (``Luttinger Theorem'') to the free model value $\r=p_F/\p$.
If $u\not=0$, there is no simple relation between the value of $p_F$ and the
density, as one can see directly in the case $\l=0$, where one can do explicit
calculations. 
 
\*
\sub(2.2) We shall begin our analysis by rewriting the {\it potential}
$\VV(a)$ as
$$\VV(a)=\VV^{(1)}(a)+u V_u(a) +\d^* V_\d(a)\;,\Eq(2.16)$$
where
$$\VV^{(1)}(a) = \l V_\l( a)+\nu N( a) - \d^* V_\d(a)\;,\Eq(2.17)$$
and
$$V_\d(a) = {1\over L\b}\sum_\kk e(k) \hat a^+_\kk \hat a_\kk^-\;.\Eq(2.18)$$
$\d^*$ is an arbitrary parameter, to be fixed later, of modulus smaller
than $1/2$; its introduction is not
really necessary, but allows to simplify the discussion of the spin
correlation function asymptotic behaviour. In terms of the Fermionic system, it will
describe the modification of the Fermi velocity due to the interaction.
 
Afterwards we ``move'' the terms $u V_u(a)$ and $\d^* V_\d(a)$ from
the interaction to the Gaussian measure. In order to describe the properties
of the new Gaussian measure, it is convenient to introduce a new set of
Grassmanian variables $\hat b^\s_{\kk,\o}$, $\o=\pm 1$, $\kk\in
\DD_{L,\b}^+$, by defining
$$\DD_{L,\b}^\o=\{\kk\in \DD_{L,\b}:\o k>0\} \cup
\{\kk\in \DD_{L,\b}:k=0, \o k_0>0\}\;,\Eq(2.19)$$
$$\hat b^\s_{\kk,\o}=\hat a^{\s\o}_{\o\kk}\;,\Eq(2.20)$$
so that, by using \equ(2.10)
$$a^\s_\xx={1\over L\b} \sum_{\kk\in {\cal D}_{L,\b}^+,\ \o=\pm 1}
\hat b^{\s\o}_{\kk,\o} e^{i\s\o\kk\cdot\xx}\;.\Eq(2.21)$$
 
It is easy to see that
$${\cal Z}_{L,\b} = e^{-L\b t_1}\int P(d b) e^{-\tilde\VV^{(1)}( b)}\;,
\Eq(2.22)$$
with $\tilde\VV^{(1)}(b)= \VV^{(1)}(a)$,
where $a$ has to be interpreted as the r.h.s. of \equ(2.21),
$$\eqalign{
P(d b) &= \Big\{ \prod_{\kk\in {\cal D}_{L,\b}^+}
{(L\b)^2\over -k_0^2-(1+\d^*)^2 e(k)^2-u^2\sin^2 k}
\prod_{\o=\pm 1}\hat b^+_{\kk,\o} \hat b^-_{\kk,\o}\Big\}\;\cdot\cr
&\cdot\;\exp \Big\{-{1\over L\b} \sum_{\kk\in {\cal D}_{L,\b}^+}
\sum_{\o,\o'}\hat b^+_{\kk,\o} T_{\o,\o'}(\kk)
\hat b^-_{\kk,\o} \Big\}\;,\cr}\Eq(2.23)$$
$$T(\kk) = \left(\matrix{-ik_0+(1+\d^*)e(k) & iu\sin k \cr
-iu\sin k & -ik_0-(1+\d^*)e(k) \cr}\right)\;,\Eq(2.24)$$
$$t_1=-{1\over L\b} \sum_{\kk\in {\cal D}_{L,\b}^+}\log
{k_0^2+(1+\d^*)e(k)^2+u^2\sin^2 k \over k_0^2+e(k)^2}\;.\Eq(2.25)$$
 
Note that $t_1$ is uniformly bounded as $L,\b\to\io$, if $|\d^*|\le 1/2$,
as we are supposing. For $\l=\n=\d^*=0$, it represents the free energy for
lattice site of $H-H_0$.

\*
\sub(2.3) For $\l=\n=0$, all the properties of the model can be analyzed in
terms of the Grassmanian measure \equ(2.23). In particular, we have
$$\int P(d b) a^{\s_1}_\xx a^{\s_2}_\yy =
{1\over L\b} \sum_{\kk\in {\cal D}_{L,\b}^+} \Big[
e^{-i\kk(\xx-\yy)} T^{-1}(\kk)_{-\s_1,\s_2}-
e^{i\kk(\xx-\yy)} T^{-1}(\kk)_{-\s_2,\s_1} \Big]\;,\Eq(2.26)$$
where $T^{-1}(\kk)$ denotes the inverse of the matrix $T(\kk)$. This matrix
is defined for any $\kk\in {\cal D}_{L,\b}$ and satisfies the symmetry
relation
$$T^{-1}(\kk)_{-\s_2,\s_1}=-T^{-1}(-\kk)_{-\s_1,\s_2}\;,\Eq(2.27)$$
so that we can write \equ(2.26) also in the form
$$\int P(d b) a^{\s_1}_\xx a^{\s_2}_\yy =
{1\over L\b} \sum_{\kk\in {\cal D}_{L,\b}}
e^{-i\kk(\xx-\yy)} T^{-1}(\kk)_{-\s_1,\s_2}\;.\Eq(2.28)$$
 
If $\l\not=0$, we shall study
the model, for $\l$ small, in terms of a perturbative expansion, based on
a multiscale decomposition of the measure \equ(2.23), by using the methods
introduced in [BG] and extended in various other papers ([BGPS], [BM1], [M1]).
In order to discuss the structure of the expansion, it is convenient to
explain first how it works in the case of the free energy for site of $H-H_0$
$$E_{L,\b}=-{1\over L\b}\log {\cal Z}_{L,\b}\;.\Eq(2.29)$$
 
Let $T^1$ be the one dimensional torus, $||k-k'||_{T^1}$ the usual
distance between $k$ and $k'$ in $T^1$ and $||k||=||k-0||$.
We introduce a {\sl scaling parameter} $\g>1$ and a positive function
$\c(\kk') \in C^{\io}(T^1\times R)$, $\kk'=(k',k_0)$, such that
$$ \c(\kk') = \c(-\kk') = \cases{
1 & if $|\kk'| <t_0 \= a_0 v_0^* /\g \;,$ \cr
0 & if $|\kk'| >a_0 v_0^*\; ,$\cr}\Eq(2.30)$$
where
$$|\kk'|=\sqrt{k_0^2+(v_0^* ||k'||_{T^1})^2}\;,\Eq(2.31)$$
$$a_0 =\min \{p_F/2, (\p-p_F)/2 \}\;,\Eq(2.32)$$
$$v_0^*=v_0 (1+\d^*)\;,\qquad v_0=\sin p_F\;.\Eq(2.33)$$
In order to give a well defined meaning
to the definition \equ(2.30), $v_0^*>0$ has to be positive.
Hence we shall suppose that
$$v_0\ge \bar v_0>0\;,\quad |\d^*|\le \fra12\;,\Eq(2.34)$$
where $\bar v_0$ is fixed once for all. All our results will be uniform in
$v_0$, under the conditions \equ(2.34), but we shall not stress this fact
anymore in the following.
 
The definition \equ(2.30) is such that the supports of $\c(k-p_F,k_0)$ and
$\c(k+p_F,k_0)$ are disjoint and the $C^\io$ function on $T^1\times R$
$$\hat f_1(\kk) \= 1- \c(k-p_F,k_0) - \c(k+p_F,k_0) \Eq(2.35)$$
is equal  to $0$, if $[v_0^* ||(|k|-p_F)||_{T^1}]^2 +k_0^2<t_0^2$.
 
We define also, for any integer $h\le 0$,
$$f_h(\kk')= \c(\g^{-h}\kk')-\c(\g^{-h+1}\kk')\; ;\Eq(2.36)$$
we have, for any $\bh<0$,
$$\c(\kk') = \sum_{h=\bh+1}^0 f_h(\kk') +\c(\g^{-\bh}\kk')\; .
\Eq(2.37)$$
Note that, if $h\le 0$, $f_h(\kk') = 0$ for $|\kk'|
<t_0\g^{h-1}$ or $|\kk'| >t_0 \g^{h+1}$, and $f_h(\kk')=
1$, if $|\kk'| =t_0\g^h$, so that
$$f_{h_1}(\kk') f_{h_2}(\kk') = 0\;, \quad \hbox{if\ \ }
|h_1-h_2|>1\;.\Eq(2.38)$$
 
We finally define, for any $h\le 0$:
$$ \hat f_h(\kk) = f_h(k-p_F,k_0) +f_h(k+p_F,k_0)\;;\Eq(2.39)$$
This definition implies that, if $h\le 0$, the support of
$\hat f_h(\kk)$ is the union of two disjoint sets, $A_h^+$ and $A_h^-$. In
$A_h^+$, $k$ is strictly positive and $||k-p_F||_{T^1}\le a_0\g^h \le a_0$,
while, in $A_h^-$, $k$ is strictly negative and $||k+p_F||_{T^1}\le a_0\g^h$.
 
The label $h$ is called the {\sl scale} or {\sl frequency} label.
Note that, if $\kk\in {\cal D}_{L,\b}$, then $|\kk\pm (p_F,0)|\ge
\sqrt{(\p\b^{-1})^2+
(v_0^*\p L^{-1})^2}$, by \equ(2.3) and the definition of ${\cal D}_{L,\b}$.
Therefore
$$\hat f_h(\kk)=0\quad \forall h< h_{L,\b} =\min \{h:t_0\g^{h+1} >
\sqrt{(\p\b^{-1})^2+(v_0^*\p L^{-1})^2} \}\;,\Eq(2.40)$$
and, if $\kk\in {\cal D}_{L,\b}$, the definitions \equ(2.35) and
\equ(2.39), together with the identity \equ(2.37), imply that
$$1=\sum_{h=h_{L,\b} }^1 \hat f_h(\kk) \; .\Eq(2.41)$$
 
We now introduce, for each scale label $h$, such that $h_{L,\b}\le h \le 1$,
a set of Grassmanian variables $b^{(h)\s}_{\kk,\o}$ and a corresponding
Gaussian measure $P(db^{(h)})$, such that, if $h=1$, then
$\kk\in {\cal D}_{L,\b}$ and
$$\int P(d b^{(1)}) b^{(1)-\s_1}_{\kk_1,\o_1} b^{(1)\s_2}_{\kk_2,\o_2}=
L\b \s_1 \d_{\s_1,\s_2} \d_{\kk_1,\kk_2}
\fra12 T^{-1}(\kk_1)_{\o_1,\o_2} \hat f_1(\kk_1)\;,\Eq(2.42)$$
while, if $h\le 0$, then $\kk\in {\cal D}_{L,\b}^+$ and
$$\int P(d b^{(h)}) b^{(h)-\s_1}_{\kk_1,\o_1} b^{(h)\s_2}_{\kk_2,\o_2}=
L\b \s_1 \d_{\s_1,\s_2} \d_{\kk_1,\kk_2}
T^{-1}(\kk_1)_{\o_1,\o_2} f_h(k_1-p_F,k_0)\;.\Eq(2.43)$$
The support properties of the r.h.s. of \equ(2.42) and \equ(2.43) allow
to impose the condition
$$b^{(h)\s}_{\kk,\o}=0,\quad \hbox{if}\;\hat f_h(\kk)=0\;.\Eq(2.44)$$
 
By using \equ(2.26) and \equ(2.27), it is easy to see that
$$\int P(d b) a^{\s_1}_\xx a^{\s_2}_\yy =\sum_{h=h_{L,\b}}^1
\sum_{\o_1,\o_2} \int P(d b^{(h)}) b^{(h)\s_1\o_1}_{\xx,\o_1}
b^{(h)\s_2\o_2}_{\yy,\o_2}\;,\Eq(2.45)$$
where, if $h\le 0$,
$$b^{(h)\s}_{\xx,\o}={1\over L\b} \sum_{\kk\in {\cal D}_{L,\b}^+}
\hat b^{(h)\s}_{\kk,\o} e^{i\s\kk\cdot\xx}\;,\Eq(2.46)$$
while, if $h=1$, a similar definition is used, with ${\cal D}_{L,\b}$
in place of ${\cal D}_{L,\b}^+$.
Note that this different definition, which is at the origin of the factor
$1/2$ in the r.h.s. of \equ(2.42), is not really necessary, but implies that
$\int P(d b^{(1)}) b^{(1)-}_{\xx,\o_1} b^{(h)+}_{\yy,\o_2}$
is bounded for $M\to\io$, a property which should otherwise be true only
for $\sum_{\o_1,\o_2}
\int P(d b^{(1)}) b^{(1)\s_1\o_1}_{\xx,\o_1} b^{(h)\s_2\o_2}_{\yy,\o_2}$.
In the following, we shall use this property in order to simplify the
discussion in some minor points.
 
The identity \equ(2.45), as it is well known, implies that, if $F(a)$ is
any function of the variables $a^\s_\xx$, then
$$\int P(d a) F(a)=\int \prod_{h=h_{L,\b} }^1 P(d b^{(h)})
F\Big(\sum_{h=h_{L,\b}}^1 a^{(h)}\Big)\;,\Eq(2.47)$$
where
$$a^{(h)\s}_\xx=\sum_{\o=\pm 1} b^{(h)\s\o}_{\xx,\o}\;.\Eq(2.48)$$
 
It is now convenient to introduce a variable which measures the distance
of the momentum from the Fermi surface, by putting
$k=k'+p_F$, with $k'\in\DD'_L=
\{k'=2(n+1/2)\pi/L, n\in \ZZZ, -[L/2]\le n \le [(L-1)/2]\}$.
Moreover, we rename the Grassmanian variables, by defining
$$\hat\psi^{(h)\s}_{\kk',\o} = \hat b^{(h)\s}_{\kk'+\pp_F,\o}\;,\quad
\psi^{(h)\s}_{\xx,\o} = {1\over L\b}\sum_{\kk'\in\DD'_{L,\b}}
e^{i\s\kk'\xx} \hat\psi^{(h)\s}_{\kk',\o}\;,\Eq(2.49)$$
where $\DD'_{L,\b}=\DD'_L\times\DD_\b$, $\kk'=(k',k_0)$ and $\pp_F=(p_F,0)$.
Note that, by \equ(2.44),
$$\hat\psi^{(h)\s}_{\kk',\o} = 0 \hbox{\ \ if \ \ } \hat f_h(\kk'+\pp_F)=0
\;.\Eq(2.50)$$
The definition \equ(2.49) allows to write \equ(2.48) in the form
$$a_\xx^{(h)\s} = \sum_\o e^{i\s\o\pp_F\xx} \psi^{(h)\s\o}_{\xx,\o}\;.
\Eq(2.51)$$
 
The measure $P(d b^{(h)})$ can be thought in a natural way as a measure on
the variables $\psi^{(h)\s}_{\xx,\o}$, that we shall denote $P(d\psi^{(h)})$.
Then, \equ(2.43) and \equ(2.49) imply that, if $h\le 1$,
$$\int P(d\psi^{(h)})\,\hat\psi^{(h)-\s_1}_{\kk'_1,\o_1}
\hat\psi^{(h)\s_2}_{\kk'_2,\o_2}= (1-\fra12 \d_{h,1})
L\b \s_1 \d_{\s_1,\s_2} \d_{\kk'_1,\kk'_2}
\tilde g^{(h)}_{\o_1,\o_2}(\kk'_1)\;,\Eq(2.52)$$
where, if $f_1(\kk')\=\hat f_1(\kk'+\pp_F)$,
$$\tilde g^{(h)}(\kk') = {f_h(\kk')\over
-k_0^2-E(k')^2-u^2\sin^2 (k'+p_F)}\;
\left(\matrix{-ik_0-E(k') & -iu\sin (k'+p_F) \cr
iu\sin  (k'+p_F) & -ik_0+E(k') \cr}\right)\;,\Eq(2.53)$$
$$E(k')= v_0^* \sin k'+(1+\d^*)(1-\cos k')\cos p_F\;.\Eq(2.54)$$
 
In the following we shall use also the notation
$$\psi^{(\le h)\s}_{\xx,\o}=\sum_{h'=h_{L,\b}}^h \psi^{(h')\s}_{\xx,\o}\quad
,\quad P(d\psi^{(\le h)})=\prod_{h'=h_{L,\b}}^h P(d\psi^{(h')})\;,\Eq(2.55)$$
which allows to write the identity \equ(2.47) as
$$\int P(d a) F(a)=\int P(d\psi^{(\le 1)})
\tilde F(\psi^{(\le 1)})\;,\Eq(2.56)$$
where $\tilde F(\psi^{(\le 1)})$ is obtained from $F(\sum_h a^{(h)})$,
by using \equ(2.51).
 
\*
\\{\bf Remark.\ }
Note that the sum over $k_0$ in \equ(2.49) can be thought as a finite sum
for any $M$, if $h\le 0$, because of the support properties of
$\hat\psi^{(h)\s}_{\kk',\o}$.
Hence, all quantities that we shall calculate will depend on $M$ only trough
the propagator $\tilde g^{(1)}(\kk')$, if $M$ is large enough.
 
\*
\sub(2.4) If we apply \equ(2.56) to ${\cal Z}_{L,\b}$ and we use \equ(2.29)
and \equ(2.22), we get
$$e^{-L\b E_{L,\b}}=e^{-L\b t_1}\int P(d\psi^{(\le 1)})
e^{-\VV^{(1)}(\psi^{(\le 1)})}\;,\Eq(2.57)$$
where
$$\VV^{(1)}(\psi^{(\le 1)})=\l V_\l(\sum_{h=h_{L,\b}}^1 a^{(h)})+
\n N(\sum_{h=h_{L,\b}}^1 a^{(h)})
-\d^* V_\d(\sum_{h=h_{L,\b}}^1 a^{(h)})\;.\Eq(2.58)$$
 
Let us now perform the integration over $\psi^{(1)}$; we get
$$e^{-L\b E_{L,\b}} = e^{-L\b (\tilde E_1+t_1)} \int P(d\psi^{(\le 0)}) \,
e^{-\bar\VV^{(0)}(\psi^{(\le 0)})}
\;,\quad \bar\VV^{(0)}(0)=0\;,\Eq(2.59)$$
$$e^{-\bar\VV^{(0)}(\psi^{(\le 0)})-L\b \tilde E_1}= \int
P(d\psi^{(+1)}) e^{-\VV^{(1)}(\psi^{(\le 0)}+\psi^{(+1)}})\;.\Eq(2.60)$$
 
This step is essentially trivial. In fact, it is easy to see that
$\tilde g^{(1)}_{\o,\o'}(\kk')$ is bounded, for $M\to\io$, uniformly in
$L,\b$, and that its Fourier transform $\tilde g^{(1)}_{\o,\o'}(\xx)$
is a bounded function with fast decaying properties (uniformly in $L,\b$).
Hence, by using standard perturbation theory, it is easy to see that
$\bar\VV^{(0)}(\psi^{(\le 0)})$ can be written in the form
$$\eqalign{
&\bar\VV^{(0)}(\psi^{(\le 0)}) = \sum_{n=1}^\io {1\over (L\b)^{2n}}
\sum_{\ss,\oo} \sum_{\kk'_1,...,\kk'_{2n}}
\prod_{i=1}^{2n} \hat\psi^{(\le 0)\s_i}_{\kk'_i,\o_i}\;\cdot\cr
&\cdot\;{\hat W}_{2n,\ss,\oo}^{(0)}(\kk'_1,...,\kk'_{2n-1})\;
\d(\sum_{i=1}^{2n}\s_i(\kk'_i+\pp_F))\;,\cr}\Eq(2.61)$$
where $\ss=(\s_1,\ldots,\s_{2n})$, $\oo=(\o_1,\ldots,\o_{2n})$ and
we used the notation
$$\d(\kk)=\d(k)\d(k_0)\;,\quad\d(k)=L \sum_{n\in\zzzz} \d_{k,2\p n}\;,\quad
\d(k_0)=\b\d_{k_0,0}\;.\Eq(2.62)$$
As we shall prove in \sec(3), the kernels $\hat
W^{(0)}_{2n,\ss,\oo}(\kk'_1,\ldots,\kk'_{2n-1})$, as well as $\tilde E_1$,
are expressed as power series of $\l,\nu$, convergent for
$\e\=Max(|\l|,|\nu|)\le\e_0$, for $\e_0$ small enough. Moreover there
exists a constant $C$, such that, uniformly in $L,\b$, $|\tilde E_1|\le
C\e$ and $|\hat W^{(0)}_{2n,\ss,\oo}|\le C^n \e^{\max(1,n-1)}$.
 
\*
{\bf Remark - } The conservation of momentum and the support
property \equ(2.50) of $\hat\psi^{(\le 0)\s}_{\kk',\o}$ imply that,
if $n=1$, only the terms with $\s_1+\s_2=0$ contribute to the sum in
\equ(2.61).
\*
 
Let us now define $\kk^*=(k,-k_0)$. It is possible to show, by using the
symmetries of the interaction and of the covariance $\tilde g^{(1)}(\kk')$,
that
$$\eqalign{
\hat W_{n,\ss,\oo}^{(0)}(\kk_1^*,\ldots,\kk_{n-1}^*)&=
(-1)^{\fra12 \sum_{i=1}^n \s_i\o_1}[\hat W_{n,\ss,\oo}^{(0)}
(\kk_1,\ldots,\kk_{n-1})]^*=\cr
&=(-1)^{\fra12 \sum_{i=1}^n \s_i\o_i}\hat W_{n,-\ss,-\oo}^{(0)}
(\kk_1,\ldots,\kk_{n-1})\;.\cr}\Eq(2.63)$$
 
\*
\sub(2.5) The integration of the fields of scale $h\le 0$ is performed
iteratively.We define a sequence of positive constants $Z_h$, $h=h_{L,\b},
\ldots,0$, a sequence of {\sl effective potentials} $\VV^{(h)}(\psi)$,
a sequence of constants $E_h$ and a sequence of functions $\s_h(\kk')$,
such that
$$Z_0=1,\quad E_0=\tilde E_1+t_1,\quad \s_0(\kk')=u\sin(k'+p_F)\;,\Eq(2.64)$$
and
$$e^{-L\b E_{L,\b}} = \int P_{Z_h,\s_h,C_h}(d\psi^{(\le h)}) \, e^{-\VV^{(h)}
(\sqrt{Z_h}\psi^{(\le h)})-L\b E_h}\;,\quad \VV^{(h)}(0)=0\;,\Eq(2.65)$$
where
$$\eqalign{
&P_{Z_{h},\s_h,C_h}(d\psi^{(\le h)}) =
\prod_{\kk':C^{-1}_h(\kk')>0}\prod_{\o=\pm1}
{d\hat\psi^{(\le h)+}_{\kk',\o}d\hat\psi^{(\le h)-}_{\kk',\o}\over \NN(\kk')}
\cdot\cr
&\exp \left\{-{1\over L\b} \sum_{\kk':C^{-1}_h(\kk')>0} \,C_h(\kk') Z_{h}
\sum_{\o,\o'=\pm1} \hat\psi^{(\le h)+}_{\kk',\o} T^{(h+1)}_{\o,\o'}
\hat\psi^{(\le h)-}_{\kk',\o'}\right\}\;,\cr} \Eq(2.66)$$
$$\NN(\kk')={C_h(\kk') Z_h\over L\b}
[k_0^2+E(k')^2+\s_h(\kk')^2]^{1/2}\;,\Eq(2.67)$$
$$C_h(\kk')^{-1}=\sum_{j=h_{L,\b}}^h f_j(\kk')\;,\Eq(2.68)$$
and the $2\times2$ matrix $T_{h}(\kk')$ is given by
$$T_{h}(\kk') = \left(\matrix{-ik_0+E(k') & i\sigma_{h-1}(\kk') \cr
-i\sigma_{h-1}(\kk') & -ik_0-E(k') \cr}\right)\;.\Eq(2.69)$$
 
We shall also prove that the $\VV^{(h)}$ can be represented as
$$\eqalign{
&\VV^{(h)}(\psi^{(\le h)}) = \sum_{n=1}^\io {1\over (L\b)^{2n}}
\sum_{\kk'_1,...,\kk'_{2n},\atop \ss,\oo}
\prod_{i=1}^{2n} \hat\psi^{(\le h)\s_i}_{\kk'_i,\o_i}\;\cdot\cr
&\cdot\;\hat W_{2n,\ss,\oo}^{(h)}(\kk'_1,...,\kk'_{2n-1})
\d(\sum_{i=1}^{2n}\s_i(\kk'_i+\pp_F))\;,\cr}\Eq(2.70)$$
with the kernels $\hat W_{2n,\ss,\oo}^{(h)}$ verifying the symmetry relations
$$\eqalign{
\hat W_{n,\ss,\oo}^{(h)}(\kk_1^*,\ldots,\kk_{n-1}^*)&=
(-1)^{\fra12 \sum_{i=1}^n \s_i\o_i}[\hat W_{n,\ss,\oo}^{(h)}
(\kk_1,\ldots,\kk_{n-1})]^*=\cr
&=(-1)^{\fra12 \sum_{i=1}^n \s_i\o_i}\hat W_{n,-\ss,-\oo}^{(h)}
(\kk_1,\ldots,\kk_{n-1})\;.\cr}\Eq(2.71)$$
 
The previous claims are true for $h=0$, by \equ(2.59), \equ(2.61), \equ(2.64)
and \equ(2.53). In order to prove them for
any $h\ge h_{L,\b}$, we must explain how $\VV^{(h-1)}(\psi)$ is
calculated, given $\VV^{(h)}(\psi)$.
It is convenient, for reasons which will be clear below, to split
$\VV^{(h)}$ as $\LL \VV^{(h)}+\RR \VV^{(h)}$, where $\RR=1-\LL$ and
$\LL$, the {\it localization operator}, is a linear operator on functions
of the form \equ(2.70), defined in the following way by its action on the
kernels $\hat W_{2n,\ss,\oo}^{(h)}$.
 
\*
1) If $2n=4$, then
$$\LL \hat W_{4,\ss,\oo}^{(h)}(\kk'_1,\kk'_2,\kk'_3)=
\hat W_{4,\ss,\oo}^{(h)}(\bk++,\bk++,\bk++)\;,\Eq(2.72)$$
where
$$\bk\h{\h'} = \left(\h{\p\over L},\h'{\p\over\b}\right)\;.\Eq(2.73)$$
Note that this definition depends on the the field variables order in
the r.h.s. of \equ(2.70), if $\sum_{i=1}^4 \s_i\not=0$.
In fact, since $\s_4 \kk'_4= -\sum_{i=1}^3
\s_i\kk'_i-\pp_F\sum_{i=1}^4 \s_i$ (modulo $(2\p,0)$),
if $\kk'_i=\bk++$ for $i=1,2,3$, $\kk'_4=\bk++$ only if $\sum_{i=1}^4 \s_i=0$.
This is apparently a problem, because the
representation \equ(2.70) is not uniquely defined (the terms which differ by
a common permutation of the $\ss$ and $\oo$ indices are equivalent).
However, it is easy to see, by using the anticommuting property of the
field variables, that the contribution to $\LL \VV^{(h)}$ of the terms with
$2n=4$ is equal to $0$, unless, after a suitable permutation of the fields,
$\ss=(+,-,+,-)$, $\oo=(+1,-1,-1,+1)$.
 
The previous discussion implies that we are free to change the order of the
field variables as we like, before applying the definition \equ(2.72); this
freedom will be useful in the construction of the main expansion in \sec(3).
 
\*
2) If $2n=2$ and, possibly after a suitable permutation of the fields,
$\ss=(+,-)$ ($\s_1+\s_2=0$, by the remark following \equ(2.62)), then
$$\eqalign{
\LL \hat W_{2,\ss,\oo}^{(h)}(\kk')&= \fra14 \sum_{\h,\h'=\pm 1}
\hat W_{2,\ss,\oo}^{(h)}(\bk\h{\h'})\;\cdot\cr
&\cdot\;\left\{ 1+
\d_{\o_1,\o_2} \left[\h {L\over \p} \left(b_L+a_L{E(k')\over v_0^*}\right) +
\h'{\b\over \p} k_0 \right]\right\}\;,\cr}\Eq(2.74)$$
where
$$a_L \fra{L}\p \sin\fra\p{L}=1\;,\qquad {\cos p_F\over v_0}(1-\cos\fra\p{L})
+b_L \fra{L}\p \sin\fra\p{L}=0\;.\Eq(2.75)$$
In order to better understand this definition, note that, if $L=\b=\io$,
$$\LL \hat W_{2,\ss,\oo}^{(h)}(\kk')=
\hat W_{2,\ss,\oo}^{(h)}(0)+\d_{\o_1,\o_2}
\left[{E(k')\over v_0^*} {\dpr \hat W_{2,\ss,\oo}^{(h)}\over \dpr k'}(0)
+ k_0 {\dpr \hat W_{2,\ss,\oo}^{(h)}\over \dpr k_0}(0)\right]\;.\Eq(2.76)$$
Hence, $\LL \hat W_{2,\ss,\oo}^{(h)}(\kk')$ has to be understood as a discrete
version of the Taylor expansion up to order $1$.
Since $a_L=1+O(L^{-2})$ and $b_L=O(L^{-2})$, this property would be true also
if $a_L=1$ and $b_L=0$; however the choice \equ(2.75) has the advantage to
share with \equ(2.76) another important property, that is
$\LL^2 \hat W_{2,\ss,\oo}^{(h)}(\kk')=\LL \hat W_{2,\ss,\oo}^{(h)}(\kk')$.
 
\*
3) In all the other cases
$$\LL \hat W_{2n,\ss,\oo}^{h}(\kk'_1,\ldots,\kk'_{2n-1})=0\;.\Eq(2.77)$$
\*
By \equ(2.72) and the remark following \equ(2.76), the operator $\LL$
satisfies the relation
$$\RR \LL =0\;.\Eq(2.78)$$
\*
By using the anticommuting properties of the Grassmanian variables (see
discussion in item 1) above) and the symmetry relations \equ(2.71),
we can write $\LL \VV^{(h)}$ in the following way:
$$\LL\VV^{(h)}(\psi^{(\le h)})=\g^h n_h F_\nu^{(\le h)}+
s_h F_\s^{(\le h)}+z_h F_\z^{(\le h)}+a_h F_\a^{(\le h)}+l_h F_\l^{(\le h)}
\;,\Eq(2.79) $$
where $n_h$, $s_h$, $z_h$, $a_h$ and $l_h$ are real numbers and
$$\eqalignno{
F_\nu^{(\le h)}&=\sum_{\o=\pm 1}{\o\over L\b}\sum_{\kk'\in {\cal D}'_{L,\b}}
\hat\psi^{(\le h)+}_{\kk',\o}\hat\psi^{(\le h)-}_{\kk',\o}\;,\cr
F_\s^{(\le h)}&=\sum_{\o=\pm 1}{i\o\over (L\b)}\sum_{\kk'\in {\cal D}'_{L,\b}}
\hat\psi^{(\le h)+}_{\kk',\o}\hat\psi^{(\le h)-}_{\kk',-\o}\;,\cr
F_\a^{(\le h)}&=\sum_{\o=\pm 1} {\o\over (L\b)}\sum_{\kk'\in
{\cal D}'_{L,\b}} {E(k')\over v_0^*} \hat\psi^{(\le h)+}_{\kk',\o}
\hat\psi^{(\le h)-}_{\kk',\o}\;,&\eq(2.80)\cr
F_\z^{(\le h)}&=\sum_{\o=\pm 1} {1\over (L\b)}
\sum_{\kk'\in {\cal D}'_{L,\b}}
(-i k_0) \hat\psi^{(\le h)+}_{\kk',\o}\hat\psi^{(\le h)-}_{\kk',\o}\;,\cr
F_\l^{(\le h)}&={1\over (L\b)^4}\sum_{\kk'_1,...,\kk'_4\in \DD'_{L,\b}}
\hat\psi^{(\le h)+}_{\kk'_1,+1}
\hat\psi^{(\le h)-}_{\kk'_2,-1} \hat\psi^{(\le h)+}_{\kk'_3,-1}
\hat\psi^{(\le h)-}_{\kk'_4,+1}\d(\kk'_1-\kk'_2+\kk'_3-\kk'_4)\;.\cr}$$
 
By using \equ(2.72) and \equ(2.74), it is easy to see that, if
$\e\=\max\{|\l|,|\n|\}$,
$$\eqalign{
l_0=4\l\sin^2(p_F+\p/L)+O(\e^2)\;,&\quad a_0=-\d^* v_0+c_0^\d\l_1+O(\e^2)\;,\cr
s_0=O(u\e)\;,\quad z_0=O(\e^2)\;,&\quad n_0=\nu+O(\e)\;,\cr}\Eq(2.81)$$
where $c_0^\d$ is a constant, bounded uniformly in $L,\b$.
 
We now {\sl renormalize} the free measure $P_{Z_h,\s_h,C_h}(d\psi^{(\le h)})$,
by adding to it part of the r.h.s. of \equ(2.79). We get
$$\eqalign{
\int P_{Z_h,\s_h,C_h}(d\psi^{(\le h)}) &\, e^{-\VV^{(h)}(\sqrt{Z_h}\psi^{(\le
h)})}=\cr
&=e^{-L\b t_h}\int P_{\tilde Z_{h-1},\s_{h-1},C_h}(d\psi^{(\le h)}) \,
e^{-\tilde\VV^{(h)}(\sqrt{Z_h}\psi^{(\le h)})}\;,\cr}\Eq(2.82)$$
where $P_{\tilde Z_{h-1},\s_{h-1},C_h}(d\psi^{(\le h)})$ is obtained from
$P_{Z_h,\s_h,C_h}(d\psi^{(\le h)})$ by substituting $Z_h$ with
$$\tilde Z_{h-1}(\kk')=Z_h[1+C_h^{-1}(\kk')z_h]\Eq(2.83)$$
and $\s_h(\kk')$ with
$$\s_{h-1}(\kk')={Z_h\over \tilde Z_{h-1}(\kk')}
[\s_h(\kk')+C_h^{-1}(\kk') s_h]\;;\Eq(2.84)$$
moreover
$$\tilde\VV^{(h)}(\sqrt{Z_h}\psi^{(\le h)}) =
\VV^{(h)}(\sqrt{Z_h}\psi^{(\le h)})- s_h Z_h
F_\s^{(\le h)}-z_h Z_h[F_\z^{(\le h)}+ v_0^* F_\a^{(\le h)}]\Eq(2.85)$$
and the factor $\exp(-L\b t_h)$ in \equ(2.82) takes into account the different
normalization of the two measures, so that
$$t_h=-{1\over L\b} \sum_{\kk':C^{-1}_h(\kk')>0}\log \left\{
[1+z_h C_h^{-1}(\kk')]^2 {k_0^2+E(k')^2+\s_{h-1}(\kk')^2\over
k_0^2+E(k')^2+\s_h(\kk')^2}\right\}\;.\Eq(2.86)$$
 
Note that
$$\LL\tilde\VV^{(h)}(\psi)=\g^h n_h F_\nu^{(\le h)}+(a_h-z_h v_0^*)
F_\a^{(\le h)}+l_h F_\l^{(\le h)}\;.\Eq(2.87)$$
 
The r.h.s of \equ(2.82) can be written as
$$e^{-L\b t_h} \int P_{Z_{h-1},\s_{h-1},C_{h-1}}(d\psi^{(\le h-1)}) \int
P_{Z_{h-1},\s_{h-1},\tilde f_h^{-1}}(d\psi^{(h)}) \, e^{-\tilde \VV^{(h)}
(\sqrt{Z_h}\psi^{(\le h)})}\; , \Eq(2.88) $$
where
$$Z_{h-1}=Z_h(1+z_h)\;, \qquad \tilde f_h(\kk')=Z_{h-1}[{C_h^{-1}(\kk')
\over \tilde Z_{h-1}(\kk')}-
{C_{h-1}^{-1}(\kk')\over Z_{h-1}}]\;.\Eq(2.89)$$
Note that $\tilde f_h(\kk')$ has the same support of $f_h(\kk')$; in fact,
by using \equ(2.38), it is easy to see that
$$\tilde f_h(\kk') = f_h(\kk')
\left[1+ {z_h f_{h+1}(\kk')\over 1+z_h f_h(\kk')}\right]\;.\Eq(2.90)$$
Moreover, by \equ(2.49),
$$\int P_{Z_{h-1},\s_{h-1},\tilde f_h^{-1}}(d\psi^{(h)})\,
\psi^{(h)-}_{\xx,\o}\psi^{(h)+}_{\yy,\o'} =
{g^{(h)}_{\o,\o'}(\xx-\yy)\over Z_{h-1}}\;,\Eq(2.91)$$
where
$$g^{(h)}_{\o,\o'}(\xx-\yy)={1\over L\b}\sum_{\kk'}e^{-i\kk'(\xx-\yy)}
\tilde f_h(\kk')[T_{h}^{-1}(\kk')]_{\o,\o'}\;,\Eq(2.92)$$
and $T_{h}^{-1}(\kk')$ is the inverse of the $T_{h}(\kk')$ defined in
\equ(2.69).
 
$T_{h}^{-1}(\kk')$ is well defined on the support of $\tilde f_h(\kk')$ and,
if we set
$$A_{h}(\kk') = \det T_h(\kk') =
-k_0^2 - E(k')^2- [\sigma_{h-1}(\kk')]^2 \; ,\Eq(2.93) $$
then
$$ T_{h}^{-1}(\kk')= {1\over A_{h}(\kk') }
\left( \matrix{-ik_0-E(k') & -i\sigma_{h-1}(\kk')\cr
i\sigma_{h-1}(\kk') & -ik_0+E(k') \cr}\right) \;.\Eq(2.94)$$
 
The propagator $g^{(h)}_{\o,\o'}(\xx)$ is an antiperiodic function of $x$ and
$x_0$, of period $L$ and $\b$, respectively. Its large distance behaviour
is given by the following lemma (see also [BM2]), where we use the definitions
$$\s_h\=\s_h(0)\;,\Eq(2.95)$$
$$d_L(x)={L\over \p} \sin({\p x\over L})\;,\quad
d_\b(x_0)={\b\over \p}  \sin({\p x_0\over \b})\;,\Eq(2.96)$$
$$\dd(\xx-\yy) = (d_L(x-y),d_\b(x_0-y_0))\;.\Eq(2.97)$$
 
\*
\sub(2.6) {\cs Lemma.} {\it Let us suppose that $h_{L,\b}\le h \le 0$ and
$$|z_h|\le \fra12\;,\quad |s_h| \le \fra12 |\s_h|\;,\quad
|\d^*|\le \fra12\;.\Eq(2.98)$$
We can write
$$g^{(h)}_{\o,\o}(\xx-\yy)=g^{(h)}_{L,\o}(\xx-\yy)+
r_1^{(h)}(\xx-\yy)+r_2^{(h)}(\xx-\yy)\;,\Eq(2.99)$$
where
$$g^{(h)}_{L,\o}(\xx-\yy)= {1\over L\b}\sum_{\kk'}{e^{-i\kk'(\xx-\yy)}\over
 -i k_0+\o v_0^* k'}\tilde f_h(\kk')\;.\Eq(2.100)$$
Moreover, given the positive integers $N, n_0, n_1$ and putting
$n=n_0+n_1$, there exist a constant $C_{N,n}$ such that
$$\eqalign{
|\dpr_{x_0}^{n_0} \bar\dpr_x^{n_1} r_1^{(h)}(\xx-\yy)|&\le C_{N,n}
{\g^{2h+n} \over 1+(\g^h|\dd(\xx-\yy))|^N}\;,\cr
|\dpr_{x_0}^{n_0} \bar\dpr_x^{n_1} r_2^{(h)}(\xx-\yy)|&\le C_{N,n} 
|{\sigma^h\over \g^h}|^2
{\g^{h+n}\over 1+(\g^h|\dd(\xx-\yy)|)^N}\;,\cr}\Eq(2.101)$$
$$|\dpr_{x_0}^{n_0} \bar\dpr_x^{n_1} g^{(h)}_{\o,-\o}(\xx-\yy)|\le C_{N,n}
|{\sigma^h\over \g^h}|
{\g^{h+n}\over 1+(\g^h|\dd(\xx-\yy)|)^N}\;.\Eq(2.102)$$
where $\bar\dpr_x$ denotes the discrete derivative.
}
 
\*
Note that $g^{(h)}_{L,\o}(\xx-\yy)$ coincides, in the limit $\b\to\io$,
with the propagator ``at scale $\g^h$'' of the Luttinger model, see [BGM],
with $\tilde f_h$ in place of $f_h$. This remark will be
crucial for studying the renormalization group flow in [BeM].
 
\*
\sub(2.7) Proof of Lemma \secc(2.6).
 
By using \equ(2.38), it is easy to see that $\s_h(\kk')=\s_h(0)$ on the
support of $f_h(\kk')$; hence, by \equ(2.83) and \equ(2.84), we have
$$\s_{h-1}(\kk')={\s_h +C_h(\kk')^{-1} s_h \over
1+z_h C_h(\kk')^{-1} }\;,\Eq(2.103)$$
implying, together with \equ(2.98), that there exist two constants
$c_1,c_2$ such that:
$$c_1 |\s_h|\le |\s_{h-1}(\kk')|\le c_2 |\s_h|\;.\Eq(2.104)$$
 
Let us now consider two integers $N_0,N_1\ge 0$, such that $N=N_0+N_1$, and
note that
$$\eqalign{
&d_L(x-y)^{N_1} d_\b(x_0-y_0)^{N_0} g^{(h)}_{\o,\o'}(\xx-\yy) =\cr
&e^{-i\p (xL^{-1}N_1+x_0\b^{-1}N_0)}(-i)^{N_0+N_1}
{1\over L\b}\sum_{\kk'}e^{-i\kk'(\xx-\yy)} \dpr_{k'}^{N_1} \dpr_{k_0}^{N_0}
\left[\tilde f_h(\kk')[T_{h}^{-1}(\kk')]_{\o,\o'}\right]\;,\cr}\Eq(2.105)$$
where $\dpr_{k'}$ and $\dpr_{k_0}$ denote the discrete derivatives.
 
If $\o=\o'$, the decomposition \equ(2.99) is related to the following
identity:
$$\eqalign{
[T_{h}^{-1}(\kk')]_{\o,\o} &= {1\over -i k_0+\o v_0^* k'} +
\left[ {1\over -i k_0+\o E(k')} - {1\over -i k_0+\o v_0^* k'} \right]+\cr
&+\left[ {i k_0+\o E(k') \over k_0^2 +E(k')^2+
[\sigma_{h-1}(\kk')]^2}-{1\over -i k_0+\o E(k')}\right]\;.\cr}\Eq(2.106)$$
 
The bounds \equ(2.101) and \equ(2.102) easily follow from \equ(2.98),
\equ(2.104), the support properties of $f_h(\kk')$ and
the observation that $\tilde f_h(\kk')$ and $\s_h(\kk')$ are smooth
functions of $\kk'$ in $R^2$, in the support of $f_h(\kk')$, so that the
discrete derivatives can be bounded as the continuous derivatives. The main
point is of course the fact that, on the support of $f_h(\kk')$,
$|-i k_0+\o E(k')|$, $|-i k_0+\o v_0^* k'|$ and $\sqrt{k_0^2 +
E(k')^2+[\sigma_{h-1}(\kk')]^2}$ are of order $\g^h$.
 
\*
\sub(2.8) We now {\it rescale} the field so that
$$\tilde\VV^{(h)}(\sqrt{Z_h}\psi^{(\le h)})=
\hat\VV^{(h)}(\sqrt{Z_{h-1}}\psi^{(\le h)})\;;\Eq(2.107)$$
it follows that
$$\LL\hat\VV^{(h)}(\psi)= \g^h\nu_h F_\nu^{(\le h)}+
\d_h F_\a^{(\le h)}+\l_h F_\l^{(\le h)}\;,\Eq(2.108)$$
where
$$\nu_h ={Z_h\over Z_{h-1}}n_h\;,\quad \d_h={Z_h\over Z_{h-1}}(a_h-v_0^*z_h)
\;,\quad \l_h=({Z_h\over Z_{h-1}})^2 l_h\;.\Eq(2.109)$$
We call the set $\vec v_h=(\nu_h,\d_h,\l_h)$ the {\it running coupling
constants}.
 
If now define
$$e^{-\VV^{(h-1)}(\sqrt{Z_{h-1}}\psi^{(\le h-1)}) -L\b \tilde E_h}
= \int P_{Z_{h-1},\s_{h-1},\tilde f_h^{-1}}(d\psi^{(h)}) \, e^{-\hat\VV^{(h)}
(\sqrt{Z_{h-1}}\psi^{(\le h)})}\;,\Eq(2.110)$$
it is easy to see that $\VV^{(h-1)}(\sqrt{Z_{h-1}}\psi^{(\le h-1)})$ is of
the form \equ(2.70) and that
$$E_{h-1} = E_h + t_h +\tilde E_h\;.\Eq(2.111)$$
It is sufficient to use the well known identity
$$L\b\tilde E_h + \VV^{(h-1)}(\sqrt{Z_{h-1}}\psi^{(\le h-1)})=
\sum_{n=1}^{\io}{1\over n!}
(-1)^{n+1}\EE^{T,n}_h(\hat\VV^{(h)}(\sqrt{Z_{h-1}}\psi^{(\le h)}))\;,
\Eq(2.112)$$
where $\EE^{T,n}_h$ denotes the {\it truncated expectation of order $n$} with
propagator $Z_{h-1}^{-1}g^{(h)}_{\o,\o'}$, see \equ(2.91), and observe that
$\psi^{(\le h)}= \psi^{(\le h-1)}+\psi^{(h)}$.
 
Moreover, the symmetry relations \equ(2.71) are still satisfied, because the
symmetry properties of the free measure are not modified by the
renormalization procedure, so that the effective potential on scale $h$ has
the same symmetries as the effective potential on scale $0$.
 
Let us now define $\tilde E_{h_{L,\b}}$, so that
$$e^{-L\b \tilde E_{h_{L,\b}}} = \int P_{Z_{h_{L,\b}-1},\s_{h_{L,\b}-1},
\tilde f^{-1}_{h_{L,\b}}}(d\psi^{(h_{L,\b})}) \, e^{-\hat\VV^{(h_{L,\b})}
(\sqrt{Z_{h_{L,\b}-1}}\psi^{(h_{L,\b})})}\;.\Eq(2.113)$$
We have
$$E_{L,\b}=\sum_{h=h_{L,\b}}^1 [\tilde E_h+t_h]\;.\Eq(2.114)$$

\*
Note that the above procedure allows us to write the running coupling
constants $\vec v_h$, $h\le 0$, in terms of $\vec v_{h'}$, $0\ge h'\ge h+1$,
and $\l,\n,u$:
$$\vec v_h=\vec\b(\vec v_{h+1},...,\vec v_0,\l,\n,u,\d^*)\;.\Eq(2.115)$$
The function $\vec\b(\vec v_{h+1},...,\vec v_0,\l,\n,u,\d^*)$ is called the
{\it Beta function}.
 
\*
\sub(2.9) Let us now explain the main motivations of the integration
procedure discussed above. In a renormalization group
approach one has to identify the relevant, marginal and irrelevant
interactions. By a power counting argument one sees
that the terms bilinear in the fields are relevant, hence one should
extract from them the relevant and marginal local contributions by
a Taylor expansion of the kernel up to order $1$ in the external momenta.
Since $\s_1+\s_2=0$ by the remark following \equ(2.62), we have to consider
only two kinds of bilinear terms: those with $\o_1=\o_2$ and those with
$\o_1=-\o_2$. It turns out that, for the bilinear terms with $\o_1=-\o_2$,
a Taylor expansion up to order $0$ is sufficient; the reason is
that the Feynman graphs contributing to such terms contain at least
one non diagonal propagator and, by lemma \secc(2.6), such propagators
are smaller than the diagonal ones by a factor $\s_h\g^{-h}$; as we shall
see, this is sufficient to improve the power counting by $1$.
 
The previous discussion implies that the regularization of the bilinear terms
produces four local terms. One of them, that proportional to $F_\nu$, is
relevant; it reflects the renormalization of the Fermi momentum and is faced
in a standard way [BG], by fixing properly the counterterm $\nu$ in the
Hamiltonian, \ie by fixing properly the chemical potential, so that the
corresponding running coupling $\n_h$ goes to $0$ for $h\to -\io$.
 
The term proportional to $F_\a$ is marginal, but, as we shall see, stays
bounded and of order $\l$ as $h\to -\io$, if $\d^*$ is of order $\l$; hence
the convergence of the flow is not related to the exact value of $\d^*$.
However, in order to get a detailed description of the spin correlation
function asymptotic behaviour, it is convenient to choose $\d^*$ so that $\d_h\to 0$
as $h\to -\io$. This choice implies that $v_0^*=v_0(1+\d^*)$ is the
``effective'' Fermi velocity of the fermion system.
 
The other two terms are marginal, but have to be treated in different ways.
The term proportional to $F_\z$ is absorbed in the free measure and produces a
field renormalization, as in the Luttinger liquid (which is indeed obtained
for $u=0$). The term proportional to $F_\s$, related to the presence of a gap
in the spectrum, is also absorbed in the free measure, since there is no free
parameter in the Hamiltonian to control its flow, as for $F_\z$. This
operation can be seen as the application of a sequence of {\it different
Bogoliubov transformations at each integration step}, to compare with the
single Bogoliubov transformation that it is sufficient to see a gap $O(u)$ at
the Fermi surface, in the $XY$ model ($\l=0$). It turns out that the gap is
deeply renormalized by the interaction, since $\s_h$ is a sort of ``mass
terms'' with a non trivial renormalization group flow.
 
Let us now consider the quartic terms, which are all marginal.
Since there are many of them, depending on the labels
$\o_i$ and $\s_i$ of each field, their renormalization group flow
seems difficult to study. However, as we have explained in \sec(2.5),
the running couplings corresponding to the quartic terms are all exactly
equal to $0$ for trivial reasons, unless, after a suitable permutation of the
fields, $\ss=(+,-,+,-)$, $\oo=(+1,-1,-1,+1)$. Hence, by a Taylor expansion of
the kernel up to order $0$ in the external momenta, all quartic terms can
be regularized, by introducing only one running coupling, $\l_h$.
 
As in the Luttinger liquid [BGPS, BM1], the flow of $\l_h$ and $\d_h$
can be controlled by using some cancellations, due to the fact that the Beta
function is ``close'' (for small $u$) to the Luttinger model Beta function.
In lemma \secc(2.6) we write the propagator as the Luttinger model propagator
plus a remainder, so that the Beta function is equal to the Luttinger
model Beta function plus a ``remainder'', which is small if $\s_h\g^{-h}$
is small.
 
Let us define
$$h^*={\rm inf}\{h: 0\ge h\ge h_{L,\b}, a_0 v_0^*\g^{\bh-1}\ge 4|\sigma_{\bh}|,
\forall \bh:0\ge\bh \ge h\}\;.\Eq(2.116)$$
Of course this definition is meaningful only if $a_0 v_0^* \g^{-1}\ge 4|\s_0|=
4|u|v_0$ (see \equ(2.64)), that is if
$$|u|\le {a_0\over 4\g}(1+\d^*)\;.\Eq(2.117)$$
If the condition \equ(2.117) is not satisfied, we shall put $h^*=1$.
 
Lemma \secc(2.6), \equ(2.86) and the definition of $h^*$ easily imply
this other Lemma.
 
\*
\sub(2.10) {\cs Lemma.} {\it
If $h> h^*\ge 0$ and the conditions \equ(2.98) are satisfied,
there is a constant $C$ such that
$$|t_h|\le C \g^{2h}\;.\Eq(2.117a)$$
Moreover, given the positive integers $N, n_0, n_1$ and putting
$n=n_0+n_1$, there exist a constant $C_{N,n}$ such that
$$|\dpr_{x_0}^{n_0} \bar\dpr_x^{n_1} g^{(h)}_{\o,\o'}(\xx;\yy)|\le C_{N,n}
{\g^{h+n}\over 1+(\g^h|\dd(\xx-\yy)|)^N}\;.\Eq(2.118)$$
}
 
\*
\sub(2.11) In \sec(3) we will see that, using the above lemmas and assuming
that the running coupling constants are bounded, the integration of the field
$\psi^{(h)}$ in \equ(2.88) is well defined in the limit $L,\b\to\io$,
for $0\ge h>h^*$.
 
The integration of the scales from $h^*$ to $h_{L,\b}$ will be performed ``in a
single step''. This is possible because we shall prove in \sec(3) that
the integration in the r.h.s. in \equ(2.82) is well defined in the limit
$L,\b\to\io$, for $h=h^*$. In order to do that, we shall use the following
lemma, whose proof is similar to the proof of lemma \secc(2.6).
 
\*
\sub(2.12) {\cs Lemma.} {\it Assume that $h^*$ is finite uniformly in $L,\b$,
so that $|{\s_{h^*-1}\g^{-h^*}}|\ge \bar\k$, for a suitable constant $\bar\k$
and define
$${\bar g^{(\le h^*)}_{\o,\o'}(\xx-\yy)\over Z_{h^*-1}}\equiv \int
P_{\tilde Z_{h^*-1},\s_{h^*-1},C_{h^*}}(d\psi^{(\le h^*)})
\psi_{\xx,\o}^{(\le h^*)-}\psi_{\yy,\o'}^{(\le h^*)+}\;.\Eq(2.119)$$
Then, given the positive integers $N, n_0, n_1$ and putting
$n=n_0+n_1$, there exist a constant $C_{N,n}$ such that
$$|\dpr_{x_0}^{n_0} \bar\dpr_x^{n_1} g^{(\le h^*)}_{\o,\o'}(\xx;\yy)|
\le C_{N,n} {\g^{h^*+n}\over 1+(\g^{h^*}|\dd(\xx-\yy)|)^N}\;.\Eq(2.120)$$}
 
\*
\sub(2.13) Comparing Lemma \secc(2.10) and Lemma \secc(2.12),
we see that the propagator of the integration of
all the scales between $h^*$ and $h_{L,\b}$ has the same bound as the propagator
of the integration of a single scale greater than $h^*$; this property is
used to perform the integration of all the scales $\le h^*$ in a single step.
In fact $\g^{h^*}$ is a momentum scale and, roughly speaking,
for momenta bigger than $\g^{h^*}$ the theory is ``essentially''
a massless theory (up to $O(\s_h\g^{-h})$ terms), while for momenta
smaller than $\g^{h^*}$ it is a ``massive'' theory with mass $O(\g^{h^*})$.
\pagina
 
\vskip2.truecm
\section(3,Analyticity of the effective potential)
 
\*
\sub(3.1) We want to study the expansion of the effective potential, which
follows from the renormalization procedure discussed in \sec(2). In order to
do that, we find it convenient to write $\VV^{(h)}$, $h\le 1$,
in terms of the variables $\psi^{(\le h)\sigma}_{\xx,\o}$.
The two contributions to $\VV^{(1)}(\psi^{(\le 1)})$, see \equ(2.58) and
\equ(2.14), become
$$\eqalign{
\l V_\l(\psi^{\le 1})&=\sum_\ss \int d\xx d\yy \,\l\,v_\l(\xx-\yy)
e^{i\pp_F\xx(\s_1+\s_4)+i\pp_F\yy(\s_2+\s_3)}\;\cdot\cr
&\cdot \psi^{(\le 1)\s_1}_{\xx,\s_1} \psi^{(\le 1)\s_2}_{\yy,\s_2}
\psi^{(\le 1)\s_3}_{\yy,-\s_3} \psi^{(\le 1)\s_4}_{\xx,-\s_4}\;,\cr
\n N(\psi^{\le 1})&=\sum_{\s_1,\s_2} \int d\xx e^{i\pp_F\xx(\s_1+\s_2)}
\,\n\,\psi^{(\le 1)\s_1}_{\xx,\s_1} \psi^{(\le 1)\s_2}_{\xx,-\s_2}\;,\cr}
\Eq(3.1)$$
where $\int d\xx$ is a shorthand for $\sum_{x\in\L}\int_{-\b/2}^{\b/2}dx_0$.
 
If we define
$$\eqalign{
&W_{2n,\ss,\oo}^{(h)}(\xx_1,\ldots,\xx_{2n}) = \cr
&={1\over (L\b)^{2n}}\sum_{\kk'_1,...,\kk'_{2n}}
e^{-i\sum_{r=1}^{2n} \s_r \kk'_r \xx_r}
\hat W_{2n,\ss,\oo}^{(h)}(\kk'_1,...,\kk'_{2n-1})
\d(\sum_{i=1}^{2n}\s_i(\kk'_i+ \pp_F))\;,\cr}\Eq(3.2)$$
we can write \equ(2.70) as
$$\VV^{(h)}(\psi^{(\le h)}) =\sum_{n=1}^\io \sum_{\ss,\oo}
\int d\xx_1\cdots d\xx_{2n} \left[
\prod_{i=1}^{2n} \psi^{(\le h)\s_i}_{\xx_i,\o_i}\right]
W_{2n,\ss,\oo}^{(h)}(\xx_1,\ldots,\xx_{2n})\;.\Eq(3.3)$$
Note that
$$W_{2n,\ss,\oo}^{(h)}(\xx_1+\xx,\ldots,\xx_{2n}+\xx)=
e^{i\pp_F\xx \sum_{r=1}^{2n} \s_r}
W_{2n,\ss,\oo}^{(h)}(\xx_1,\ldots,\xx_{2n})\;,\Eq(3.4)$$
hence $W_{2n,\ss,\oo}^{(h)}(\xx_1,\ldots,\xx_{2n})$ is translation invariant
if and only if $\sum_{r=1}^{2n} \s_r=0$.
 
The representation of $\LL \VV^{(h)}(\psi^{(\le h)})$ in terms of the
$\psi^{(\le h)\sigma}_{\xx,\o}$ variables is obtained by substituting in
the r.h.s. of \equ(2.79) the $\xx$-space representations of the definitions
\equ(2.80). We have
$$\eqalignno{
F_\nu^{(\le h)}&=\sum_{\o=\pm 1}\o\;
\int d\xx \;\psi^{(\le h)+}_{\xx,\o}\psi^{(\le h)-}_{\xx,\o}\;,\cr
F_\s^{(\le h)}&=\sum_{\o=\pm 1}i\o\;
\int d\xx \;\psi^{(\le h)+}_{\xx,\o}\psi^{(\le h)-}_{\xx,-\o}\;,\cr
F_\a^{(\le h)}&=\sum_{\o=\pm 1} i\o
\int d\xx \;\psi^{(\le h)+}_{\xx,\o}
[\bar\dpr_1\psi^{(\le h)-}_{\xx,\o} +{i\cos p_F\over 2 v_0}
\bar\dpr_1^2\psi^{(\le h)-}_{\xx,\o}]=&\eq(3.5)\cr
&=\sum_{\o=\pm 1} i\o \int d\xx \;
[-\bar\dpr_1\psi^{(\le h)+}_{\xx,\o} +{i\cos p_F\over 2 v_0}
\bar\dpr_1^2\psi^{(\le h)+}_{\xx,\o}]
\psi^{(\le h)-}_{\xx,\o}\;,\cr
F_\z^{(\le h)}&=\sum_{\o=\pm 1}\int d\xx\;
\psi^{(\le h)+}_{\xx,\o}\dpr_0\psi^{(\le h)-}_{\xx,\o}=
-\sum_{\o=\pm 1}\int d\xx\;
\dpr_0\psi^{(\le h)+}_{\xx,\o}\psi^{(\le h)-}_{\xx,\o}\;,\cr
F_\l^{(\le h)}&=\int d\xx\;
\psi^{(\le h)+}_{\xx,+1}\psi^{(\le h)-}_{\xx,-1}
\psi^{(\le h)+}_{\xx,-1}\psi^{(\le h)-}_{\xx,+1}\;,\cr}$$
where
$\dpr_0$ is the derivative w.r.t. $x_0$,
$\bar\dpr_1$ is the symmetric discrete derivative w.r.t. $x$, that is,
given a function $f(\xx)$,
$$\bar\dpr_1 f(\xx) =[f(x+1,x_0)-f(x-1,x_0)]/2\;,\Eq(3.6)$$
and $\bar\dpr_1^2$ (which is not the square of $\bar\dpr_1$, but has the
same properties) is defined by the equation
$$\bar\dpr_1^2 f(\xx) =f(x+1,x_0)+f(x-1,x_0)-2f(x,x_0)\;.\Eq(3.7)$$
 
Let us now discuss the action of the operator $\LL$ and $\RR=1-\LL$ on the
effective potential in the $x$-space representation, by considering the terms
for which $\LL\not=0$.
 
\*
1)If $2n=4$, by \equ(2.72),
$$\LL \int d\xxx W(\xxx) \prod_{i=1}^{4} \psi^{(\le h)\s_i}_{\xx_i,\o_i}=
\int d\xxx W(\xxx) \prod_{i=1}^{4} \big[G_{\s_i}(\xx_i-\xx_4)
\psi^{(\le h)\s_i}_{\xx_4,\o_i}\big]\;,\Eq(3.8)$$
where $\xxx=(\xx_1,\ldots,\xx_4)$, $W(\xxx)=W_{4,\ss,\oo}^{(h)}
(\xx_1,\xx_2,\xx_3,\xx_4)$ and
$$G_\s(\xx)=e^{i\s\bk++\xx}=e^{i\s\p(\fra{x}L+\fra{x_0}\b)}\;.\Eq(3.9)$$
 
Note that, as we have discussed in \sec(2.5), the r.h.s. of \equ(3.8) is
always equal to $0$, unless, after a suitable permutation of the fields,
$\ss=(+,-,+,-)$, $\oo=(+1,-1,-1,+1)$. In this last case the function
$W(\xxx) \prod_{i=1}^{4} G_{\s_i}(\xx_i-\xx_4)=W(\xxx)
G_+(\xx_1-\xx_2+\xx_3-\xx_4)$ is translation invariant and
periodic in the space and time components of all variables $\xx_k$,
of period $L$ and $\b$, respectively. It follows that
the quantities $G_{\s_i}(\xx_i-\xx_4) \psi^{(\le h)\s_i}_{\xx_4,\o_i}$
in the r.h.s. of \equ(3.8) can be substituted with
$G_{\s_i}(\xx_i-\xx_k) \psi^{(\le h)\s_i}_{\xx_k,\o_i}$, $k=1,2,3$.
Hence we have four equivalent representations of the localization operation,
which differ by the choice of the {\it localization point}.
The freedom in the choice of the localization point will be useful
in the following.
 
If the localization point is chosen as in \equ(3.8), we have
$$\eqalign{
&\RR \int d\xxx W(\xxx) \prod_{i=1}^{4} \psi^{(\le h)\s_i}_{\xx_i,\o_i}=\cr
&=\int d\xxx W(\xxx) \left[\prod_{i=1}^{4} \psi^{(\le h)\s_i}_{\xx_i,\o_i}-
\prod_{i=1}^{4} G_{\s_i}(\xx_i-\xx_4)
\psi^{(\le h)\s_i}_{\xx_4,\o_i}\right]\;.\cr}\Eq(3.10)$$
 
The term in square brackets in the above equation can be written as
$$\eqalign{
&\psi^{(\le h)\s_1}_{\xx_1,\o_1} \psi^{(\le h)\s_2}_{\xx_2,\o_2}
D^{1,1(\le h)\s_3}_{\xx_3,\xx_4,\o_3} \psi^{(\le h)\s_4}_{\xx_4,\o_4}+\cr
&+G_{\s_3}(\xx_3-\xx_4)\psi^{(\le h)\s_1}_{\xx_1,\o_1}
D^{1,1(\le h)\s_2}_{\xx_2,\xx_4,\o_2} \psi^{(\le h)\s_3}_{\xx_4,\o_3}
\psi^{(\le h)\s_4}_{\xx_4,\o_4}+\cr
&+G_{\s_3}(\xx_3-\xx_4) G_{\s_2}(\xx_2-\xx_4)
D^{1,1(\le h)\s_1}_{\xx_1,\xx_4,\o_1}\psi^{(\le h)\s_2}_{\xx_4,\o_2}
\psi^{(\le h)\s_3}_{\xx_4,\o_3} \psi^{(\le h)\s_4}_{\xx_4,\o_4}\;,\cr}
\Eq(3.11)$$
where
$$D_{\yy,\xx,\o}^{1,1(\le h)\sigma}=\psi^{(\le h)\sigma}_{\yy,\o}
-G_\s(\yy-\xx)\psi^{(\le h)\sigma}_{\xx,\o}\;.\Eq(3.12)$$
Similar expressions can be written, if the localization point is chosen in
a different way.
 
Note that the decomposition \equ(3.11) corresponds to the following identity:
$$\eqalign{
\RR \hat W_{\t,\bP }^{(h)}(\kk'_1,\kk'_2,\kk'_3)&=
\left[\hat W_{\t,\bP }^{(h)}(\kk'_1,\kk'_2,\kk'_3)-
\hat W_{\t,\bP }^{(h)}(\kk'_1,\kk'_2,\bk++)\right]+\cr
&+\left[\hat W_{\t,\bP }^{(h)}(\kk'_1,\kk'_2,\bk++)-
\hat W_{\t,\bP }^{(h)}(\kk'_1,\bk++,\bk++)\right]+\cr
&+\left[\hat W_{\t,\bP }^{(h)}(\kk'_1,\bk++,\bk++)-
\hat W_{\t,\bP }^{(h)}(\bk++,\bk++,\bk++)\right]\;,\cr}\Eq(3.13)$$
and that the $i$-th term in the r.h.s. of \equ(3.13) is equal to $0$ for
$\kk'_i=\bk++$.
 
The field $D_{\yy,\xx,\o}^{1,1(\le h)\sigma}$ is antiperiodic
in the space and time components of $\xx$ and $\yy$, of period $L$ and $\b$,
and is equal to $0$ if $\xx=\yy$ modulo $(L,\b)$. This means that it is
dimensionally equivalent to the product of $d(\xx,\yy)$ (see \equ(2.97)) and
the derivative of the field, so that the bound of its contraction with another
field variable on a scale $h'<h$ will produce a ``gain'' $\g^{-(h-h')}$ with
respect to the contraction of $\psi^{(\le h)\sigma}_{\yy,\o}$.
 
If we insert \equ(3.11) in the r.h.s. of \equ(3.10), we can decompose the l.h.s
in the sum of three terms, which differ from the term which $\RR$ acts on
mainly because one $\psi^{(\le h)}$ field is substituted with a
$D^{1,1(\le h)}$ field and some of the other $\psi^{(\le h)}$ fields are
``translated'' in the localization point. All three terms share the
property that the field whose $\xx$ coordinate is equal to the localization
point is not affected by the action of $\RR$.
 
In our approach, the regularization effect of $\RR$ will be exploited
trough the decomposition \equ(3.11). However, for reasons that will become
clear in the following, it is convenient to start the analysis by using
another representation of the expression resulting from the insertion of
\equ(3.11) in \equ(3.10). If
$\psi_{\xx_i}\=\psi^{(\le h)\s_i}_{\xx_i,\o_i}$, we can write, if the
localization point is $\xx_4$,
$$\eqalign{
&\RR \int d\xxx \prod_{i=1}^{4} \psi_{\xx_i} W(\xxx)=\cr
&=\int d\xxx \prod_{i=1}^{4} \psi_{\xx_i}
\Big[ W(\xxx)-\d(\xx_3-\xx_4) \int d\yy_3 W(\xx_1,\xx_2,\yy_3,\xx_4)
G_{\s_3}(\yy_3-\xx_4)\Big]+\cr
&+\int d\xxx \prod_{i=1}^{4} \psi_{\xx_i} \d(\xx_3-\xx_4) \int d\yy_3
\Big[ W(\xx_1,\xx_2,\yy_3,\xx_4) G_{\s_3}(\yy_3-\xx_4) -\cr
&-\d(\xx_2-\xx_4)
\int d\yy_2 W(\xx_1,\yy_2,\yy_3,\xx_4) G_{\s_3}(\yy_3-\xx_4)
G_{\s_2}(\yy_2-\xx_4)\Big]+\cr
&+\int d\xxx \prod_{i=1}^{4} \psi_{\xx_i} \d(\xx_2-\xx_4)  \d(\xx_3-\xx_4)
\int d\yy_2 \int d\yy_3
\Big[ W(\xx_1,\yy_2,\yy_3,\xx_4) G_{\s_3}(\yy_3-\xx_4)\cdot\cr
&\cdot G_{\s_2}(\yy_2-\xx_4) -\d(\xx_1-\xx_4)
\int d\yy_1 W(\yy_1,\yy_2,\yy_3,\xx_4) \prod_{i=1}^3
G_{\s_i}(\yy_i-\xx_4)\Big]\;,\cr}\Eq(3.14)$$
where $\d(\xx)$ is the antiperiodic delta function, that is
$$\d(\xx)={1\over L\b}\sum_{\kk'\in\DD'_{L,\b}} e^{i\s\kk'\xx}\;.\Eq(3.15)$$
Similar expressions are obtained, if the localization point is chosen in a
different way.
 
In the new representation, the action of $\RR$ is seen as the decomposition of
the original term in the sum of three terms, which are still of the form
\equ(3.3), but with a different kernel, containing suitable delta functions.
 
\*
2)If $2n=2$ and, possibly after a suitable permutation of the fields,
$\ss=(+,-)$, $\o_1=\o_2=\o$, by \equ(2.74),
$$\eqalign{
&\LL \int d\xx_1 d\xx_2 W_{2,\ss,\oo}^{(h)}(\xx_1-\xx_2)
\psi^{(\le h)+}_{\xx_1,\o} \psi^{(\le h)-}_{\xx_2,\o}=\cr
&=\int d\xx_1 d\xx_2 W_{2,\ss,\oo}^{(h)}(\xx_1-\xx_2)
\psi^{(\le h)+}_{\xx_1,\o} T_{\xx_2,\xx_1,\o}^{1(\le h)-}\cr
&=\int d\xx_1 d\xx_2 W_{2,\ss,\oo}^{(h)}(\xx_1-\xx_2)
T_{\xx_1,\xx_2,\o}^{1(\le h)+}\psi^{(\le h)-}_{\xx_2,\o} \;,\cr
}\Eq(3.16)$$
with
$$\eqalign{
&T_{\yy,\xx,\o}^{1(\le h)\s}=
\psi^{(\le h)\s}_{\xx,\o} c_\b(y_0-x_0) [c_L(y-x)+b_L d_L(y-x)]+\cr
&+[\bar\dpr_1\psi^{(\le h)\s}_{\xx,\o} +{i\cos p_F\over 2 v_0}
\bar\dpr_1^2\psi^{(\le h)\s}_{\xx,\o}]c_\b(y_0-x_0) a_L d_L(y-x)+\cr
&+\dpr_0\psi^{(\le h)\s}_{\xx,\o} d_\b(y_0-x_0) c_L(y-x)\;,\cr
}\Eq(3.17)$$
where $d_L(x)$ and $d_\b(x_0)$ are defined as in \equ(2.96) and
$$c_L(x)=\cos(\p xL^{-1})\;,\quad c_\b(x_0)=\cos(\p x_0\b^{-1})\;.\Eq(3.18)$$
 
As in the item 1), we define the localization point as the $\xx$ coordinate
of the field which is left unchanged $\LL$. We are free to choose it equal
to $\xx_1$ or $\xx_2$. This freedom affects also the action of $\RR$, which
can be written as
$$\eqalign{
&\RR \int d\xx_1 d\xx_2 W_{2,\ss,\oo}^{(h)}(\xx_1-\xx_2)
\psi^{(\le h)+}_{\xx_1,\o} \psi^{(\le h)-}_{\xx_2,\o}=\cr
&=\int d\xx_1 d\xx_2 W_{2,\ss,\oo}^{(h)}(\xx_1-\xx_2)
\psi^{(\le h)+}_{\xx_1,\o} D_{\xx_2,\xx_1,\o}^{2(\le h)-}\cr
&=\int d\xx_1 d\xx_2 W_{2,\ss,\oo}^{(h)}(\xx_1-\xx_2)
D_{\xx_1,\xx_2,\o}^{2(\le h)+}\psi^{(\le h)-}_{\xx_2,\o} \;,\cr
}\Eq(3.19)$$
with
$$D_{\yy,\xx,\o}^{2(\le h)\s}=\psi^{(\le h)\s}_{\yy,\o}-
T_{\yy,\xx,\o}^{1(\le h)\s}\;.\Eq(3.20)$$
Hence the effect of $\RR$ can be described as the replacement of a
$\psi^{(\le h)\s}$ field with a $D^{2(\le h)\s}$ field, with a gain in the
bounds (see discussion in item 1) above) of a factor $\g^{-2(h-h')}$.
 
Also in this case, it is possible to write the regularized term in the form
\equ(3.3). We get
$$\eqalignno{
&\RR \int d\xx d\yy W_{2,\ss,\oo}^{(h)}(\xx-\yy)
\psi^{(\le h)+}_{\xx,\o} \psi^{(\le h)-}_{\yy,\o}
=\int d\xx d\yy
\psi^{(\le h)+}_{\xx,\o} \psi^{(\le h)-}_{\yy,\o}
\Big\{ W_{2,\ss,\oo}^{(h)}(\xx-\yy)-\cr
&-\d(\yy-\xx) \int d\zz W_{2,\ss,\oo}^{(h)}(\xx-\zz)
c_\b(z_0-x_0) [c_L(z-x)+b_L d_L(z-x)]-\cr
&-[-\bar\dpr_1 \d(\yy-\xx) +{i\cos p_F\over 2 v_0}
\bar\dpr_1^2 \d(\yy-\xx) ]\int d\zz W_{2,\ss,\oo}^{(h)}(\xx-\zz)
c_\b(z_0-x_0) a_L d_L(z-x)-\cr
&+\dpr_0 \d(\yy-\xx) \int d\zz W_{2,\ss,\oo}^{(h)}(\xx-\zz)
d_\b(z_0-x_0) c_L(z-x)\Big\}\;.&\eq(3.21)\cr}$$
 
\*
3)If $2n=2$ and, possibly after a suitable permutation of the fields,
$\ss=(+,-)$, $\o_1=-\o_2=\o$, by \equ(2.74),
$$\eqalign{
&\LL \int d\xx_1 d\xx_2 W_{2,\ss,\oo}^{(h)}(\xx_1-\xx_2)
\psi^{(\le h)+}_{\xx_1,\o} \psi^{(\le h)-}_{\xx_2,-\o}=\cr
&=\int d\xx_1 d\xx_2 W_{2,\ss,\oo}^{(h)}(\xx_1-\xx_2)
\psi^{(\le h)+}_{\xx_1,\o} T_{\xx_2,\xx_1,-\o}^{0(\le h)-}\cr
&=\int d\xx_1 d\xx_2 W_{2,\ss,\oo}^{(h)}(\xx_1-\xx_2)
T_{\xx_1,\xx_2,\o}^{0(\le h)+}\psi^{(\le h)-}_{\xx_2,-\o} \;,\cr
}\Eq(3.22)$$
where
$$T_{\yy,\xx,\o}^{0(\le h)\sigma}=
c_\b(y_0-x_0) c_L(y-x)\psi^{(\le h)\sigma}_{\xx,\o}\;.\Eq(3.23)$$
 
Therefore
$$\eqalign{
&\RR \int d\xx_1 d\xx_2 W_{2,\ss,\oo}^{(h)}(\xx_1-\xx_2)
\psi^{(\le h)+}_{\xx_1,\o} \psi^{(\le h)-}_{\xx_2,-\o}=\cr
&=\int d\xx_1 d\xx_2 W_{2,\ss,\oo}^{(h)}(\xx_1-\xx_2)
\psi^{(\le h)+}_{\xx_1,\o} D_{\xx_2,\xx_1,-\o}^{1,2(\le h)-}\cr
&=\int d\xx_1 d\xx_2 W_{2,\ss,\oo}^{(h)}(\xx_1-\xx_2)
D_{\xx_1,\xx_2,\o}^{1,2(\le h)+}\psi^{(\le h)-}_{\xx_2,-\o} \;,\cr
}\Eq(3.24)$$
where
$$D_{\yy,\xx,\o}^{1,2(\le h)\sigma}=\psi^{(\le h)\sigma}_{\yy,\o}
-T_{\yy,\xx,\o}^{0(\le h)\sigma}\;.\Eq(3.25)$$
Hence the effect of $\RR$ can be described as the replacement of a
$\psi^{(\le h)\s}$ field with a $D^{1,2(\le h)\s}$ field, with a gain in the
bounds (see discussion in item 1) above) of a factor $\g^{-(h-h')}$.
As before, we can also write
$$\eqalign{
&\RR \int d\xx d\yy W_{2,\ss,\oo}^{(h)}(\xx-\yy)
\psi^{(\le h)+}_{\xx,\o} \psi^{(\le h)-}_{\yy,-\o}=
\int d\xx d\yy \psi^{(\le h)+}_{\xx,\o} \psi^{(\le h)-}_{\yy,-\o}\;\cdot\cr
&\cdot\;\Big\{ W_{2,\ss,\oo}^{(h)}(\xx-\yy)-
\d(\yy-\xx) \int d\zz W_{2,\ss,\oo}^{(h)}(\xx-\zz)
c_\b(z_0-x_0) c_L(z-x) \Big\}\;.\cr}\Eq(3.26)$$

\*
\sub(3.2) By using iteratively the ``single scale expansion'' \equ(2.112),
starting from $\hat\VV^{(1)}=\VV^{(1)}$, we can write the effective
potential $\VV^{(h)}(\sqrt{Z_h}\psi^{(\le h)})$, for $h\le 0$,
in terms of a {\it tree expansion}, similar to that described, for
example, in [BGPS].
 
\insertplot{300pt}{150pt}%
{\ins{30pt}{85pt}{$r$}\ins{50pt}{85pt}{$v_0$}\ins{130pt}{100pt}{$v$}%
\ins{35pt}{-2pt}{$h$}\ins{55pt}{-2pt}{$h+1$}\ins{135pt}{-2pt}{$h_v$}%
\ins{215pt}{-2pt}{$0$}\ins{235pt}{-2pt}{$+1$}\ins{255pt}{-2pt}{$+2$}}%
{fig51}{\eqg(1)}
 
\vglue.5truecm
We need some definitions and notations.
 
\0 1) Let us consider the family of all trees which can be constructed
by joining a point $r$, the {\it root}, with an ordered set of $n\ge 1$
points, the {\it endpoints} of the {\it unlabeled tree} (see Fig.
\graf(1)), so that $r$ is not a branching point. $n$ will be called the
{\it order} of the unlabeled tree and the branching points will be called
the {\it non trivial vertices}.
The unlabeled trees are partially ordered from the root to the endpoints in
the natural way; we shall use the symbol $<$ to denote the partial order.
 
Two unlabeled trees are identified if they can be superposed by a suitable
continuous deformation, so that the endpoints with the same index coincide.
It is then easy to see that the number of unlabeled trees with $n$ end-points
is bounded by $4^n$.
 
We shall consider also the {\it labeled trees} (to be called simply trees in
the following); they are defined by associating some labels with the unlabeled
trees, as explained in the following items.
 
\0 2) We associate a label $h\le 0$ with the root and we denote $\TT_{h,n}$ the
corresponding set of labeled trees with $n$ endpoints. Moreover, we introduce
a family of vertical lines, labeled by an an integer taking values in
$[h,2]$, and we represent any tree $\t\in\TT_{h,n}$ so that, if $v$ is an
endpoint or a non trivial vertex, it is contained in a vertical line with
index $h_v>h$, to be called the {\it scale} of $v$, while the root is on the
line with index $h$. There is the constraint that, if $v$ is an endpoint,
$h_v>h+1$.
 
The tree will intersect in general the vertical lines in set of
points different from the root, the endpoints and the non trivial vertices;
these points will be called {\it trivial vertices}. The set of the {\it
vertices} of $\t$ will be the union of the endpoints, the trivial vertices
and the non trivial vertices.
Note that, if $v_1$ and $v_2$ are two vertices and $v_1<v_2$, then
$h_{v_1}<h_{v_2}$.
 
Moreover, there is only one vertex immediately following
the root, which will be denoted $v_0$ and can not be an endpoint;
its scale is $h+1$.
 
Finally, if there is only one endpoint, its scale must be equal to $+2$ or
$h+2$.
 
\0 3) With each endpoint $v$ of scale $h_v=+2$ we associate one of the two
contributions to $\VV^{(1)}(\psi^{(\le 1)})$, written as in \equ(3.1) and a set
$\xx_v$ of space-time points (the corresponding integration variables),
two for $\l V_\l(\psi^{(\le 1)})$, one for $\n N(\psi^{(\le 1)})$;
we shall say that the
endpoint is of type $\l$ or $\n$, respectively. With each endpoint $v$ of
scale $h_v\le 1$ we associate one of the four local terms that we obtain if
we write $\LL V^{(h_v-1)}$ (see \equ(2.108)) by using the expressions
\equ(3.5) (there are four terms since $F_\a$ is the sum of two different
local terms), and one space-time point $\xx_v$; we shall say that the
endpoint is of type $\n$, $\d_1$, $\d_2$, $\l$, with an obvious correspondence
with the different terms.
 
Given a vertex $v$, which is not an endpoint, $\xx_v$ will denote the family
of all space-time points associated with one of the endpoints following $v$.
 
Moreover, we impose the constraint that, if $v$ is an endpoint and $\xx_v$ is
a single space-time point (that is the corresponding term is local),
$h_v=h_{v'}+1$, if $v'$ is the non trivial vertex immediately preceding $v$.
 
\0 4) If $v$ is not an endpoint, the {\it cluster } $L_v$ with frequency $h_v$
is the set of endpoints following the vertex $v$; if $v$ is an endpoint, it is
itself a ({\it trivial}) cluster. The tree provides an organization of
endpoints into a hierarchy of clusters.
 
\0 5) The trees containing only the root and an endpoint of scale $h+1$ will
be called the {\it trivial trees}; note that they do not belong to $\TT_{h,1}$,
if $h\le 0$, and can be associated with the four terms in the local part of
$\hat\VV^{(h)}$.
 
\0 6) We introduce a {\it field label} $f$ to distinguish the field variables
appearing in the terms associated with the endpoints as in item 3);
the set of field labels associated with the endpoint $v$ will be called $I_v$.
Analogously, if $v$ is not an endpoint, we shall
call $I_v$ the set of field labels associated with the endpoints following
the vertex $v$; $\xx(f)$, $\s(f)$ and $\o(f)$ will denote the space-time
point, the $\s$ index and the $\o$ index, respectively, of the
field variable with label $f$.
 
If $h_v\le 0$, one of the field variables belonging to $I_v$ carries also a
discrete derivative $\bar\dpr_1^m$, $m\in\{1,2\}$, if the corresponding local
term is of type $\d_m$, see \equ(3.5).
Hence we can associate with each field label $f$ an integer $m(f)\in\{0,1,2\}$,
denoting the order of the discrete derivative. Note that $m(f)$ is not
uniquely determined, since we are free to use the first or the second
representation of $F^{(\le h_v-1)}_\a$ in \equ(3.5); we shall use this freedom
in the following.
 
\*
By using \equ(2.112), it is not hard to see that, if $h\le 0$,
the effective potential can be written in the following way:
$$\VV^{(h)}(\sqrt{Z_h}\psi^{(\le h)}) + L\b \tilde E_{h+1}=
\sum_{n=1}^\io\sum_{\t\in\TT_{h,n}}
V^{(h)}(\t,\sqrt{Z_h}\psi^{(\le h)})\Eq(3.27)\;,$$
where, if $v_0$ is the first vertex of $\t$ and $\t_1,..,\t_s$ ($s=s_{v_0}$)
are the subtrees of $\t$ with root $v_0$,\\
$V^{(h)}(\t,\sqrt{Z_h}\psi^{(\le h)})$ is defined inductively by the relation
$$\eqalign{
&\qquad V^{(h)}(\t,\sqrt{Z_h}\psi^{(\le h)})=\cr
&{(-1)^{s+1}\over s!} \EE^T_{h+1}[\bar
V^{(h+1)}(\t_1,\sqrt{Z_{h}}\psi^{(\le h+1)});..; \bar
V^{(h+1)}(\t_{s},\sqrt{Z_{h}}\psi^{(\le h+1)})]\;,\cr}\Eq(3.28)$$
and $\bar V^{(h+1)}(\t_i,\sqrt{Z_{h}}\psi^{(\le h+1)})$
 
\0 a) is equal to $\RR\hat \VV^{(h+1)}(\t_i,\sqrt{Z_{h}}\psi^{(\le h+1)})$ if
the subtree $\t_i$ is not trivial (see \equ(2.107) for the definition of
$\hat \VV^{(h)}$);
 
\0 b) if $\t_i$ is trivial and $h\le -1$, it
is equal to one of the terms in the r.h.s. of \equ(2.108) with scale $h+1$ or,
if $h=0$, to one of the terms contributing to $\hat\VV^{(1)}(\psi^{\le 1})$.
 
If $h=0$, the r.h.s. of \equ(3.28) can be written more explicitly in the
following way. Given $\t\in\TT_{0,n}$, there are $n$ endpoints of scale $2$
and only another one vertex, $v_0$, of scale $1$; let us call $v_1,\ldots,
v_n$ the endpoints. We choose, in any set $I_{v_i}$, a subset
$Q_{v_i}$ and we define $P_{v_0}=\cup_i Q_{v_i}$; then we can write (recall
that $Z_0=1$)
$$V^{(0)}(\t,\sqrt{Z_0}\psi^{(\le 0)})=\sum_{P_{v_0}}
V^{(0)}(\t,P_{v_0})\;,\Eq(3.29)$$
$$V^{(0)}(\t,P_{v_0})=\sqrt{Z_0}^{|P_{v_0}|}\int d\xx_{v_0} \tilde\psi^{\le 0}
(P_{v_0}) K_{\t,P_{v_0}}^{(1)}(\xx_{v_0})\;,\Eq(3.30)$$
$$K_{\t,P_{v_0}}^{(1)}(\xx_{v_0})={1\over n!} \EE^T_{1}[
\tilde\psi^{(1)}(P_{v_1}\bs Q_{v_1}),\ldots,
\tilde\psi^{(1)}(P_{v_n}\bs Q_{v_n})]
\prod_{i=1}^n K^{(2)}_{v_i}(\xx_{v_i})\;,\Eq(3.31)$$
where we used the definitions
$$\tilde\psi^{(h)}(P_v)= \prod_{f\in P_v}\bar\dpr_1^{m(f)}
\psi^{(h)\s(f)}_{\xx(f),\o(f)}\;,\Eq(3.32)$$
$$K^{(2)}_{v_i}(\xx_{v_i})=e^{i\pp_F\sum_{f\in I_{v_i}}\xx(f)\s(f)}
\cases{ \l v_\l(\xx-\yy)
& if $v_i$ is of type $\l$ and $\xx_{v_i}=(\xx,\yy)$,\cr
\n & if $v_i$ is of type $\n$,}\Eq(3.33)$$
and we suppose that the order of the (anticommuting) field variables
in \equ(3.32) is suitable chosen in order to fix the sign as in \equ(3.31).
 
Note that the terms with $P_{v_0}\not=\emptyset$ in the r.h.s. of \equ(3.29)
contribute to $L\b \tilde E_1$, while the others contribute to
$\VV^{(0)}(\sqrt{Z_0}\psi^{(\le 0)})$.
 
The potential $\hat\VV^{(0)}(\sqrt{Z_{-1}}\psi^{(\le 0)})$, needed to iterate
the previous procedure, is obtained, as explained in \sec(2.5) and \sec(2.8),
by decomposing $\VV^{(0)}$ in the sum of $\LL\VV^{(0)}$ and $\RR\VV^{(0)}$, by
moving afterwards some local terms to the free measure and finally by
rescaling the fields variables. The representation we get for
$\VV^{(-1)}(\sqrt{Z_{-1}}\psi^{(\le -1)})$ depends on the representation
we use for $\RR V^{(0)}(\t,P_{v_0})$. We choose to use that based on
\equ(3.14), \equ(3.21) and \equ(3.26), where the regularization is seen, for
each term in the r.h.s. of \equ(3.29) with $P_{v_0}\not=\emptyset$,
as a modification of the kernel
$$W_{\t,P_{v_0}}^{(0)}(\xx_{P_{v_0}})=\int d(\xx_{v_0}\bs\xx_{P_{v_0}})
K_{\t,P_{v_0}}^{(1)}(\xx_{v_0})\;,\Eq(3.34)$$
where $\xx_{P_{v_0}}=\cup_{f\in P_{v_0}} \xx(f)$.
In order to remember this choice, we write
$$\RR V^{(0)}(\t,P_{v_0})=\sqrt{Z_0}^{|P_{v_0}|}\int d\xx_{v_0}
\tilde\psi^{(\le 0)}(P_{v_0})[\RR K_{\t,P_{v_0}}^{(1)}(\xx_{v_0})]\;.
\Eq(3.35)$$
 
It is then easy to get, by iteration of the previous procedure, a simple
expression for $V^{(h)}(\t,\sqrt{Z_h}\psi^{(\le h)})$, for any
$\t\in\TT_{h,n}$.
 
We associate with any vertex $v$ of the
tree a subset $P_v$ of $I_v$, the {\it external fields} of $v$.
These subsets must satisfy various constraints. First of all, if $v$ is not
an endpoint and $v_1,\ldots,v_{s_v}$ are the vertices immediately following it,
then $P_v \subset \cup_i P_{v_i}$; if $v$ is an endpoint, $P_v=I_v$. We shall
denote $Q_{v_i}$ the intersection of $P_v$ and $P_{v_i}$; this definition
implies that $P_v=\cup_i Q_{v_i}$. The subsets $P_{v_i}\bs Q_{v_i}$,
whose union will be made, by definition, of the {\it internal fields} of $v$,
have to be non empty, if $s_v>1$.
 
Given $\t\in\TT_{h,n}$, there are many possible choices of the subsets $P_v$,
$v\in\t$, compatible with all the constraints; we shall denote $\PP_\t$ the
family of all these choices and $\bP$ the elements of $\PP_\t$. Then we can
write
$$V^{(h)}(\t,\sqrt{Z_h}\psi^{(\le h)})=\sum_{\bP\in\PP_\t}V^{(h)}(\t,\bP)
\;;\Eq(3.36)$$
$V^{(h)}(\t,\bP)$ can be represented as in \equ(3.30), that is as
$$V^{(h)}(\t,\bP)=\sqrt{Z_h}^{|P_{v_0}|}\int d\xx_{v_0} \tilde\psi^{(\le h)}
(P_{v_0}) K_{\t,\bP}^{(h+1)}(\xx_{v_0})\;,\Eq(3.37)$$
with $K_{\t,\bP}^{(h+1)}(\xx_{v_0})$ defined inductively (recall that $h_{v_0}
=h+1$) by the equation, valid for any $v\in\t$ which is not an endpoint,
$$\eqalign{
K_{\t,\bP}^{(h_v)}(\xx_v)&={1\over s_v !}
\left({Z_{h_v}\over Z_{h_v-1}}\right)^{|P_v|\over 2}
\prod_{i=1}^{s_v} [K^{(h_v+1)}_{v_i}(\xx_{v_i})]\;\cdot\cr
&\cdot\;\tilde\EE^T_{h_v}[ \tilde\psi^{(h_v)}(P_{v_1}\bs Q_{v_1}),\ldots,
\tilde\psi^{(h_v)}(P_{v_{s_v}}\bs Q_{v_{s_v}})]\;,\cr}\Eq(3.38)$$
where $\tilde\EE^T_h$ denotes
the truncated expectation with propagator $g^{(h)}$ (without the scaling
factor $Z_{h-1}$, which is present in the definition of $\EE^T_h$ used
in \equ(2.112)) and $Z_1\=1$.
Moreover, if $v$ is an endpoint and
$h_v=2$, $K^{(h_v)}_{v}(\xx_v)$ is defined by \equ(3.33), otherwise
$$K^{(h_v)}_v(\xx_v)=
\cases{ \l_{h_v-1} & if $v$ is of type $\l$ ,\cr
i\o\d_{h_v-1} & if $v$ is of type $\d_1$, $d_2$ and $\o(f)=\o$
for both $f\in I_v$,\cr
\o\g^{h_v-1}\n_{h_v-1}& if $v$ is of type $\n$ and $\o(f)=\o$
for both $f\in I_v$.\cr}\Eq(3.39)$$
If $v$ is not an endpoint, $K^{(h_v)}_{v}=\RR K^{(h_v)}_{\t_i,\bP_i}$,
where $\t_1,\ldots,\t_{s_v}$ are the subtrees of $\t$ with root $v$,
$\bP_i=\{P_v,v\in\t_i\}$ and the action of $\RR$ is defined using
the representation \equ(3.14), \equ(3.21) and \equ(3.26) of the regularization
operation, seen as a modification of the kernel
$$W_{\t,\bP}^{(h_v)}(\xx_{P_v})=\int d(\xx_v\bs\xx_{P_v})
K_{\t,\bP}^{(h_v)}(\xx_v)\;,\Eq(3.40)$$
where $\xx_{P_v}=\cup_{f\in P_v} \xx(f)$.
Finally we suppose again that the order of the (anticommuting) field variables
is suitable chosen in order to fix the sign as in \equ(3.37).

{\bf Remark - }
The definitions \equ(3.14), \equ(3.21) and \equ(3.26) of $\RR$
are sufficient, even if they are restricted to external fields with
$m(f)=0$, because we can use the freedom in the definition of $m(f)$,
see item 6) above, so that the external fields of $v$ have always $m(f)=0$,
if $v$ is a vertex where the $\RR$ operation is acting on.
This last claim follows from the observation that,
since the truncated expectation in \equ(3.38)
vanishes if $s_v>1$ and $P_{v_i}\bs Q_{v_1}=\emptyset$ for some $i$, at least
one of the fields associated with the endpoints of type $\d_1$ or $\d_2$, the
only ones which have fields with $m(f)>0$, has to be an internal field; hence,
if one of the two fields is external, we can put $m(f)=0$ for it.
If $s_v=1$ the previous argument should not work, but in this case the only
vertex immediately following $v$ can be an endpoint of type $\d_1$ or $\d_2$
only if $v=v_0$, see item 2 above; however this is not a problem since the
action of $\RR$ on a local term is equal to $0$.
 
Note also that the kernel
$K_{\t,\bP}^{(h_v)}(\xx_v)$ is translation invariant, if
$\sum_{f\in P_v}\s(f)$ $=0$; in general, it satisfies the relation
$$K_{\t,\bP}^{(h_v)}(\xx_v+\xx)=
e^{i\pp_F \xx \sum_{f\in P_v}\s(f)}K_{\t,\bP}^{(h_v)}(\xx_v)\;.\Eq(3.41)$$
 
There is a simple interpretation of $V^{(h)}(\t,\bP)$ as the sum of a family
$\cal G_\bP$ of connected Feynman graphs build with single scale propagators
of different scales, connecting the space-time points associated with the
endpoints of the tree. A graph $g\in {\cal G}_\bP$ is build by
contracting, for any $v\in\t$, all the internal fields in couples in
all possible ways, by using the propagator $g^{h_v}$ , so that we get a
connected Feynman graph, if we represent as single points all the clusters
associated with the vertices immediately following $v$.
These graphs have the
property that the set of lines connecting the endpoints of the cluster $L_v$
and having scale $h'\ge h_v$ is a connected subgraph; by the way this
property is indeed another constraint on the possible choices of $\bP$.
We shall call these graphs {\it compatible} with $\bP$.
 
\*
\sub(3.3)
The representation \equ(3.37) of $V^{(h)}(\t,\bP)$ is based on the choice of
representing the regularization as acting on the kernels. If we use instead
the representation of $\RR$ based on \equ(3.10), \equ(3.11), \equ(3.19) and
\equ(3.24), some field variables have to be substituted with new ones,
depending on two space-time points and containing possibly some derivatives.
As we shall see, these new variables allow to get the right dimensional
bounds, at the price of making much more involved the combinatorics. Hence,
it is convenient to introduce a label $r_v(f)$ to keep trace of the
regularization in the vertices of the tree where $f$ is associated with
an external field and the action
of $\RR$ turns out to be {\it non trivial}, that is $\RR\not=1$.
 
There are many vertices, where $\RR=1$ by definition, that is the vertices
with more than $4$ external fields, the endpoints and $v_0$. For these
vertices all external fields will be associated with a label $r_v(f)=0$.
 
Moreover, since $\LL\RR=0$, the action of $\RR$ is trivial even in most
trivial vertices $v$ with $|P_v|\le 4$. This happens if the vertex (trivial or
not) $\tilde v$ immediately following $v$ has the same number of external
fields as $v$, since then the kernels associated with $v$ and $\tilde v$
are identical, up to a rescaling constant.
In particular, this remark implies that, given
the non trivial vertex $v$ and the non trivial vertex $v'$ immediately
preceding $v$ on the tree, there are at most two vertices $\bar v$, such that
$v'<\bar v \le v$ and the action of $\RR$ is non trivial.
For the same reason, given an endpoint $v$ of scale $h_v=+2$ of type $\l$
(hence not local), there are at most two
vertices between $v$ and the non trivial vertex $v'$ immediately
preceding $v$, where the action of $\RR$ is non trivial. Since the number
of endpoints is $n$ and the
number of non trivial vertices is bounded by $n-1$, the number of vertices
where the action of $\RR$ is non trivial is bounded by $2(2n-1)$.
 
Let us now consider one of these vertices, which all have $4$ or $2$ external
fields.
If $|P_v|=2$ and the $\o$ indices of the external fields are equal,
we keep trace of the regularization by
labeling the field variable, which is substituted with a $D^2$ field, see
\equ(3.19), with $r_v(f)=2$ and the other with $r_v(f)=0$. In principle
we are free to decide which variable is labeled with $r_v(f)=2$, that is
how we fix the localization point; we make a
choice in the following way. If there is no non trivial vertex $v'$ such that
$v_0\le v'<v$, we make an arbitrary choice, otherwise we put $r_v(f)=2$ for
the field which is an internal field in the nearest non trivial vertex
preceding $v$.
In other words, we try to avoid that a field affected by the
regularization stays external in the vertices preceding $v$.
 
If $|P_v|=2$ and the $\o$ indices of the external fields are different,
we label the field variable, which is
substituted with a $D^{1,2}$ field, see \equ(3.24),
with $r_v(f)=1$ and the other with $r_v(f)=0$; which variable is labeled
with $r_v(f)=1$ is decided as in the previous case.
 
If $|P_v|=4$, first of all we choose the
localization point in the following way. If there is a vertex $v'$ such that
$v_0\le v'<v$ and $P_{v'}$ contains one and only one $f\in P_v$, we chose
$\xx(f)$ as the localization point in $v$; in the other cases, we make an
arbitrary choice. After that, we split the kernel associated with $v$ into
three terms as in \equ(3.14); then we distinguish the three terms by putting
$r_v(f)=1$ for the external field which is substituted with a $D^{1,1(\le h)}$
field, when the delta functions are eliminated, and $r_v(f)=0$ for the others.
 
\*
The previous definitions imply that, given $f\in I_{v_0}$, it is possible
that there are
many different vertices in the tree, such that $r_v(f)\not=0$, that is
many vertices where the corresponding field variable appears as an external
field and the action of $\RR$ is non trivial.
As a consequence, the expressions given in \sec(3.1) for the regularized
potentials would not be sufficient and we should consider more general
expressions, containing as external fields more general variables.
Even worse, there is the risk that field derivatives of arbitrary order have
to be considered; this event would produce ``bad'' factorials in the bounds.
Fortunately, we can prove that this phenomenon can be easily controlled,
thanks to our choice of the localization point, see above, by a
more careful analysis of the regularization procedure, that we shall
keep trace of by changing the definition of the $r_v(f)$ labels.
 
Let us suppose first that $|P_v|=4$ and that there is $f\in P_v$, such that
$r_{\bar v}(f)\not=0$ for some $\bar v>v$. We want to show that the action of
$\RR$ on $v$ is indeed trivial; hence we can put $r_v(f)=0$ for all
$f\in P_v$, in agreement with the fact that the contribution to the
effective potential associated with $v$ is dimensionally irrelevant.
First of all, note that it is not possible that $|P_{\bar v}|=2$, as a
consequence of the choice of the localization point in the vertices with
two external fields, see above. On the other hand,
if $|P_{\bar v}|=4$, the fact that the action of $\RR$ in the vertex $v$ is
equal to the identity follows from the observation following \equ(3.13) and
the definition \equ(2.72).
 
Let us now consider the vertices $v$ with $P_v=(f_1,f_2)$. We can exclude
as before that $r_{\bar v}(f_i)\not=0$ for $i=1$ or $i=2$ or both
and $|P_{\bar v}|=2$.
The same conclusion can be reached, if there is no vertex $\bar v>v$,
such that $|P_{\bar v}|=4$, the action of $\RR$ on $\bar v$ is non trivial
and both $f_1$ and $f_2$ belong to the set of its
external fields; this claim easily follows from the criterion for the choice
of the localization point in the vertices with $4$ external fields.
 
If, on the contrary, $f_1$ and $f_2$ are both labels of external fields of a
vertex $\bar v>v$, such that $|P_{\bar v}|=4$ and the action of $\RR$ is non
trivial, we have to distinguish two possibilities.
If there is a non trivial vertex $v'$ such that $v_0\le v'<v$, and one of
the external fields of $v$, let us say of label $f_1$, is an internal field,
our choice of the localization points imply that both $r_v(f_1)$ and
$r_{\bar v}(f_1)$ are different from $0$, while $r_v(f_2)=r_{\bar v}(f_2)=0$.
If there is no non trivial vertex $v'<v$ with the previous property,
that is if $f_1$ and $f_2$ are both labels of external fields down to $v_0$
(hence all vertices between $v$ and $v_0$ are trivial) or they become
together labels of internal fields in some vertex $v'<v$, we are still free
to choose as we want the localization points in $v$ and $\bar v$; we decide
to choose them equal.
 
The previous discussion implies that, as a consequence of our
prescriptions, a field variable can be affected by the regularization only
once, except in the case considered in the last paragraph. However, also in
this case, it is easy to see that everything works as we did not apply to
the variable with label $f_1$ the regularization in the vertex $\bar v$.
In fact, the first or second order zero (modulo $(L,\b)$) in the difference
$\xx(f_1)-\xx(f_2)$, related to the regularization in the vertex $v$,
see \sec(3.1),
cancels the contribution of the term proportional to the delta function,
related with the regularization of $\bar v$, see \equ(3.14). This apparent
lack of regularization in $\bar v$ is compensated by the fact that
$\xx(f_1)-\xx(f_2)$ is of order $\g^{-h_{\bar v}}$, hence smaller than the
factor $\g^{-h_v}$ sufficient for the regularization of $v$ (together with
the improving effect of the field derivative). Hence there is a gain with
respect to the usual bound of a factor $\g^{-(h_{\bar v}-h_v)}$, sufficient to
regularize the vertex $\bar v$.
 
\*
\sub(3.4)
There is in principle another problem. Let us suppose that we decide to
represent all the non trivial $\RR$ operations as acting on
the field variables. Let us suppose also that
the field variable with label $f$ is substituted, by the action of $\RR$
on the vertex $v$, with a $D_{\yy,\xx}^{1,i}$ or a $D_{\yy,\xx}^2$ field,
where $\yy=\xx(f)$ and $\xx=\xx(f')$ is the corresponding localization point.
At first sight it seems possible that even the variable with label $f'$ can
be substituted with a $D^{1,i}$ or a $D^2$ field by the action of $\RR$ on
a vertex $\bar v>v$. If this happens, the point $\xx(f')$ can not be considered
as fixed and there is an ``interference'' between the two regularization
operations, or even more than two, since this phenomenon could involve an
ordered chain of vertices. This interference would not produce bad factorials
in the bounds, but would certainly make more involved our expansion. However,
we can show that, thanks to our localization prescription, this problem is
not really present.
 
Let us suppose first that $|P_v|=2$. In this case, if the field with label
$f'$ is external in some vertex $\bar v>v$, with $|P_{\bar v}|$ equal to $2$
or $4$, we are sure that $\xx(f')$ is the localization point in $\bar v$, see
\sec(3.3), hence the corresponding filed can not be affected by the action
of $\RR$ on $\bar v$. The same conclusion can be reached, if $|P_v|=4$ and
$|P_{\bar v}|=2$
 
If $|P_v|=|P_{\bar v}|=4$ and the field with label $f'$ is substituted,
by the action of $\RR$ on the vertex $\bar v$, with a $D^{1,i}$
or a $^2$ field, we know that the same can not be true for the field with
label $f$, since the action of $\RR$ on $v$ is trivial.
 
The previous discussion implies that the field with label $f'$ can
be affected by the regularization (if $|P_v|=|P_{\bar v}|=4$) only by
changing its $\xx$ label, but this is not a source of any problem.
 
\*
\sub(3.5)
In this section we want to discuss the representation of the fields
$D^{1,i(\le h)\s}_{\yy,\xx,\o}$, $i=1,2$, and $D^{2(\le h)\s}_{\yy,\xx,\o}$
introduced in \sec(3.1), which allows to exploit the regularization properties
of the $\RR$ operation.
In order to do that, we extend the definition
of the fields $\psi^{(\le h)\s}_{\xx,\o}$ to $\RRR^2$, by using \equ(2.49);
we get functions with values in the Grassman algebra, antiperiodic in $x_0$
and $x$ with periods $\b$ and $L$, respectively.
 
Let us choose a family of positive functions $\c_{\h,\h'}(\xx)$,
$\h,\h'\in\{-1,0,+1\}$, on $\RRR^2$, such that
$$\eqalign{
\c_{\h,\h'}(\xx) &= \cases{1 & if $|x-\h|\le 1/4$ and $|x_0-\h'|\le 1/4$\cr
0 & if $|x-\h|\ge 3/4$ or $|x_0-\h'|\ge 3/4$\cr}\cr
\sum_{\h,\h'} \c_{\h,\h'}(\xx) &=1\quad \hbox{if\ } \xx\in [-1,1]\times
[-1,1]\;.\cr}\Eq(3.42)$$
 
Given $\xx,\yy\in\L\times[-\b/2,\b/2]$, if $\c_{\h,\h'}(\tilde\yy-
\tilde\xx)>0$, where $\tilde\xx=(x/L,x_0/\b)$ and $\tilde\yy=(y/L,y_0/\b)$,
we can define $\bar\yy=\yy-(\h L,\h'\b)$, so that $|x_0-\bar y_0|\le 3\b/4$
and $|x-\bar y|\le 3L/4$. We
see immediately that $D^{1,1(\le h)\s}_{\yy,\xx,\o}=
(-1)^{\h+\h'} D^{1,1(\le h)\s}_{\bar\yy,\xx,\o}$ and we can write
$$D^{1,1(\le h)\s}_{\bar\yy,\xx,\o}=
[\psi^{(\le h)\s}_{\bar\yy,\o}-\psi^{(\le h)\s}_{\xx,\o}]+
[1-G_\s(\bar\yy-\xx)] \psi^{(\le h)\s}_{\xx,\o}\;.\Eq(3.43)$$
It is easy to see that, if $|y_0|\le 3\b/4$ and $|y|\le 3L/4$,
$$1-G_\s(\yy)={1\over L}\bh_1(\tilde\yy)d_L(y)+{1\over \b}
\bh_2(\tilde\yy)d_\b(y_0)\;,\quad \tilde\yy=(y/L,y_0/\b)\;,\Eq(3.44)$$
where $\bh_i(\yy)$, $i=1,2$, are suitable functions, uniformly smooth
in $L$ and $\b$. Moreover
$$\psi^{(\le h)\s}_{\bar\yy,\o}-\psi^{(\le h)\s}_{\xx,\o}=
(\bar\yy-\xx)\cdot \int_0^1 \; dt\; \Dpr \psi^{(\le h)\s}_{\xxi(t),\o}
\;,\quad \xxi(t)=\xx+t(\bar\yy-\xx)\;,\Eq(3.45)$$
where $\Dpr=(\dpr_1,\dpr_0)$ is the gradient, and it is easy to see that,
if $|y_0|\le 3\b/4$ and $|y|\le 3L/4$,
$$\yy=\big( \bh_3(\tilde\yy)d_L(y), \bh_4(\tilde\yy)d_\b(y_0)\big)\;,
\Eq(3.46)$$
where $\bh_i(\yy)$, $i=3,4$, are other suitable functions, uniformly smooth
in $L$ and $\b$.
 
Hence we can write
$$\eqalignno{
&D^{1,1(\le h)\s}_{\yy,\xx,\o}=\sum_{\h,\h'} \Big\{
\left[ {1\over L} h_{1,\h,\h'}(\tilde\yy,\tilde\xx) d_L(y-x)+
{1\over \b} h_{2,\h,\h'}(\tilde\yy,\tilde\xx) d_\b(y_0-x_0) \right]
\psi^{(\le h)\s}_{\xx,\o}+&\eq(3.47)\cr
&+h_{3,\h,\h'}(\tilde\yy,\tilde\xx) d_L(y-x)
\int_0^1 \; dt\; \dpr_1 \psi^{(\le h)\s}_{\xxi(t),\o}
+h_{4,\h,\h'}(\tilde\yy,\tilde\xx) d_\b(y_0-x_0) \int_0^1 \; dt\;
\dpr_0\psi^{(\le h)\s}_{\xxi(t),\o}\Big\}\;,\cr}$$
where
$$h_{i,\h,\h'}(\tilde\yy,\tilde\xx)=(-1)^{\h+\h'}
\c_{\h,\h'}(\tilde\yy-\tilde\xx) \bh_i((\bar y-x)/L,(\bar y_0-x_0)/\b)
\;,\quad i=1,4,\Eq(3.48)$$
are smooth functions with support
in the region $\{|y-x-\h L|\le 3L/4,|y_0-x_0-\h'\b|\le 3\b/4\}$, such that their
derivatives of order $n$ are bounded by a constant (depending on $n$)
times $\g^{nh_{L,\b}}$.
 
A similar expression is valid for $D^{1,2(\le h)\s}_{\yy,\xx,\o}$. Let us now
consider $D^{2(\le h)\s}_{\yy,\xx,\o}$, see \equ(3.20). We can write
$$D^{2(\le h)\s}_{\yy,\xx,\o}=(-1)^{\h+\h'} \tilde D^{2(\le h)\s}_{\bar\yy,
\xx,\o}+ h(\tilde\yy-\tilde\xx) d_L(y-x)
\bar\dpr_1^2\psi^{(\le h)\s}_{\xx,\o}\;,\Eq(3.49)$$
where $h(\yy-\xx)$ is a uniformly smooth function and
$$\eqalign{
&\tilde D^{2(\le h)\s}_{\bar\yy,\xx,\o}=
\psi^{(\le h)\s}_{\bar\yy,\o}-\psi^{(\le h)\s}_{\xx,\o}-
(\bar\yy-\xx)\cdot \Dpr\psi^{(\le h)\s}_{\xx,\o}-\cr
&-\psi^{(\le h)\s}_{\xx,\o} \{ [ c_\b(\bar y_0-x_0) c_L(\bar y-x)-1]+
b_L c_\b(\bar y_0-x_0) d_L(\bar y-x) \}-\cr
&-\bar\dpr_1\psi^{(\le h)\s}_{\xx,\o} \{ [c_\b(\bar y_0-x_0)-1] d_L(\bar y-x)
+[d_L(\bar y-x)-(\bar y-x)] \}-\cr
&-(\bar y-x) [\bar\dpr_1\psi^{(\le h)\s}_{\xx,\o}-
\dpr_1\psi^{(\le h)\s}_{\xx,\o}]-
\dpr_0\psi^{(\le h)\s}_{\xx,\o} [d_\b(\bar y_0-x_0) c_L(y-x)-(\bar y_0-x_0)]
\;.\cr}\Eq(3.50)$$
Note that
$$\bar\dpr_1\psi^{(\le h)\s}_{\xx,\o} -\dpr_1\psi^{(\le h)\s}_{\xx,\o}
= {i\s\over L\b}\sum_{\kk'\in\DD'_{L,\b}}
e^{i\s\kk'\xx} (\sin k'-k') \hat\psi^{(h)\s}_{\kk',\o}\Eq(3.51)$$
behaves dimensionally as $\dpr_1^3\psi^{(\le h)\s}_{\xx,\o}$, hence we
shall define
$$\bar\dpr_1^3\psi^{(\le h)\s}_{\xx,\o}=
\bar\dpr_1\psi^{(\le h)\s}_{\xx,\o} -\dpr_1\psi^{(\le h)\s}_{\xx,\o}
\;.\Eq(3.52)$$
It is now easy to show that there exist functions $h_{\un,\h,\h'}(\yy,\xx)$,
with $\un=(n_1,\ldots,n_6)$, and $h_{i,j,\h,\h'}(\yy,\xx)$, $i,j=0,1$, smooth
uniformly in $L$ and $\b$, such that
$$\eqalignno{
D^{2(\le h)\s}_{\yy,\xx,\o}&= \sum_{\h,\h'} \Big\{
\sum_\un h_{\un,\h,\h'}(\tilde\yy,\tilde\xx)
d_L(y-x)^{n_1} d_\b(y_0-x_0)^{n_2} L^{-n_3} \b^{-n_4}
\bar\dpr_1^{n_5}\dpr_0^{n_6}\psi^{(\le h)\s}_{\xx,\o} +\cr
&+\sum_{i,j} h_{i,j,\h,\h'}(\tilde\yy,\tilde\xx) d_i(\yy-\xx) d_i(\yy-\xx)
\int_0^1 dt (1-t) \dpr_i\dpr_j \psi^{(\le
h)\s}_{\xxi(t),\o}\Big\}\;,&\eq(3.53)\cr}$$
the sum over $\un$ being constrained by the conditions
$$n_1+n_2\le 2\;,\quad 3\ge \sum_{i=3}^6 n_i \ge 2\;.\Eq(3.54)$$
 
\*
\sub(3.6)
In order to exploit the regularization properties of formulas like \equ(3.47)
or \equ(3.53), one has to prove that the ``zeros'' $d_L(y-x)$ and
$d_\b(y_0-x_0)$ give a contribution to the bounds of order $\g^{-h'}$, with
$h'\ge h$, if $h$ is the scale at which the zero was produced by the action of
$\RR$. In \sec(3.7) we shall realize this task by ``distributing'' the zeros
along a path connecting a family of space-time points associated with a
subset of field variables.
Let $\xx_0=\xx,\xx_1,\ldots,\xx_n=\yy$ be the family of points
connected by the path; it is easy to show that
$$d_L(y-x)=\sum_{r=1}^n d_L(x_r-x_{r-1}) e^{-i\fra{\p}L
(x_r+x_{r-1}-x_n-x_0)}\;.\Eq(3.55)$$
A similar expression is valid for $d_\b(y_0-x_0)$.
 
It can happen that one of the terms in the r.h.s. of \equ(3.55) or the
analogous expansion for $d_\b(y_0-x_0)$ depends
on the same space-time points as the integration variables in the r.h.s. of
a term like \equ(3.21) or \equ(3.26).
We want to study the effect of this event.
Let us call $W(\xx-\yy)$ the kernel appearing in the
l.h.s. of \equ(3.21) or \equ(3.26), $W_R(\xx-\yy)$ its regularization, that
is the quantity appearing in braces in the corresponding r.h.s.,
and let us define
$$I_{n_1,n_2} =
\int d\xx d\yy \psi^{(\le h)+}_{\xx,\o} \psi^{(\le h)-}_{\yy,\o}
W_R(\xx-\yy) [e^{-i\p\fra{y}L}d_L(y-x)]^{n_1} [e^{-i\p\fra{y_0}{\b}}
d_\b(y_0-x_0)]^{n_2}\;.\Eq(3.56)$$
In the following we shall meet such expressions for values of $n_1$ and $n_2$,
such that $1\le n_1+n_2 \le 2$.
 
If $W(\xx-\yy)$ is the kernel appearing in the
l.h.s. of \equ(3.26), it is easy to see that, if $n_1+n_2 \ge 1$,
$$I_{n_1,n_2} =
\int d\xx d\yy \psi^{(\le h)+}_{\xx,\o} \psi^{(\le h)-}_{\yy,\o}
W(\xx-\yy) [e^{-i\p\fra{y}L}d_L(y-x)]^{n_1} [e^{-i\p\fra{y_0}{\b}}
d_\b(y_0-x_0)]^{n_2}\;,\Eq(3.57)$$
that is the presence of the zeros simply erases the effect of the
regularization.
 
Let us now suppose that $W(\xx-\yy)$ is the kernel appearing in the
l.h.s. of \equ(3.21) and $W_R(\xx-\yy)$ its regularization. We have
$$\eqalign{
I_{1,0}&=
\int d\xx d\yy \psi^{(\le h)+}_{\xx,\o}
W(\xx-\yy) d_L(y-x) \Big\{ D^{1,3(\le h)-}_{\yy,\xx,\o}-\cr
&-c_\b(y_0-x_0)[
\fra12 \bar\dpr_1^2 (e^{-i\p\fra{x}L}\psi^{(\le h)-}_{\xx,\o})+
{i\cos p_F\over v_0}\bar\dpr_1 (e^{-i\p\fra{x}L}\psi^{(\le h)-}_{\xx,\o})
]\Big\}\;,\cr}\Eq(3.58)$$
 
$$I_{0,1}= \int d\xx d\yy \psi^{(\le h)+}_{\xx,\o}W(\xx-\yy) d_\b(y_0-x_0)
D^{1,4(\le h)-}_{\yy,\xx,\o}\;,\Eq(3.59)$$
where
$$D^{1,3(\le h)-}_{\yy,\xx,\o}= e^{-i\p\fra{y}L}\psi^{(\le h)-}_{\yy,\o}-
c_\b(y_0-x_0)e^{-i\p\fra{x}L}\psi^{(\le h)-}_{\xx,\o}\;.\Eq(3.60)$$
$$D^{1,4(\le h)-}_{\yy,\xx,\o}=
e^{-i\p\fra{y_0}\b}\psi^{(\le h)-}_{\yy,\o}-
c_L(y-x) e^{-i\p\fra{x_0}\b}\psi^{(\le h)-}_{\xx,\o}\;.\Eq(3.61)$$
Moreover
$$\eqalignno{
&I_{2,0}=
\int d\xx d\yy \psi^{(\le h)+}_{\xx,\o}
W(\xx-\yy) d_L(y-x) \Big\{d_L(y-x)e^{-2i\p\fra{y}L}\psi^{(\le h)-}_{\yy,\o}-
{c_\b(y_0-x_0)\over a_L}\;\cdot\cr
&\cdot\;[\bar\dpr_1 (e^{-2i\p\fra{x}L}\psi^{(\le h)-}_{\xx,\o})+
{i\cos p_F\over v_0}
\big(e^{-2i\p\fra{x}L}\psi^{(\le h)-}_{\xx,\o}+
\fra12 \bar\dpr_1^2 (e^{-2i\p\fra{x}L}\psi^{(\le h)-}_{\xx,\o})\big)
]\Big\}\;,&\eq(3.62)\cr}$$

$$I_{0,2}=\int d\xx d\yy \psi^{(\le h)+}_{\xx,\o}
W(\xx-\yy) d_\b(y_0-x_0)^2 e^{-2i\p\fra{y_0}\b}\psi^{(\le h)-}_{\yy,\o}
\;,\Eq(3.63)$$
 
$$I_{1,1}=\int d\xx d\yy \psi^{(\le h)+}_{\xx,\o}
W(\xx-\yy) d_L(y-x) d_\b(y_0-x_0) e^{-i\p\fra{y}L-i\p\fra{y_0}\b}
\psi^{(\le h)-}_{\yy,\o}\;.\Eq(3.64)$$

Note that no cancellations are possible for $\xx=\yy$ modulo $(L,\b)$
between the various terms contributing to $I_{n_1,n_2}$;
hence they will be bounded separately.
 
Note also that the fields
$D^{1,3(\le h)-}_{\yy,\xx,\o}$ and $D^{1,4(\le h)-}_{\yy,\xx,\o}$
have a zero of first order for $\xx=\yy$ modulo $(L,\b)$ and can be
represented by
expressions analogous to the r.h.s. of \equ(3.47). Moreover, the
terms contributing to $I_{0,1}$ and $I_{1,0}$ and containing  these
fields can also be written in a form analogous to \equ(3.26).
 
Finally, we want to stress the fact that the integrands in the previous
expressions of $I_{n_1,n_2}$, $1\le n_1+n_2 \le 2$, have a zero of order
at most two for $\xx=\yy$ modulo $(L,\b)$, that is a zero of order not
higher of the zero introduced in the r.h.s. of \equ(3.56).
As it will be more clear in \sec(3.7), this property would be lost if one
uses the representation \equ(3.19) of the regularization operation, before
performing the "decomposition of the zeros"; one should get in this case a
zero of order four and the iteration
of the procedure of decomposition of the zeros would produce zeros of
arbitrary order and, as a consequence, bad combinatorial factors in the
bounds.
 
\*
\sub(3.7) We are now ready to describe in more detail our expansion. First
of all, we insert the decomposition \equ(3.14) of $V^{(h)}(\t,\psi^{(\le
h)})$ in the vertices with $|P_v|=4$, by following the prescription for
the choice of the localization
point described in \sec(3.3). The discussion of \sec(3.3) allows also
to define a new label $r(f)$, to be called the $\RR$-label,
for any $f\in I_{v_0}$, by putting
 
\0 (i) $r(f)=0$, if $r_v(f)=0$ for any $v$ such that $f\in P_v$;
 
\0 (ii) $r(f)=(i,v)$, if there exists one and only one vertex $v$,
such that $f\in P_v$ and $r_v(f)=i\not=0$;
 
\0 (iii) $r(f)=(2,v,\bar v)$, if there are two vertices $v$ and $\bar v$,
such that $v<\bar v$, $f\in P_v \subset P_{\bar v}$, $|P_v|=2$, $|P_{\bar
v}|=4$, $r_v(f)=2$, $r_{\bar v}(f)=1$; see discussion in the last two
paragraphs of \sec(3.3).
 
Then, we can write
$$V^{(h)}(\t,\sqrt{Z_h}\psi^{(\le h)})=\sum_{\bP\in\PP_\t,\rr}
V^{(h)}(\t,\bP,\rr)\;,\Eq(3.65)$$
where $\rr=\{r(f),f\in I_{v_0}\}$ and the sum over $\rr$ must be understood
as the sum over the possible choices of $\rr$ compatible with $\bP$.
 
We can also write
$$V^{(h)}(\t,\bP,\rr)=\sqrt{Z_h}^{|P_{v_0}|}\int d\xx_{v_0}
K_{\t,\bP,\rr}^{(h)}(\xx_{v_0})\tilde\psi^{(\le h)}(P_{v_0})\;,\Eq(3.66)$$
with $K_{\t,\bP,\rr}^{(h)}(\xx_{v_0})$ defined inductively as in \equ(3.38).
 
Let us consider first the action of $\RR$ on $V^{(h)}(\t,\bP,\rr)$. We can
write for $\RR V^{(h)}(\t,\bP,\rr)$ an expression similar to \equ(3.66), if
we continue to use for the $\RR$ operation the representation based on
\equ(3.14),\equ(3.21) and \equ(3.26), which affects the kernels leaving the
fields unchanged. We shall use the notation
$$\RR V^{(h)}(\t,\bP,\rr)=\int d\xx_{v_0}\tilde\psi^{(\le h)}(P_{v_0})
[\RR K^{(h)}_{\t,\bP,\rr}(\xx_{v_0})]\;.\Eq(3.67)$$
Moreover, we define $\rr'$ so that $r'(f)=r(f)$ except for the field labels
$f\in P_{v_0}$, for which $r'(f)$ takes into account also the regularization
acting on $v_0$.
 
However, we can use for the $\RR$ operation also the representation
based on \equ(3.10), \equ(3.11), \equ(3.19) and \equ(3.24), which
can be derived from the previous one by integrating the
$\d$-functions; the effect is to replace one of the external fields
with one of the fields $D^{1,i(\le h)\s}$, $i=1,2$ or $D^{2(\le h)\s}$.
We can describe the result by writing
$$\RR V^{(h)}(\t,\bP,\rr)=\int d\xx_{v_0}[\RR \tilde\psi^{(\le h)}(P_{v_0})]
K^{(h)}_{\t,\bP,\rr}(\xx_{v_0})\;.\Eq(3.68)$$
 
The discussion in \sec(3.3) and \sec(3.5) implies that there is a finite set
$A_{v_0}$, such that
$$[\RR \tilde\psi^{(\le h)}(P_{v_0})]=\sum_{\a\in A_{v_0}}
h_\a(\tilde\xx_{P_{v_0}}) d^{n_1(\a)}_L d^{n_2(\a)}_\b \prod_{f\in P_{v_0}}
[\hat\partial_{j_\a(f)}^{q_\a(f)}\psi]^{(\le h)\s(f)}_{\xx_\a(f),\o(f)}\;,
\Eq(3.69)$$
where $\tilde\xx_{P_{v_0}}=(L^{-1} x_{P_{v_0}}, \b^{-1}x_{0 P_{v_0}})$,
$d^{n_1(\a)}_L$ and $d^{n_2(\a)}_\b$ are powers of the functions \equ(2.96),
with argument the difference of two points belonging to $\xx_{P_{v_0}}$,
and $\hat\partial_j^q$, $q=0,1,2$, $j=1,\ldots,m_q$, is a family of
operators acting on the field variables, which are
dimensionally equivalent to derivatives of order $q$. In particular $m_0=1$,
$\hat\partial_1^0$ is the identity and the action
of $\RR$ is trivial, that is $|A_{v_0}|=1$, $h_\a=1$, $n_1(\a)=n_2(\a)=0$
and $q_\a(f)=0$ for any $f\in P_{v_0}$, except in the following cases.
 
\*
\0 1) If $|P_{v_0}|=4$ and $r(f)=0$ for any $f\in P_{v_0}$,
there is $\bar f\in P_{v_0}$,
such that the action of $\RR$ over the fields consists in replacing one of the
field variables with a $D^{1,1(\le h)\s}_{\yy,\xx,\o}$ field, where
$\yy=\xx(\bar f)$ and $\xx=\xx(f)$ for some other $f\in P_{v_0}$, see
\equ(3.11); moreover, one or two of the other fields change their space-time
point. We write $D^{1,1(\le h)\s}_{\yy,\xx,\o}$ in the representation
\equ(3.47); the resulting expression is of the form \equ(3.69), with
$A_{v_0}$ consisting of four different terms, such that $d_L=d_L(y-x)$,
$d_\b=d_\b(y_0-x_0)$, $n_1(\a)+n_2(\a)=1$ and, for all $f\not=\bar f$,
$q_\a(f)=0$, while $q_\a(\bar f)=1$.
Moreover, if $f\not=\bar f$, $\xx_\a(f)$ is
a single point belonging to $\xx_{P_{v_0}}$, not necessarily coinciding with
$\xx(f)$, while, if $f=\bar f$, $\xx_\a(f)$ is equal to $\xx$ or to the couple
$(\xx,\yy)$ (using the previous definitions). The precise values of
$\xx_\a(\bar f)$ and $[\hat\partial^1_{j_\a(\bar f)}\psi]^{(\le h)
\s(\bar f)}_{\xx_\a(\bar f),\o(\bar f)}$,
together with the functions $h_\a$, can be deduced from \equ(3.47).
 
\0 2) If $P_{v_0}=(f_1,f_2)$ and $\o(f_1)=\o(f_2)$,
the action of $\RR$ consists in replacing one of the external fields,
of label, let us say, $f_1$, with a $D^{2(\le h)\s}_{\yy,\xx,\o}$ field, where
$\yy=\xx(f_1)$ and $\xx=\xx(f_2)$, if $f_2$ is the second field label.
By using the representation \equ(3.53) of $D^{2(\le h)\s}_{\yy,\xx,\o}$,
we get an expression of the form \equ(3.69)
consisting of many different terms, such that $d_L=d_L(y-x)$,
$d_\b=d_\b(y_0-x_0)$, $n_1(\a)+n_2(\a)\le 2$, $q_\a(f_1)=2$, $q_\a(f_2)=0$,
$\xx_\a(f_2)=\xx(f_2)$. The values of $\xx_\a(f_1)$ and
$[\hat\partial^2_{j_\a(f_1)}\psi]^{(\le h)\s(f_1)}_{\xx_\a(f_1),\o(f_1)}$,
together with the functions $h_\a$, can be deduced from \equ(3.53).
 
\0 3) If $P_{v_0}=(f_1,f_2)$ and $\o(f_1)=-\o(f_2)$,
the action of $\RR$ consists in replacing one of the external fields,
of label, let us say, $f_1$, with a $D^{1,2(\le h)\s}_{\yy,\xx,\o}$ field,
where $\yy=\xx(f_1)$ and $\xx=\xx(f_2)$, if $f_2$ is the second field label.
By using the analogous of the representation \equ(3.47) for
$D^{1,2(\le h)\s}_{\yy,\xx,\o}$, we get an expression of the form \equ(3.69)
consisting of four different terms, such that $n_1(\a)+n_2(\a)=1$,
$q_\a(f_1)=1$, $q_\a(f_2)=0$, $\xx_\a(f_2)=\xx(f_2)$.
 
Let us now consider the action of $\LL$ on $V^{(h)}(\t,\sqrt{Z_h}
\psi^{(\le h)})$. We get an expansion similar to that based on \equ(3.68),
that we can write, by using \equ(2.79), \equ(3.65) and translation invariance,
in the form
$$\eqalign{
\LL V^{(h)}(\t,\sqrt{Z_h}\psi^{(\le h)})&=\g^h n_h(\t) Z_h F_\nu^{(\le h)}+
s_h(\t) Z_h F_\s^{(\le h)}+z_h(\t) Z_h F_\z^{(\le h)}+\cr &+a_h(\t) Z_h
F_\a^{(\le h)}+l_h(\t) Z_h^2 F_\l^{(\le h)}\;,\cr}\Eq(3.70)$$
where
$$\eqalign{
n_h(\t)&= {\g^{-h}\over L\b} \sum_{\bP\in\PP_\t,\rr\atop P_{v_0}=(f_1,f_2),
\o(f_1)=\o(f_2)=+1} \int d\xx_{v_0} h_1(\tilde\xx_{P_{v_0}})
K_{\t,\bP,\rr}^{(h)}(\xx_{v_0})\;,\cr
s_h(\t)&= {1\over L\b} \sum_{\bP\in\PP_\t,\rr\atop P_{v_0}=(f_1,f_2),\o(f_1)
=-\o(f_2)=+1} \int d\xx_{v_0} h_2(\tilde\xx_{P_{v_0}})
K_{\t,\bP,\rr}^{(h)}(\xx_{v_0})\;,\cr
z_h(\t)&= {1\over L\b} \sum_{\bP\in\PP_\t,\rr\atop P_{v_0}=(f_1,f_2),\o(f_1)
=\o(f_2)=+1} \int d\xx_{v_0} h_3(\tilde\xx_{P_{v_0}})
d_\b(x(f_2)-x(f_1)) K_{\t,\bP,\rr}^{(h)}(\xx_{v_0})\;,\cr
a_h(\t)&= {1\over L\b} \sum_{\bP\in\PP_\t,\rr\atop P_{v_0}=(f_1,f_2),\o(f_1)
=\o(f_2)=+1} \int d\xx_{v_0} h_4(\tilde\xx_{P_{v_0}})
d_L(x(f_2)-x(f_1)) K_{\t,\bP,\rr}^{(h)}(\xx_{v_0})\;,\cr
l_h(\t)&= {1\over L\b} \sum_{\bP\in\PP_\t,\rr\atop |P_{v_0}|=4,
\ss=(+,-,+,-), \oo=(+1,-1,-1,+1)}
\int d\xx_{v_0} h_5(\tilde\xx_{P_{v_0}}) K_{\t,\bP,\rr}^{(h)}(\xx_{v_0})\;,\cr
}\Eq(3.71)$$
$h_i(\tilde\xx_{P_{v_0}})$, $i=1,\ldots,5$, being bounded functions, whose
expressions can be deduced from \equ(3.8), \equ(3.16) and \equ(3.22), also
taking into account the permutations needed to order the field variables
as in the r.h.s. of \equ(3.70).
 
The constants $n_h$, $s_h$, $z_h$, $a_h$ and $l_h$, which characterize the
local part of the effective potential, can be obtained from \equ(3.71) by
summing over $n\ge 1$ and $\t\in\TT_{h,n}$. Finally, the constant $\tilde
E_{h+1}$ appearing in the l.h.s. of \equ(3.27) can be written in the form
$$\tilde E_{h+1}=\sum_{n=1}^\io\sum_{\t\in\TT_{h,n}} \tilde E_{h+1}(\t)\;,
\Eq(3.72)$$
where
$$\tilde E_{h+1}(\t) = {1\over L\b}
\sum_{\bP\in\PP_\t,\rr\atop P_{v_0}=\emptyset} \int d\xx_{v_0}
K_{\t,\bP,\rr}^{(h)}(\xx_{v_0})\;.\Eq(3.73)$$
 
\*
\sub(3.8)
We want now to iterate the previous procedure, by using equation \equ(3.38),
in order to suitably take into account the non trivial $\RR$ operations in
the vertices $v\not=v_0$. We shall focus our discussion on $\RR V^{(h)}
(\t,\bP,\rr)$, but the following analysis applies also to $\LL V^{(h)}(\t,
\bP,\rr)$ and $\tilde E_{h+1}(\t)$.
 
Let us consider the truncated expectation in the r.h.s. of \equ(3.38) and
let us put $s=s_v$, $P_i\=P_{v_i}\bs Q_{v_i}$. Moreover we order in an
arbitrary way the sets $P_i^\pm\=\{f\in P_i,\s(f)=\pm\}$, we call $f_{ij}^\pm$
their elements and we define $\xx^{(i)}=\cup_{f\in P_i^-}\xx(f)$,
$\yy^{(i)}=\cup_{f\in P_i^+}\xx(f)$, $\xx_{ij}=\xx(f^-_{i,j})$,
$\yy_{ij}=\xx(f^+_{i,j})$. Note that $\sum_{i=1}^s |P_i^-|=\sum_{i=1}^s
|P_i^+|\=n$, otherwise the truncated expectation vanishes. A couple
$l\=(f^-_{ij},f^+_{i'j'})\=(f^-_l,f^+_l)$ will be called a line joining the
fields with labels $f^-_{ij},f^+_{i'j'}$ and $\o$ indices $\o^-_l,\o^+_l$
and connecting the points
$\xx_l\=\xx_{i,j}$ and $\yy_l\=\yy_{i'j'}$, the {\it endpoints} of $l$;
moreover we shall put $m_l\=m(f^-_l)+m(f^+_l)$.
Then, it is well known (see [Le], [BGPS], for example) that, up to a sign,
if $s>1$,
$$\tilde\EE^T_{h}(\tilde\psi^{(h)}(P_1),...,\tilde\psi^{(h)}(P_s))=
\sum_{T}\prod_{l\in T} \bar\dpr_1^{m_l} g^{(h)}_{\o^-_l,\o^+_l}(\xx_l-\yy_l)\int
dP_{T}(\tt) \det G^{h,T}(\tt)\;,\Eq(3.74)$$
where $T$ is a set of lines forming an {\it anchored tree graph} between the
clusters of points $\xx^{(i)}\cup\yy^{(i)}$, that is $T$ is a set of lines,
which becomes a tree graph if one identifies all the points in the same
cluster. Moreover $\tt=\{t_{i,i'}\in [0,1],
1\le i,i' \le s\}$, $dP_{T}(\tt)$ is a probability measure with support
on a set of $\tt$ such that $t_{i,i'}=\uu_i\cdot\uu_{i'}$ for some family of
vectors $\uu_i\in \RRR^s$ of unit norm. Finally $G^{h,T}(\tt)$ is a
$(n-s+1)\times (n-s+1)$ matrix, whose elements are given by
$G^{h,T}_{ij,i'j'}=t_{i,i'} \bar\dpr_1^{m(f^-_{ij})+m(f^+_{i'j'})}
g^{(h)}_{\o^-_l,\o^+_l}(\xx_{ij}-\yy_{i'j'})$ with $(f^-_{ij},
f^+_{i'j'})$ not belonging to $T$.
 
If $s=1$, the sum over $T$ is empty, but we
shall still use equation \equ(3.74), by interpreting the r.h.s. as $1$, if
$P_1$ is empty (which is possible, for $s=1$), and as $\det G^h({\bf 1})$
otherwise.
 
\*
Inserting \equ(3.74) in the r.h.s. of \equ(3.38) (with $v=v_0$) we obtain,
up to a sign,
$$\eqalign{
&\RR V^{(h)}(\t,\bP,\rr)={1\over s_{v_0}!}\sqrt{Z_{h}}^{|P_{v_0}|}
\sum_{T_{v_0}}\int d\xx_{v_0}\int dP_{T_{v_0}}(\tt)
[\RR \tilde\psi^{(\le h)}(P_{v_0})] \;\cdot\cr
&\cdot\;\Big[ \prod_{l\in T_{v_0}} \bar\dpr_1^{m_l}
g^{(h+1)}_{\o^-_l,\o^+_l}(\xx_l-\yy_l) \Big]
\det G^{h+1,T_{v_0}}(\tt)\sqrt{Z_{h+1}\over Z_h}^{|P_{v_0}|}
\prod_{i=1}^{s_{v_0}} [K^{(h+2)}_{v_i}(\xx_{v_i})]\cr}\Eq(3.75)$$
 
Let us now consider the contribution to the r.h.s. of \equ(3.75) of
one of the terms in the representation \equ(3.69) of $\RR \tilde\psi^{(\le h)}
(P_{v_0})$ with $n_1(\a)+n_2(\a)>0$.
For each choice of $T_{v_0}$, we decompose the factors
$d^{n_1(\a)}_L(y-x)$ and $d^{n_2(\a)}_\b(y_0-x_0)$, by using equation
\equ(3.55) and the analogous equation for $d_\b(y_0-x_0)$, with $\xx_0=\xx$,
$\xx_n=\yy$ and the other points $\xx_r$, $r=1,\ldots,n-1$, chosen in the
following way.
 
Let us consider the unique subset $(l_1,\ldots,l_m)$ of $T_{v_0}$, which
selects a path joining the cluster containing $\xx_0$ with the cluster
containing $\xx_n$, if one identifies all the points in the same cluster.
Let $(\bar v_{i-1},\bar v_i)$, $i=1,m$, the couple of
vertices whose clusters of points are joined by $l_i$. We shall put
$\xx_{2i-1}$, $i=1,m$, equal to the endpoint of $l_i$ belonging to
$\xx_{\bar v_{i-1}}$ and $\xx_{2i}$ equal to the endpoint of $l_i$ belonging
to $\xx_{\bar v_i}$. This definition implies that there are two points of
the sequence $\xx_r$, $r=0,\ldots,n=2m+1$,
possibly coinciding, in any set $\xx_{\bar v_i}$, $i=0,\ldots,m$; these two
points are the space-time points of two different fields belonging to $P_{\bar
v_i}$.
Since $n\le 2s_{v_0}-1$, this decomposition will
produce a finite number of different terms ($\le (2s_{v_0}-1)^2$, since
$n_1(\a)+n_2(\a)\le 2$), that we shall distinguish with a label $\a'$
belonging to a set $B_{v_0}$, depending on $\a\in A_{v_0}$ and $T_{v_0}$.
These terms can be described in the following way.
 
Each term is obtained from the one chosen in the r.h.s. of \equ(3.75) by
adding a factor $\exp\{i\p L^{-1}n_1(\a)(x+y)+i\p \b^{-1}n_2(\a)(x_0+y_0)\}$.
Moreover each propagator $g^{(h+1)}_{\o^-_l,\o^+_l}(\xx_l-\yy_l)$ is
multiplied by a factor $d^{b_{\a'}(l)}_{j_{\a'}(l)}(\xx_l,\yy_l)$, where
$d^b_j$, $d=0,1,2$, $j=1,\ldots,m_b$ is a family of functions so defined.
If $b=0$, $m_0=1$ and $d_1^0=1$. If $b=1$, $m_b=2$ and $j$ distinguishes,
up to the sign, the two functions
$$e^{-i\fra{\p}L (x_l+y_l)}d_L(x_l-y_l)\;,\quad
e^{-i\fra{\p}\b (x_{0,l}+y_{0,l})}d_\b(x_{l0}-y_{l0})\;.\Eq(3.76)$$
If $b=2$, $j$ distinguishes the three possibilities, obtained by taking the
product of two factors equal to one of the terms in \equ(3.76).
Finally each one of the vertices
$v_1,\ldots,v_{s_{v_0}}$ is multiplied by a similar factor
$d^{b_{\a'}(v_i)}_{j_{\a'}(v_i)}(\xx_i,\yy_i)$.
 
Note that the definitions were chosen so that
$|d_j^b(\xx,\yy)|\le |\dd(\xx-\yy)|^b$.
Moreover there is the constraint that
$$\sum_{l\in T_{v_0}} b_{\a'}(l) + \sum_{i=1}^{s_{v_0}} b_{\a'}(v_i)
=n_1(\a)+n_2(\a)\;.\Eq(3.77)$$
 
The previous discussion implies that \equ(3.75) can be written in the form
$$\eqalign{
&\RR V^{(h)}(\t,\bP,\rr)={1\over s_{v_0}!}\sqrt{Z_{h}}^{|P_{v_0}|}
\sum_{\a\in A_{v_0}}\sum_{T_{v_0}}\sum_{\a'\in B_{v_0}}
\int d\xx_{v_0} \int dP_{T_{v_0}}(\tt)\;\cdot\cr
&\cdot\; h_\a(\tilde\xx_{P_{v_0}})\Big[\prod_{f\in P_{v_0}}
(\hat\partial_{j_\a(f)}^{q_\a(f)}\psi)^{(\le h)\s(f)}_{\xx_\a(f),\o(f)}\Big]
\Big[\prod_{l\in T_{v_0}} d^{b_{\a'}(l)}_{j_{\a'}(l)}(\xx_l,\yy_l)
\bar\dpr_1^{m_l} g^{(h+1)}_{\o^-_l,\o^+_l}(\xx_l-\yy_l)\Big]\;\cdot\cr
&\cdot\; \det G^{h+1,T_{v_0}}(\tt) \sqrt{Z_{h+1}\over Z_h}^{|P_{v_0}|}
\Big[\prod_{i=1}^{s_{v_0}} d^{b_{\a'}(v_i)}_{j_{\a'}(v_i)}(\xx_i,\yy_i)
K^{(h+2)}_{v_i}(\xx_{v_i})\Big]\;,\cr}\Eq(3.78)$$
where the function $h_\a(\tilde\xx_{P_{v_0}})$ has be redefined in order to
absorb the factor \hfill\break
$\exp\{i\p L^{-1}n_1(\a)(x+y)+i\p\b^{-1}n_2(\a)(x_0+y_0)\}$.
 
\*
\sub(3.9) We are now ready to begin the iteration of the previous
procedure, by considering those among the vertices $v_1,\ldots,v_{s_{v_0}}$,
where the action of $\RR$ is non trivial.
It turns out that we can not simply repeat the arguments used for $v_0$, but
we have to consider some new situations and introduce some new prescriptions,
which will be however sufficient to complete the iteration up to the endpoints,
without any new problem.
 
Let us select a term in the r.h.s. of \equ(3.78) and one of the vertices
immediately following $v_0$, let us say $\bar v$, where the action of $\RR$
is non trivial.
We have to consider a few different cases.
 
\*
\0 A) Suppose that $b(\bar v)=0$ (we shall omit the dependence on $\a$
and $\a'$).
In this case the action of $\RR$ is exploited following
essentially the same procedure as for $v_0$. If $\RR$ is different from the
identity, we move its action on the external fields of $\bar v$, by using
the analogous of \equ(3.69), by taking into account that some of
the external fields of $\bar v$ are internal fields of $v_0$, hence they are
involved in the calculation of the truncated expectation \equ(3.74). This
means that, if $f$ is the label of an internal field with $q(f)>0$, the
corresponding (non trivial) $\hat\dpr_{j(f)}^{q(f)}$ operator acts on the
quantities in the r.h.s. of \equ(3.78), which depend on $f$, that is
$d_{j(l)}^{b(l)}(\xx_l,\yy_l) g^{(h+1)}_{\o^-_l,\o^+_l}(\xx_l-\yy_l)$ or the
matrix elements of
$\det G^{h+1,T_{v_0}}$, which are obtained by contracting the field with
label $f$ with another internal field. For example, if $\xx(f)=\xx_l$ and
$\hat\dpr_{j(f)}^{q(f)}$ is the operator associated with the third term
in the r.h.s. of \equ(3.47), we must substitute
$d_{j(l)}^{b(l)}(\xx_l,\yy_l) \bar\dpr_1^{m_l} g^{(h+1)}_{\o^-_l,\o^+_l}
(\xx_l-\yy_l)$ with
$$\int_0^1 dt \partial_1 [d_{j(l)}^{b(l)}(\xxi(t)-\yy_l)
\bar\dpr_1^{m_l} g^{(h+1)}_{\o^-_l,\o^+_l}(\xxi(t)-\yy_l)]\;,\Eq(3.79)$$
with $\xxi(t)=\xx'+t(\bar\xx_l-\xx')$, for some $\xx'\in\xx_{\bar v}$,
$\bar\xx_l$ being defined in terms of $\xx_l$ as $\bar y$ is defined in terms
of $y$ in \sec(3.5) (that is $\bar\xx_l$ and $\xx_l$ are equivalent
representation of the same point on the space-time torus).
 
There is apparently another problem, related to the possibilities that
the operators $\hat\dpr_{j(f)}^{q(f)}$ related with the action of
$\RR$ on $\bar v$ do not commute with the functions $h_\a$ and the field
variables introduced by the action of $\RR$ on $v_0$.
However, the discussion in \sec(3.4) implies that this can not happen,
because of our prescription for the choice of the localization points.
This argument is of general validity, hence we will not consider anymore
this problem in the following.
 
\*
\0 B) If $b(\bar v)>0$, we shall proceed in a different
way, in order to avoid growing powers of the factors $d_L$ and
$d_\b$, which should produce at the end bad combinatorial factors in the
bounds. We need to distinguish four different cases.
 
\0 B1) If $|P_{\bar v}|=4$, we
do not use the decomposition \equ(3.47) for the field changed by the action of
$\RR$ in a $D^{1,1}$ field, but we simply write it as the sum of the two
terms in the r.h.s. of \equ(3.12) (in some cases the second term does not
really contributes, because the argument of the factor $d_{j(\bar v)}^{
b(\bar v)}$ is the same as the argument of the delta function in the
representation \equ(3.14) of the $\RR$ action, but this is not true in general).
We still get a representation of the form
\equ(3.69) for $[\RR \tilde\psi^{(\le h)}(P_{\bar v})]$, but with the property
that $q(f)=0$ for any $\a\in A_{\bar v}$ and any $f\in P_{\bar v}$. This
procedure works, because we do not need to exploit the regularization property
of $\RR$ in this case, as the following analysis will make clear.
 
\0 B2) If $|P_{\bar v}|=2$, and the $\o$-labels of the external fields
are different, the action of $\RR$, after the insertion of the zero, is indeed
trivial, as explained in \sec(3.6), see \equ(3.57). Hence we
do not make any change in the external fields.
 
\0 B3) If $|P_{\bar v}|=2$, the $\o$-labels of the external fields are equal
and $b(\bar v)=2$, the presence of the factor $d_{j(\bar v)}^{b(\bar v)}$
does not allow to use for the action of $\RR$ on the external fields
the representation \equ(3.69), because that factor depends on the space-time
labels of the external fields.
However, we can use the representation following from
the equations \equ(3.62),\equ(3.63),\equ(3.64), by considering the
different terms in the r.h.s. as different contributions (in any case no
cancellations among such terms are possible).
 
Note that this representation has the same properties of the representation
\equ(3.69) and can be written exactly in the same form, by suitable defining
the various quantities. In particular, it is still true that
$n_1(\a)+n_2(\a)\le 2$.
 
Of course, we have to take also into account that some of
the external fields of $\bar v$ are internal fields of $v_0$, but this can
be done exactly as in item A).
 
\0 B4) Finally, if $|P_{\bar v}|=2$, the $\o$-labels of the external fields
are equal and $b_{\bar v}=1$, we use for the action of $\RR$ on the external
fields the representation following from the equations \equ(3.58) and
\equ(3.59), after writing for the fields $D^{1,3}$ and $D^{1,4}$
the analogous of the decomposition \equ(3.47).
 
\*
The above procedure can be iterated, by decomposing the factors $d_{j(v)}^{
b(v)}$ coming from the previous steps of the iteration
along the spanning tree associated with the clusters $L_v$, up to the
endpoints. The final result can be described in the following way.
 
Let us call a {\it zero} each factor equal to one of the two terms in
\equ(3.76). Each zero produced by the action of $\RR$ on the vertex $v$
is distributed along a tree graph $S_v$ on the set $x_v$, obtained by
putting together an anchored tree graph $T_{\bar v}$ for each non trivial
vertex $\bar v\ge v$ and adding a line for the couple of space-time points
belonging to the set $\xx_{\bar v}$ for each (not local) endpoint
$\bar v\ge v$ with
$h_{\bar v}=2$ of type $\l$ or $u$. At the end we have many terms, which
are characterized, for what concerns the zeros, by a tree graph $T$ on the
set $x_{v_0}$ and not more than two zeros on each line $l\in T$; the very
important fact that there are at most two zeros on each line follows from
the considerations in item B) of \sec(3.9).
 
\*
\sub(3.10)
The final result can be written in the following way:
$$\eqalign{
\RR V^{(h)}(\t,\bP,\rr)&=\sqrt{Z_{h}}^{|P_{v_0}|}
\sum_{T\in {\bf T}} \sum_{\a\in A_T}
\int d\xx_{v_0} W_{\t,\bP,\rr,T,\a}(\xx_{v_0})\;\cdot\cr
&\cdot\; \Big\{ \prod_{f\in P_{v_0}}
[\hat\partial_{j_\a(f)}^{q_\a(f)}\psi]^{(\le h)\s(f)}_{\xx_\a(f),\o(f)}\Big\}
\;,\cr}\Eq(3.80)$$
where
$$\eqalign{
&W_{\t,\bP,\rr,T,\a}(\xx_{v_0})=h_\a(\tilde\xx_{v_0})
\Big[\prod_{v\,\hbox{\ottorm not e.p.}}
\Big(Z_{h_v}/Z_{h_v-1}\Big)^{|P_v|/2}\Big]\;\cdot\cr
&\cdot \Big[\prod_{i=1}^n d_{j_\a(v^*_i)}^{b_\a(v^*_i)}(\xx_i,\yy_i)
K^{h_i}_{v^*_i}(\xx_{v^*_i})\Big]
\Big\{\prod_{v\,\hbox{\ottorm not e.p.}}{1\over s_v!} \int
dP_{T_v}(\tt_v) \;\cdot\cr
&\cdot\; \det G_\a^{h_v,T_v}(\tt_v)
\Big[\prod_{l\in T_v} \hat\partial^{q_\a(f^-_l)}_{j_\a(f^-_l)}
\hat\partial^{q_\a(f^+_l)}_{j_\a(f^+_l)} [d^{b_\a(l)}_{j_\a(l)}(\xx_l,\yy_l)
\bar\dpr_1^{m_l} g^{(h_v)}_{\o^-_l,\o^+_l}(\xx_l-\yy_l)]\Big]\Big\}\;,\cr}
\Eq(3.81)$$
${\bf T}$ is the set of the tree graphs on $\xx_{v_0}$, obtained by
putting together an anchored tree graph $T_v$ for each non trivial vertex $v$
and adding a line (which will be by definition the only element of $T_v$)
for the couple of space-time points belonging to the set
$\xx_v$ for each (not local) endpoint $v$ with $h_v=2$ of type $\l$ or $u$;
$A_T$ is a
set of indices which allows to distinguish the different terms produced by
the non trivial $\RR$ operations and the iterative decomposition of the zeros;
$v^*_1,\ldots,v^*_n$ are the endpoints of $\t$, $f^-_l$ and $f^+_l$ are the
labels of the two fields forming the line $l$, ``e.p.''  is an
abbreviation of ``endpoint''. Moreover
$G_\a^{h_v,T_v}(\tt_v)$ is obtained from the matrix $G^{h_v,T_v}(\tt_v)$,
associated with the vertex $v$ and $T_v$, see \equ(3.74), by substituting
$G^{h_v,T_v}_{ij,i'j'}=t_{v,i,i'} \bar\dpr_1^{m(f^-_{ij})+m(f^+_{i'j'})}
g^{(h_v)}_{\o^-_l,\o^+_l}(\xx_{ij}-\yy_{i'j'})$ with
$$G^{h_v,T_v}_{\a,ij,i'j'}=t_{v,i,i'}
\hat\partial_{j_\a(f^-_{ij})}^{q_\a(f^-_{ij})}
\hat\partial_{j_\a(f^+_{ij})}^{q_\a(f^+_{ij})}
\bar\dpr_1^{m(f^-_{ij})+m(f^+_{i'j'})}
g^{(h_v)}_{\o^-_l,\o^+_l}(\xx_{ij}-\yy_{i'j'})\;.\Eq(3.82)$$
Finally, $\hat\partial_j^q$, $q=0,1,2,3$, $j=1,\ldots,m_q$, is a family of
operators, implicitly defined in the previous sections, which are
dimensionally equivalent to derivatives of order $q$; for each $\a\in
A_T$, there is an operator $\hat\partial_{j_\a(f)}^{q_\a(f)}$ associated
with each $f\in I_{v_0}$.
 
It would be very difficult to give a precise description of the various
contributions to the sum over $A_T$, but fortunately we only need to know
some very general properties, which easily follows from the discussion in the
previous sections.
 
\*
\0 1) There is a constant $C$ such that, $\forall T\in {\bf T}_\t$,
$|A_T|\le C^n$ and,
$\forall\a\in A_T$, $|h_a(\tilde\xx_{v_0})|\le C^n$.
 
\0 2) For any $\a\in A_T$, the following inequality is satisfied
$$\Big[\prod_{f\in I_{v_0}} \g^{h_\a(f) q_\a(f)} \Big]
\Big[\prod_{l\in T} \g^{-h_\a(l) b_\a(l)} \Big] \le
\prod_{v\,\hbox{\ottorm not e.p.}} \g^{-z(P_v)}\;,\Eq(3.83)$$
where $h_\a(f)=h_{v_0}-1$ if $f\in P_{v_0}$, otherwise it is the scale of
the vertex where the field with label $f$ is contracted;
$h_\a(l)=h_v$, if
$l\in T_v$ and
$$z(P_v)=\cases{
1 & if $|P_v|=4$,\cr
1 & if $|P_v|=2$ and $\sum_{f\in P_v} \o(f)\not=0\;,$\cr
2 & if $|P_v|=2$ and $\sum_{f\in P_v} \o(f)=0\;,$\cr
0 & otherwise.\cr}\Eq(3.84)$$
 
\*
\sub(3.11)
In order to prove \equ(3.83), let us suppose first that there is no vertex
with two external fields and equal $\o$ indices;
hence $q_\a(f)\le 1$, $\forall f\in I_{v_0}$,
and $b_\a(l)\le 1$, $\forall l\in T$.
Let us choose $f\in I_{v_0}$, such that $q_\a(f)=1$; by analyzing the
procedure described in \sec(3.8) and \sec(3.9), one can easily see that there
are three vertices $v'<v\le \bar v$ and a line $l\in T_{\bar v}$, such that
 
\*
\0 (i) the field with label $f$ is affected by the action of $\RR$ on the
vertex $v$;
 
\0 (ii) $h_{v'}=h_\a(f)$ and $b_\a(l)=1$;
 
\0 (iii) if $v\le \tilde v < \bar v$ and $\tilde l\in T_{\tilde v}$, then
$b_\a(\tilde l)=0$;
 
\0 (iv) if $v'<\tilde v\le \bar v$ and $f\not=\tilde f\in P_{\tilde v}$, then
$q_\a(\tilde f)=0$.
\*
 
(ii) follows from the definition of $h_\a(f)$ and from the remark
that the zero produced by the action of $\RR$ on $v$ is moved by the process
of distribution of the zeros along $T$ in some vertex $\bar v\ge v$. The
property (iii) characterizes $\bar v$; in fact the procedure described in item
B1) and B2) of \sec(3.9) guarantees that no zero can be produced by the action
of $\RR$ in the vertices between $v$ and $\bar v$, if the zero in $\bar v$
``originated'' from the regularization in $v$.
(iv) follows from the previous remark and from the fact that the action of
$\RR$ is trivial in all the vertices between $v'$ and $v$, see \sec(3.3).
 
The previous considerations imply that we can associate each factor
$\g^{h_\a(f)}$ in the l.h.s. of \equ(3.83) with a factor $\g^{-b_\a(l)}$,
by forming disjoint pairs; with each pair we can associate two vertices
$v'$ and $\bar v$ and the path on $\t$ containing all the vertices $v'<
\tilde v\le \bar v$. Since each vertex with four external fields or two
external fields and different $\o$ indices certainly belongs to one of these
paths, the inequality \equ(3.83) then follows from the trivial identity
$$\g^{-(b_\a(l)-h_\a(f))}=\g^{-(h_{\bar v}-h_{v'})}=
\prod_{v'<\tilde v\le \bar v} \g^{-1}\;.\Eq(3.85)$$
 
In order to complete the proof, we have now to consider also the possibility
that there is some vertex with two external fields and equal $\o$ indices,
where the action of $\RR$ is non trivial. This means that there is some
$f\in I_{v_0}$, such that $q_\a(f)=2$ or even (see B4) in \sec(3.9))
$q_\a(f)=1$, if there is a zero associated with a line of the spanning
tree related with the vertex where $f$ is affected by the regularization.
One can proceed essentially in the same way, but has to consider a few
different situations, since the value of $q_\a(f)$ is not fixed and, if
$q_\a(f)=2$, there are two zeros to associate with a
single factor $\g^{2 h_\a(f)}$ in the l.h.s. of \equ(3.83).
We shall not give the details, which have essentially to formalize the
claim that each order one derivative couples with a order one zero, so
that the corresponding factors in the l.h.s. of \equ(3.83) contribute
a factor $\g^{-1}$ to all vertices between the vertex where the derivative
takes its action and the vertex where the zero is ``sitting''.
 
\*
Let us now introduce, given any set $P\subset I_{v_0}$, the notation
$$q_\a(P)= \sum_{f\in P} q_\a(f)\;,\quad m(P)=\sum_{f\in P} m(f)\;.
\Eq(3.86)$$
Note that, by the remark at the end of \sec(3.2), $m(P_v)=0$ for any
$v\not=v_0$ which is not an endpoint of type $\d_1$ or $\d_2$ and that also
$m(P_{v_0})=0$ for all the terms in the r.h.s. of \equ(3.80).
 
We also define
$$\eqalign{
|\vec v_h|&=\cases{\sup \{|\l|,|\n|\},& if $h=+1$,\cr
\sup \{|\l_h|,|\d_h|,|\n_h|\},& if $h=\le 0$.\cr}\cr
\e_h&=\sup_{h'>h}|\vec v_{h'}|\;.\cr}\Eq(3.87)$$
Moreover, we suppose that the condition \equ(2.117) is satisfied,
so that $h^*\ge 0$. We shall prove the following theorem.
 
\*
\sub(3.12) {\cs Theorem.} {\it Let $h> h^*\ge 0$, with $h^*$ defined by
\equ(2.116). If the bounds \equ(2.98) are satisfied and, for some constants
$c_1$,
$$\sup_{h'>h}\Big|{Z_{h'}\over Z_{h'-1}}\Big|\le e^{c_1\e_h^2},\quad
\sup_{h'>h}\Big|{\s_{h'}\over \s_{h'-1}}\Big|\le e^{c_1\e_h},\Eq(3.88)$$
there exists a constant $\bar\e$ (depending on $c_1$) such that,
if $\e_h\le \bar\e$,
then, for a suitable constant $c_0$, independent of $c_1$, as well as of
$u$, $L$ and $\b$,
$$\eqalign{
\sum_{\t\in \TT_{h,n}} &\sum_{\bP\atop |P_{v_0}|=2m} \sum_{\rr}
\sum_{T\in {\bf T}} \sum_{\a\in A_T\atop q_\a(P_{v_0})=k}
\int d\xx_{v_0} |W_{\t,\bP,\rr,T,\a}(\xx_{v_0})|\le\cr
&\le L\b \g^{-hD_k(P_{v_0})} (c_0\e_h)^n\;,\cr}\Eq(3.89)$$
where
$$D_k(P_{v_0})=-2+m+k\;.\Eq(3.90)$$
Moreover
$$\sum_{\t\in \TT_{h,n}} \left[|n_h(\t)| + |z_h(\t)| + |a_h(\t)|+
|l_h(\t)|\right]\le (c_0\e_h)^n\;,\Eq(3.91)$$
$$\sum_{\t\in \TT_{h,n}} |s_h(\t)| \le |\s_h| (c_0\e_h)^n\;,\Eq(3.92)$$
$$\sum_{\t\in \TT_{h,n}} |\tilde E_{h+1}(\t)| \le \g^{2h}(c_0\e_h)^n\;.
\Eq(3.93)$$
}
 
\*
\sub(3.13) An important role in the proof of Theorem \secc(3.12)
plays the estimation of $\det G_\a^{h_v,T_v}(\tt_v)$, that we shall now
discuss, by referring to \sec(3.8) and \sec(3.10) for the notation.
 From now on $C$ will denote a generic constant independent of $u$,
$L$ and $\b$.
 
Given a vertex $v$ which is not an endpoint
and an anchored tree graph $T_v$ (empty, if $v$ is trivial), we consider the
set of internal fields which do not belong to the any line of $T_v$ and
the corresponding sets $\tilde P^{\s,\o}$ of field labels with $\s(f)=\s$
and $\o(f)=\o$. The sets $\cup_\o \tilde P^{-,\o}$ and
$\cup_\o \tilde P^{+,\o}$ label the rows and the columns, respectively,
of the matrix $G_\a^{h_v,T_v}(\tt_v)$, hence they contain the same number of
elements; however, $|\tilde P^{-,\o}|$ can be different from
$|\tilde P^{+,\o}|$, if $h\le 0$. We introduce an integer
$\r(T_v)$, that we put equal to $1$, if $|\tilde P^{-,\o}|\not=|\tilde
P^{+,\o}|$, equal to $0$ otherwise.
We want to prove that
$$\eqalign{
&|\det G_\a^{h_v,T_v}(\tt_v)|
\le \left( {|\s_{h_v}|\over \g^{h_v}}\right)^{\r(T_v)}
C^{\sum_{i=1}^{s_v}|P_{v_i}|-|P_v|-2(s_v-1)}\;\cdot\cr
&\cdot\;
\g^{{h_v\over 2}\left(\sum_{i=1}^{s_v}|P_{v_i}|-|P_v|-2(s_v-1)\right)}
\g^{h_v  \sum_{i=1}^{s_v}\left[ q_\a(P_{v_i}\bs Q_{v_i})+m(P_{v_i}\bs Q_{v_i})
\right] }\;\cdot\cr
&\cdot\; \g^{-h_v \sum_{l\in T_v}\left[q_\a(f^+_l)+q_\a(f^-_l)+
m(f^+_l)+m(f^-_l)\right]}\;.\cr}\Eq(3.94)$$

In order to prove this inequality, we shall suppose, for simplicity,
that all the operators $\hat\dpr^{q(f)}_{j(f)}$ and $\hat\dpr_1^{m(f)}$
acting on the fields with
field label $f\in\cup_{\s,\o}\tilde P^{\s,\o}$ are equal to the identity.
It is very easy to modify the following argument, in order to prove
that each operator $\hat\dpr^{q(f)}_{j(f)}$ or $\hat\dpr_1^{m(f)}$
gives a contribution to the bound proportional to $\g^{h_v q(f)}$ or
$\g^{h_v m(f)}$, so proving \equ(3.94) in the general case.
 
The proof is based on the well known {\it Gram-Hadamard inequality}, stating
that, if $M$ is a square matrix with elements $M_{ij}$ of the form
$M_{ij}=<A_i,B_j>$, where $A_i$, $B_j$ are vectors in a Hilbert space with
scalar product $<\cdot,\cdot>$, then
$$|\det M|\le \prod_i ||A_i||\cdot ||B_i||\;.\Eq(3.95)$$
where $||\cdot||$ is the norm induced by the scalar product.
 
Let $\HH=\RRR^s\otimes \HH_0$, where $\HH_0$ is the Hilbert space of complex
four dimensional vectors $F(\kk')=(F_1(\kk'),\ldots,F_4(\kk')$), $F_i(\kk')$
being a function on the set $\DD'_{L,\b}$, with scalar product
$$<F,G>=\sum_{i=1}^4 {1\over\b L}\sum_{\kk'} F^*_i(\kk') G_i(\kk')\;.
\Eq(3.96)$$
If $h_v\le 0$, it is easy to verify that
$$G^{h_v,T_v}_{ij,i'j'}=t_{i,i'} g^{(h_v)}_{\o^-_l,\o^+_l}(\xx_{ij}-\yy_{i'j'})
=<\uu_i\otimes A^{(h_v)}_{\xx(f^-_{ij}),\o(f^-_{ij})},
\uu_{i'}\otimes B^{(h_v)}_{\xx(f^+_{i'j'}),\o(f^+_{i'j'})}>\;,\Eq(3.97)$$
where $\uu_i\in \RRR^s$, $i=1,\ldots,s$, are the vectors such that
$t_{i,i'}=\uu_i\cdot\uu_{i'}$, and
$$\eqalign{
A^{(h)}_{\xx,\o}(\kk')&=e^{i\kk'\xx}{\sqrt{\tilde f_h(\kk')}\over
\sqrt{-A_h(\kk')}} \cdot \cases{
(-ik_0+E(k'),0,-i\s_{h-1}(\kk'),0),& if $\o=+1$,\cr
(0,i\s_{h-1}(\kk'),0,\s_{h-1}),& if $\o=-1$,\cr}\cr
B^{(h)}_{\xx,\o}&=e^{i\kk'\yy}{\sqrt{\tilde f_h(\kk')}\over
\sqrt{-A_h(\kk')}} \cdot \cases{
(1,1,0,0),& if $\o=+1$,\cr
(0,0,1,(ik_0-E(k'))/\s_{h-1}),& if $\o=-1$.\cr}\cr}\Eq(3.98)$$
 
Let us now define $n_+=|\tilde P^{-,+}|$, $m_+=|\tilde P^{+,+}|$,
$m=|\tilde P^{-,+}|+|\tilde P^{-,-}|=|\tilde P^{+,+}|+|\tilde P^{+,-}|$;
by using \equ(3.95) and \equ(3.98), it is easy to see, by proceeding
as in \sec(2.7), that, if the conditions \equ(2.98) hold,
$$|\det G_\a^{h_v,T_v}(\tt_v)|\le C^m\g^{h_v n_+}|\s_{h_v}|^{m-n_+}
\left({\g^{h_v}\over|\s_{h_v}|}\right)^{m-m_+}= C^m\g^{h_v m}
\left({|\s_{h_v}|\over\g^{h_v}}\right)^{m_+-n_+}\;.\Eq(3.99)$$
Since $2m=\sum_{i=1}^{s_v}|P_{v_i}|-|P_v|-2(s_v-1)$ and
$\sum_{i=1}^{s_v}q_\a(P_{v_i}/Q_{v_i})
-\sum_{l\in T_v}[q_\a(f^+_l)+q_\a(f^-_l)]=0$, we get the inequality
\equ(3.94), if $m_+\ge n_+$, by using \equ(2.116).
The case $m_+<n_+$ can be treated in a similar
way, by exchanging the definitions of
$A^{(h)}_{\xx,\o}(\kk')$ and $B^{(h)}_{\xx,\o}(\kk')$.
 
\*
\sub(3.14) Proof of Theorem \secc(3.12).
 
By using \equ(3.81) and \equ(3.94) we get
$$\eqalign{
&\int d\xx_{v_0} |W_{\t,\bP,\rr,T,\a}(\xx_{v_0})|\le C^n J_{\t,\bP,\rr,T,\a}
\prod_{v\,\hbox{\ottorm not e.p.}}
\Big\{ \Big(Z_{h_v}/Z_{h_v-1}\Big)^{|P_v|/2}\cdot\cr
&\cdot C^{\sum_{i=1}^{s_v}|P_{v_i}|-|P_v|-2(s_v-1)}
\left( {|\s_{h_v}|\over \g^{h_v}}\right)^{\r(T_v)}
\g^{{h_v\over 2}\left(\sum_{i=1}^{s_v}|P_{v_i}|-|P_v|-2(s_v-1)\right)}
\;\cdot\cr &\cdot\;
\g^{h_v  \sum_{i=1}^{s_v}\left[ q_\a(P_{v_i}\bs Q_{v_i})+m(P_{v_i}\bs Q_{v_i})
\right] } \g^{-h_v \sum_{l\in T_v}\left[q_\a(f^+_l)+q_\a(f^-_l)+
m(f^+_l)+m(f^-_l)\right]}\Big\},\cr}\Eq(3.100)$$
where
$$\eqalign{
J_{\t,\bP,\rr,T,\a}&=\int d\xx_{v_0}
\Big| \Big[ \prod_{i=1}^n d_{j_\a(v^*_i)}^{b_\a(v^*_i)}(\xx_i,\yy_i)
K^{h_i}_{v^*_i}(\xx_{v^*_i})\Big]\cdot\cr
&\cdot \Big\{ \prod_{v\,\hbox{\ottorm not e.p.}} {1\over s_v!}
\Big[\prod_{l\in T_v} \hat\partial^{q_\a(f^-_l)}_{j_\a(f^-_l)}
\hat\partial^{q_\a(f^+_l)}_{j_\a(f^+_l)} [d^{b_\a(l)}_{j_\a(l)}(\xx_l,\yy_l)
\hat\dpr_1^{m_l}g^{(h_v)}_{\o^-_l,\o^+_l}(\xx_l-\yy_l)]\Big]\Big\}\Big|\;.\cr}
\Eq(3.101)$$
 
In \sec(3.15) we will prove that
$$\eqalign{
J_{\t,\bP,\rr,T,\a}&\le C^n L \b (\e_h)^n
\prod_{v\,\hbox{\ottorm not e.p.}} \Big[{1\over s_v!} C^{2(s_v-1)}
\g^{h_v n_\n(v)}
\Big(\prod_{l\in \bar T_v}\big|{\s_{h_v}\over\g^{h_v}}\big|\Big)\;\cdot\cr
&\cdot\;\g^{-h_v\sum_{l\in T_v}b_\a(l)}
\g^{-h_v(s_v-1)}\g^{h_v\sum_{l\in T_v}\left[q_\a(f^+_l)+q_\a(f^-_l)+
m(f^+_l)+m(f^-_l)\right]}\Big]\;,\cr} \Eq(3.102)$$
where $n_\n(v)$ is the number
of vertices of type $\n$ with scale $h_v+1$ and $\bar T_v$ is the
subset of the lines of $T_v$ corresponding to {\it non diagonal}
propagators, that is propagators with different $\o$ indices.
 
It is easy to see that
$$\sum_{v\ \hbox{\ottorm not e.p.}} h_v \sum_{i=1}^{s_v} q_\a(P_{v_i}\bs
Q_{v_i}) +h\ q_\a(P_{v_0})=\sum_{f\in I_{v_0}} h_\a(f) q_\a(f)\Eq(3.103)$$
and, by using also the remark after \equ(3.86), that
$$\eqalign{
&\sum_{\bar v\ge v} \left\{ \fra12 \Big(\sum_{i=1}^{s_{\bar v}}
|P_{\bar v_i}|-|P_{\bar v}|\Big) -2(s_{\bar v-1})+ n_\n(\bar v) +
\sum_{i=1}^{s_{\bar v}} m(P_{\bar v_i}\bs Q_{\bar v_i}) \right\}=\cr
&=\fra12 (|I_v|-|P_v|)+m(I_v\bs P_v)+\sum_{\bar v\ge v}n_\n(\bar v)
-2(n_v-1)=-\fra12 |P_v|+2\;.\cr}\Eq(3.104)$$
 
By inserting \equ(3.102) in \equ(3.100) and using \equ(3.83), \equ(3.103),
\equ(3.104), we find
$$\eqalign{
&\int d\xx_{v_0} |W_{\t,\bP,\rr,T,\a}(\xx_{v_0})|\le
C^n L\b\e_h^n \g^{-h D_k(P_{v_0})}
\prod_{v\in V_2}{|\s_{h_v}|\over\g^{h_v}}\;\cdot\cr
&\cdot\; \prod_{v\,\hbox{\ottorm not e.p.}} \left\{ {1\over s_v!}
C^{\sum_{i=1}^{s_v}|P_{v_i}|-|P_v|}
\Big(Z_{h_v}/Z_{h_v-1}\Big)^{|P_v|/2}
\g^{-[-2+{|P_v|\over 2}+z(P_v)]}\right\}\;,\cr}\Eq(3.105)$$
where $V_2$ is the set of vertices, which are not endpoints, such that
$\r(T_v)+|\tilde T_v|>0$, while the vertices $\bar v>v$
do not enjoy this property.
 
Let us now consider a vertex $v$, which is not an endpoint, such that
$|P_v|=2$ and $\sum_{f\in P_v} \o(f)\not =0$. We want to show that
there is a vertex $\bar v\ge v$, such that $\bar v\in V_2$.
In order to prove this claim, we note that, if $v^*$ is an endpoint, then
$\sum_{f\in P_{v^*}} \s(f)\o(f)=0$, while $\sum_{f\in P_v} \s(f)\o(f)\not=0$.
Since all diagonal propagators join two fields with equal $\o$ indices
and opposite $\s$ indices, given any Feynman graph connecting the endpoints
of the cluster $L_v$, at least one of its lines has to be a non diagonal
propagator, so that at least one of the vertices $\bar v\ge v$ must belong to
$V_2$.
 
Moreover, if $v\in V_2$,
$${|\sigma_{h_v}|\over \g^{h_v}}={|\sigma_h|\over \g^h}
{|\sigma_{h_v}|\over |\sigma_h|}
\g^{h-h_v} \le {|\sigma_h|\over \g^h} \g^{(h-h_v)(1-c_1\e_h)}\le
C\g^{(h-h_v)(1/2)}\;,\Eq(3.106)$$
if $\e_h\le \bar\e$ and $\bar\e\le 1/(2c_1)$. We have
used the second inequality in \equ(3.88) and the definition
\equ(2.116), implying that $|\s_h|\le {a_0\over 4\g}\g^h$, if $h\ge h^*$.
 
It follows that
$$\prod_{v\in V_2}{|\s_{h_v}|\over\g^{h_v}}
\le C^n \prod_{v\,\hbox{\ottorm not e.p.}}
\g^{-{1\over 2}\tilde z(P_v)}\;,\Eq(3.107)$$
where
$$\tilde z(P_v)=\cases{
1 & if $|P_v|=2$ and $\sum_{f\in P_v} \o(f)\not=0\;,$\cr
0 & otherwise,\cr}\Eq(3.108)$$
so that
$$-2+{|P_v|\over 2}+z(P_v)+{\tilde z(P_v)\over 2}\ge \fra12\;,\quad
\forall v\, \hbox{\ottorm not e.p.}\;.\Eq(3.109)$$
Hence \equ(3.105) can be changed in
$$\eqalign{
&\int d\xx_{v_0} |W_{\t,\bP,\rr,T,\a}(\xx_{v_0})|\le
C^n L\b\e_h^n \g^{-h D_k(P_{v_0})}\;\cdot\cr
&\cdot\; \prod_{v\,\hbox{\ottorm not e.p.}} \left\{ {1\over s_v!}
C^{\sum_{i=1}^{s_v}|P_{v_i}|-|P_v|}
\Big(Z_{h_v}/Z_{h_v-1}\Big)^{|P_v|/2}
\g^{-[-2+{|P_v|\over 2}+z(P_v)+{\tilde z(P_v)\over 2}]}
\right\}\;,\cr}\Eq(3.110)$$
 
In order to complete the proof of the bound \equ(3.89), we have to perform the
sums in the r.h.s. of \equ(3.89). The number of unlabeled trees is $\le
4^n$; fixed an unlabeled tree, the number of terms in the sum over the
various labels of the tree is bounded by $C^n$, except the sums over the scale
labels and the sets $\bP$. The number of addenda in the sums over $\a$ and
$\rr$ is again bounded by $C^n$, since the action of $\RR$ can be non trivial
at most two times between two consecutive non trivial vertices (see \sec(3.3))
and the number of non trivial vertices is of order $n$.
 
Regarding the sum over $T$, it is empty if $s_v=1$. If $s_v>1$ and
$N_{v_i}\=|P_{v_i}|-|Q_{v_i}|$,
the number of anchored trees with $d_i$ lines branching from the vertex
$v_i$ can be bounded, by using Caley's formula, by
$${(s_v-2)!\over (d_1-1)!...(d_{s_v}-1)!} N_{v_1}^{d_1}...
N_{v_{s_v}}^{d_{s_v}}\;;$$
hence the number of addenda in $\sum_{T\in {\bf T}}$ is bounded by
$\prod_{v\,\hbox{not \ottorm e.p.}} s_v!\;
C^{\sum_{i=1}^{s_v}|P_{v_i}|-|P_v|}$.
 
In order to bound the sums over the scale labels and $\bP$ we first use
the inequality, following from \equ(3.109) and the first inequality in
\equ(3.88), if $c_1\e_h^2\le 1/16$,
$$\eqalign{
&\prod_{v\,\hbox{\ottorm not e.p.}}
\Big(Z_{h_v}/Z_{h_v-1}\Big)^{|P_v|/2}
\g^{-{1\over 2} [-2+{|P_v|\over 2}+z(P_v)+{\tilde z(P_v)\over 2}]}]\le\cr
&\le [\prod_{\tilde v} \g^{-{1\over 40}(h_{\tilde v}-h_{\tilde v'})}]
[\prod_{v\,\hbox{\ottorm not e.p.}}\g^{-{|P_v|\over 40}}]\;,\cr}\Eq(3.111)$$
where $\tilde v$ are the non trivial vertices, and $\tilde v'$ is the
non trivial vertex immediately preceding $\tilde v$ or the root. The
factors $\g^{-{1\over 40}(h_{\tilde v}-h_{\tilde v'})}$ in the r.h.s.
of \equ(3.111) allow to bound the sums over the scale labels by $C^n$.
 
Finally the sum over $\bP$ can be bound by using the following combinatorial
inequality, trivial for $\g$ large enough, but valid for any $\g>1$ (see
[BGPS], \$3). Let $\{p_v, v\in \t\}$ a set of integers such that
$p_v\le \sum_{i=1}^{s_v} p_{v_i}$ for all $v\in\t$ which are not endpoints;
then
$$\prod_{v\,\hbox{\ottorm not e.p.}} \sum_{p_v} \g^{-{p_v\over 40}}
\le C^n\;.\Eq(3.112)$$
It follows that
$$\sum_{\bP\atop |P_{v_0}|=2m}\prod_{v\,\hbox{\ottorm not e.p.}}
\g^{-{|P_v|\over 40}}\le \prod_{v\,\hbox{\ottorm not e.p.}} \sum_{p_v}
\g^{-{p_v\over 40}} \le C^n\;.\Eq(3.113)$$
 
\*
The proof of the bounds \equ(3.91) and \equ(3.93) is very similar. For
the terms contributing to $n_h$ one gets a bound like \equ(3.89), with
$m=1$ and $k=0$, but the factor $\g^{-h D_k(P_{v_0})}=\g^h$ is compensated
by the factor $\g^{-h}$ appearing in the definition of $n_h(\t)$, see
\equ(3.71). For the terms contributing to $z_h$ and $a_h$ $D_k(P_{v_0})=0$
($m=k=1$), as well as for those contributing to $l_h$ ($m=2$, $k=0$).
Finally, for the terms contributing to $\tilde E_{h+1}$, $D_k(P_{v_0})=2$.
For the terms contributing to $s_h$, $D_k(P_{v_0})=-1$, but each term has
also at least one small factor $|\s_h|\g^{-h}$ in its bound, since $|V_2|\ge
1$, see \equ(3.106); so we get the bound \equ(3.92).
 
\*
\sub(3.15) Proof of \equ(3.102).
 
We shall refer to the definitions and the discussion in \sec(3.7) and
\sec(3.9). Let us consider the factor in the r.h.s. of \equ(3.101)
associated with the line $l\in T_v$ and let us suppose that $\xx_l\in
\xx^{(i)}$, $\yy_l\in\xx^{(i')}$. By using \equ(3.47), \equ(3.53) and the
similar expressions for the other difference fields produced by the
regularization, we can write
$$\eqalign{
&\hat\partial^{q_\a(f^-_l)}_{j_\a(f^-_l)}
\hat\partial^{q_\a(f^+_l)}_{j_\a(f^+_l)}[d^{b_\a(l)}_{j_\a(l)}(\xx_l,\yy_l)
\bar\dpr_1^{m_l} g^{(h_v)}_{\o^-_l,\o^+_l}(\xx_l-\yy_l)]=\cr
&=\int_0^1 dt_l\int_0^1 ds_l \tilde\partial^{q_\a(f^-_l)}_{j_\a(f^-_l)}
\tilde\partial^{q_\a(f^+_l)}_{j_\a(f^+_l)}
[d^{b_\a(l)}_{j_\a(l)}(\xx'_l(t_l),\yy'_l(s_l)) \bar\dpr_1^{m_l}
g^{(h_v)}_{\o^-_l,\o^+_l}(\xx'_l(t_l)-\yy'_l(s_l))]\;,\cr}\Eq(3.114)$$
where, depending on $\a$, there are essentially two different possibilities
for the operators $\tilde\partial^{q_\a}_{j_\a}$ and the space-time
points $\xx'_l(t_l)$, $\yy'_l(s_l)$.
Let us consider, for example, $f_l^-$; then the first
possibility is that $\tilde\partial^{q_\a}_{j_\a}$ is a derivative of order
$q_\a$ and
$$\xx'_l(t_l)=\tilde\xx_l+t_l(\bar\xx_l-\tilde\xx_l)\;,\hbox{for some}\
\tilde\xx_l\in\xx^{(i)}\;,\Eq(3.115)$$
$\bar\xx_l$ being defined in terms of $\xx_l$ as $\bar y$ is defined in terms
of $y$ in \sec(3.5) (that is $\bar\xx_l$ and $\xx_l$ are equivalent
representation of the same point on the space-time torus).
The second possibility is that $\tilde\partial^{q_\a}_{j_\a}$
is a local operator of the form $L^{-n_1} \b^{-n_2} \bar\dpr_1^{n_3}
\dpr_0^{n_4}$, with $q_\a\le \sum_{i=1}^4 n_i \le q_\a+1$, and
$\xx'_l(t_l)=\tilde\xx_l\in\xx^{(i)}$. Note that, by \equ(2.40),
$L^{-n_1} \b^{-n_2}\le \g^{h_{L,\b}(n_1+n_2)}\le \g^{h_v(n_1+n_2)}$.
 
By proceeding as in the proof of lemma (2.6) and using \equ(2.105) it is very
easy to show that, for any $N>1$,
$$\eqalign{
&\left| \tilde\partial^{q_\a(f^-_l)}_{j_\a(f^-_l)}
\tilde\partial^{q_\a(f^+_l)}_{j_\a(f^+_l)}
[d^{b_\a(l)}_{j_\a(l)}(\xx'_l(t_l),\yy'_l(s_l)) \bar\dpr_1^{m_l}
g^{(h_v)}_{\o^-_l,\o^+_l}(\xx'_l(t_l)-\yy'_l(s_l))]\right| \le\cr
&\le C {\g^{h_v[1+q_\a(f^+_l)+q_\a(f^-_l)+m(f^-_l)+m(f^+_l)-b_\a(l)]}\over
1+[\g^{h_v}|\dd(\xx'_l(t_l)-\yy'_l(s_l))|]^N}
\Big({|\s_{h_v}|\over \g^{h_v}}\Big)^{\r_l}\;,\cr}\Eq(3.116)$$
where $\dd(\xx)$ is defined in \equ(2.97) and $\r_l=1$ if $\o(f_l^-)\not=
\o(f_l^+)$, $\r_l=0$ otherwise. We used here the fact that, if $h_v=+1$,
then $q_\a(f^-_l)=q_\a(f^+_l)=0$, which allows to avoid the problems connected
with the singularity of the time derivatives of the scale $1$ propagator at
$x'_{l,0}(t_l)-y'_{l,0}(s_l)=0$.
 
Let us now consider the contribution of the endpoints to the r.h.s. of
\equ(3.101) and recall (see \sec(3.10)) that $T_{v^*_i}$ is empty, if
$|\xx_{v^*_i}|=1$, hence $b_\a(v^*_i)=0$,
while, if $\xx_{v^*_i}=(\xx_i,\yy_i)$, $T_{v^*_i}$ contains the line $l_i$
connecting $\xx_i$ with $\yy_i$ and $h_{v^*_i}=2$.
By using \equ(3.33) and \equ(3.39), we get, if $h_i\=h_{v^*_i}$ and
$S_\n\=\{i:v^*_i\ \hbox{is of type}\ \n\}$,
$$\eqalign{
&\left| \Big[ \prod_{i=1}^n d_{j_\a(v^*_i)}^{b_\a(v^*_i)}(\xx_i,\yy_i)
K^{h_i}_{v^*_i}(\xx_{v^*_i})\Big]\right|\le \cr
&\le C^n \e_h^n \prod_{i: |\xx_{v^*_i}|=2} {1\over [1+|\dd(\xx_i-\yy_i)|]^N}
\prod_{i\in S_\n}\g^{(h_i-1)}\;.\cr}\Eq(3.117)$$
 
Let us now remark that, after the insertion of the bounds \equ(3.116) and
\equ(3.117) in the r.h.s. of \equ(3.101), by possibly changing the constant
$C$, we can substitute $\int d\xx_{v_0}$, which is there a shorthand for
$\prod_{\xx\in \xx_{v_0}} \sum_{x\in \L}\int dx_0$, with the real integral
over $(\TTT_{L,\b})^{|\xx_{v_0}|}$, where $\TTT_{L,\b}$ is the space-time
torus $[-L/2,L/2]\times [-\b/2,\b/2]$. Moreover, equation \equ(3.115) can be
thought, and we shall do that, as defining an interval on $\TTT_{L,\b}$,
when $t_l$ spans the interval $[0,1]$; this is possible thanks to the
introduction of the partition \equ(3.42) in \sec(3.5).
 
Hence, in order to complete the proof of \equ(3.102), we have to show that,
fixed a point $\bar\xx\in \xx_{v_0}$, the interpolation parameters associated
with the regularization operations and an integer $N\ge 3$,
$$\int_{\Xi} d(\xx_{v_0}\bs \bar\xx)
\prod_{v\in\t} \prod_{l\in T_v} {1\over
1+[\g^{h_v}|\dd(\xx'_l(t_l)-\yy'_l(s_l))|]^N}\le \prod_{v\in\t} C
\g^{-h_v(s_v-1)}\;,\Eq(3.118)$$
where $\Xi$ denotes the subset of $(\TTT_{L,\b})^{|\xx_{v_0}\bs \bar\xx|}$
satisfying all the constraints associated with the interpolated points of
the form \equ(3.115).
 
Let us call $\tilde T=\cup_v \tilde T_v$, where $\tilde T_v$ is the set of
lines connecting $\xx'_l(t_l)$ with $\yy'_l(s_l)$, for any $l\in T_v$. $\tilde
T$ is not a tree in general; however, for any $v$, $\tilde T_v$ is still an
anchored tree graph between the clusters of points $\xx^{(i)}$, $i=1,\ldots,
s_v$. Hence, the proof of \equ(3.118) becomes trivial, if we can show that
$$d(\xx_{v_0}\bs \bar\xx)=\prod_{l\in \tilde T} d\rr_l\;,\Eq(3.119)$$
where $\rr_l=\xx'_l(t_l)-\yy'_l(s_l)$.
 
In order to prove \equ(3.119), we can proceed, for example, as
in [BM1].
Let us consider first a vertex $v$ with $|T_v|>0$, which is maximal with
respect to the tree order; hence either $v$ is a non local endpoint with
$h_v=2$ or it is a non trivial vertex with no vertex $v'$ with $|T_{v'}|>0$
following it. In this case $\tilde T_v=T_v$, that is no line depends on
the interpolation parameters, and $\tilde T_v$ is a tree on the set $\xx_v$,
so that we get immediately the identity
$$d\xx_v=d\bar\xx^{(v)} \prod_{l\in \tilde T_v} d\rr_l\;,\Eq(3.120)$$
where $\bar\xx^{(v)}$ is an arbitrary point of $\xx_v$.
If we use \equ(3.120) for the family $S_0$ of all maximal vertices with
$|T_v|>0$, we get
$$d\xx_{v_0}=\prod_{v\in S_0} \Big[d\bar\xx^{(v)}\prod_{l\in \tilde T_v}
d\rr_l \Big]\;.\Eq(3.121)$$
Let us now consider a line $\bar l\in\tilde T$, which connects two clusters
of points $\xx_{v_1}$ and $\xx_{v_2}$, with $v_i\in S_0$, $i=1,2$. By
\equ(3.115)
$$\rr_{\bar l}=\xx'_{\bar l}(t_{\bar l})-\yy'_l(s_{\bar l})=
t_{\bar l}\xx_{\bar l}+(1-t_{\bar l})\bar\xx_{\bar l}-\yy'_{\bar l}
(s_{\bar l})\;,\Eq(3.122)$$
implying that
$$\bar\xx^{(v_1)}=\rr_{\bar l}+\bar\xx^{(v_1)}-\rr_{\bar l}=
\rr_{\bar l}+t_{\bar l}(\bar\xx^{(v_1)}-\xx_{\bar l})+
(1-t_{\bar l})(\bar\xx^{(v_1)}-\bar\xx_{\bar l})
+\yy'_{\bar l}(s_{\bar l})\;.\Eq(3.123)$$
Since $\yy'_{\bar l}(s_{\bar l})$ depends only on the variables $\xx_{v_2}$
and $(\bar\xx^{(v_1)}-\xx_{\bar l})$ and $(\bar\xx^{(v_1)}-\bar\xx_{\bar l})$
both depend only on $\{\rr_l,l\in \tilde T_{v_1}\}$, we get
$$\prod_{i=1}^2 \Big[d\bar\xx^{(v_i)} \prod_{l\in \tilde T_{v_i}} d\rr_l\Big]
=d\rr_{\bar l} d\bar\xx^{(v_2)}
\prod_{i=1}^2 \prod_{l\in \tilde T_{v_i}} d\rr_l\;.\Eq(3.124)$$
 
By iterating this procedure, one gets \equ(3.119).
 
\*
\sub(3.16) As we have discussed in \sec(2.13), it is not necessary to perform
the scale decomposition of the Grassmanian integration up to the last scale
$h_{L,\b}$, but we can stop it to the scale $h^*$, defined in \equ(2.116).
Hence, we redefine $\tilde E_{h^*}$, so that
$$e^{-L\b \tilde E_{h^*}} = \int P_{Z_{h^*-1},\s_{h^*-1},
C_{h^*}}(d\psi^{(\le h^*)}) \, e^{-\hat\VV^{(h^*)}
(\sqrt{Z_{h^*-1}}\psi^{(\le h^*)})}\;,\Eq(3.125)$$
implying that
$$E_{L,\b}=\sum_{h=h^*}^1 [\tilde E_h+t_h]\;.\Eq(3.126)$$
 
Thanks to Lemma \secc(2.12), we can proceed as in the proof of Theorem
\secc(3.12) to prove the following Theorem.
 
\*
\sub(3.17)
{\cs Theorem.} {\it There exists a constant $\bar\e$ such that, if $\e_{h^*}
\le\bar\e$ and, for $h=h^*$, \equ(2.98) holds and the
bounds \equ(3.88) are satisfied, then
$$\sum_{\t\in \TT_{h^*-1,n}} |\tilde E_{h^*}(\t)| \le \g^{2h^*}(C\e_{h^*})^n
\;.\Eq(3.127)$$
}

\*
\sub(3.18)
Theorems \secc(3.12) and \secc(3.17), together with \equ(3.126) and
\equ(2.117a), imply that the expansion defining $E_{L,\b}$ is convergent,
uniformly in $L,\b$. With some more work (essentially trivial, but
cumbersome to describe) one can also prove that $\lim_{L,\b\to\io}
E_{L,\b}$ does exist.

\pagina
\vskip1cm
\centerline{\titolo References}
\*
\halign{\hbox to 1.2truecm {[#]\hss} &
        \vtop{\advance\hsize by -1.25 truecm \0#}\cr
A& {I. Affleck: Field theory methods and quantum critical phenomena.
Proc. of Les Houches summer school on Critical phenomena, Random Systems,
Gauge theories, North Holland (1984). }\cr
 
B& {R.J. Baxter:
Eight-Vertex Model in Lattice Statistics.
{\it Phys. Rev. Lett.} {\bf 26}, 832--833 (1971). }\cr
BG& {G. Benfatto, G. Gallavotti:
Perturbation Theory of the Fermi Surface in Quantum Liquid. A General
Quasiparticle Formalism and One-Dimensional Systems.
{\it J. Stat. Phys.} {\bf 59}, 541--664 (1990). }\cr
BGM& {G. Benfatto, G. Gallavotti, V. Mastropietro:
Renormalization Group and the Fermi Surface in the Luttinger Model.
{\it Phys. Rev. B} {\bf 45}, 5468--5480 (1992). }\cr
BGPS& {G. Benfatto, G. Gallavotti, A. Procacci, B. Scoppola:
Beta Functions and Schwinger Functions for a Many Fermions System in One
Dimension.
{\it  Comm. Math. Phys.} {\bf  160}, 93--171 (1994). }\cr
BM1& {F. Bonetto, V. Mastropietro: Beta Function and Anomaly of the Fermi
Surface for a $d=1$ System of Interacting Fermions in a Periodic Potential.
{\it Comm. Math. Phys.} {\bf  172}, 57--93 (1995). }\cr
BM2& {F. Bonetto, V. Mastropietro:
Filled Band Fermi Systems.
{\it Mat. Phys. Elect. Journal} {\bf 2}, 1--43 (1996). }\cr
BeM& {G.Benfatto, V. Mastropietro:
Renormalization group, hidden symmetries and approximate Ward identities
in the $XYZ$ model, II. {\it preprint} (2000). }\cr
EFIK& {F.Essler, H.Frahm, A. Izergin, V.Korepin:
Determinant representation for correlation functions of spin-${1\over 2}$
XXX and XXZ Heisenberg Magnets.
{\it Comm. Math. Phys.} {\bf  174}, 191--214 (1995). }\cr
GS& {G. Gentile, B. Scoppola:
Renormalization group and the ultraviolet
problem in the Luttinger model,
{\it Comm. Meth. Phys.} {\bf 154}, 153--179 (1993). }\cr
JKM& {J.D. Johnson, S. Krinsky, B.M.McCoy:
Vertical-Arrow Correlation Length in the Eight-Vertex Model and the
Low-Lying Excitations of the $XYZ$ Hamiltonian.
{\it Phys. Rev. A} {\bf 8}, 2526--2547 (1973). }\cr
LP& {A. Luther, I.Peschel: Calculation of critical exponents in two
dimensions from quantum field theory in one dimension.
{\it Phys. Rev. B} {\bf 12}, 9, 3908--3917 (1975). }\cr
Le& {A. Lesniewski:
Effective action for the Yukawa 2 quantum field Theory.
{\it Comm. Math. Phys.} {\bf 108}, 437-467 (1987). }\cr
LSM& {E. Lieb, T. Schultz, D. Mattis:
Two Soluble Models of an Antiferromagnetic Chain.
{\it Ann. of Phys.} {\bf 16}, 407--466 (1961). }\cr
LSM1& {E. Lieb, T. Schultz, D. Mattis:
Two-dimensional Ising model as a soluble problem of many fermions.
{\it Journ. Math. Phys.} Rev. Modern Phys. 36, 856--871 (1964). }\cr
M1& {V. Mastropietro: Small denominators and anomalous behaviour in
the Holstein-Hubbard model.
{\it Comm. Math. Phys} 201, 1, 81-115 (1999). }\cr
M2& {V. Mastropietro: Renormalization group for the XYZ model,
Renormalization group for the XYZ model.
{\it Letters in Mathematical physics}, 47, 339-352 (1999). }\cr
Mc& {B.M. McCoy:
Spin Correlation Functions of the $X-Y$ Model.
{\it Phys. Rev.} {\bf 173}, 531--541 (1968). }\cr
MD& {W.Metzner, C Di Castro:
Conservation laws and correlation functions in the Luttinger liquids.
{\it Phys. Rev. B} {\bf 47}, 16107--16123 (1993). }\cr
NO& {J.W. Negele, H. Orland:
Quantum many-particle systems.
Addison-Wesley, New York (1988). }\cr
S& {S.B. Suterland:
Two-Dimensional Hydrogen Bonded Crystals.
{\it J. Math. Phys.} {\bf 11}, 3183--3186 (1970). }\cr
Sp& {H.Spohn:
Bosonization, vicinal surfaces and Hydrodynamic fluctuation theory.
\hfill\break {\it cond-mat/9908381} (1999). }\cr
Spe& {T.Spencer: A mathematical approach to universality in
two dimensions. {\it preprint} (1999). }\cr
YY& {C.N. Yang, C.P. Yang:
One dimensional chain of anisotropic spin-spin interactions, I and II.
{\it Phys. Rev.} {\bf 150}, 321--339 (1966). }\cr
}
\bye